\documentclass{LMCS}

\def\dOi{9(2:24)2013}
\lmcsheading%
{\dOi}
{1--53}
{}
{}
{Apr.~\phantom03, 2012}
{Sep.~24, 2013}
{}

\usepackage{hyperref}
\usepackage{latexsym}
\usepackage{amssymb}  
\usepackage{amsmath}
\usepackage{stmaryrd}
\usepackage{ifthen}
\usepackage{array,boxedminipage}
\usepackage{xcolor}
\usepackage[tight,figbotcap,TABTOPCAP]{subfigure}

\usepackage{csp}
\usepackage{digitlncscom2}

\usepackage{booktabs}
\usepackage{enumerate}
\usepackage{fancybox}
\usepackage{algorithmic}
\usepackage{algorithm}

\usepackage{graphics}
\long\def\beginpgfgraphicnamed#1#2\endpgfgraphicnamed{\includegraphics{#1}}

\def\eg{{e.g.,\ }}
\def\ie{{i.e.,\ }}

\def\st{ \hspace{0.1cm} | \hspace{0.1cm}}

\renewcommand{\defemph}[1]{\textsl{#1}}
\newcommand{\length}{\mathsf{length}}
\newcommand{\TD}{\mathcal{T}^{\Downarrow}}

\def\parA{\smash{\parallel{\!\!\lower5 pt\hbox{$\scriptstyle A$}}}}
\def\parA{\smash{\raise 3 pt\hbox{$\scriptstyle\parallel{\llap{\lower 7 pt\hbox{$\scriptscriptstyle A$}}}$}}}

\let\specoverline\overline
\def \CSP {\mathsf{\overline{CSP}}}
\def \TCSP {\mathsf{CSP}}
\newcommand{\SFS}{\specoverline{\mathsf{SFS}}}

\newcommand{\true}{\mathrm{true}}
\newcommand{\false}{\mathrm{false}}

\def \absfair {\mathit{Fair}}
\def \absmedium {\mathit{Medium}}
\def \abssend {\mathit{Send}}
\def \network {\mathit{Network}}
\def \system {\mathit{System}}
\def \cherr {\mathit{error}}
\def \chin {\mathit{in}}
\def \chout {\mathit{out}}

\newcommand{\semslap}[1]{#1}

\newcommand{\traces}[1]{\mathsf{traces}(#1)}
\newcommand{\inftraces}[1]{\mathsf{traces^{\hspace{0.02cm}\omega}}(#1)}
\newcommand{\alltraces}[1]{\mathsf{traces^{\infty}}(#1)}
\newcommand{\btraces}[1]{\mathsf{traces_\bot}(#1)}
\newcommand{\divergences}[1]{\mathsf{divergences}(#1)}
\def\concat{\cat}
\newcommand{\renam}[1]{[#1]}
\def\ticks{\langle \tick \rangle}
\newcommand{\tup}[1]{{\langle{#1}\rangle}}
\def\restr{ \upharpoonright }
\def\closure{ \hspace{0.1cm} \uparrow }
\newcommand{\powerset}[1]{\mathcal{P}(#1)}
\newcommand{\modulus}[1]{| \hspace{0.07cm}  #1 \hspace{0.07cm} |}
\newcommand{\ol}[1]{#1}
\def\renamomega{\mathrel{R}}

\def \src  {\mathit{src}}
\def \dest {\mathit{dest}}
\newcommand{\thena}[1]{ \overset{#1}{\then} }
\def \gpath  {\mathsf{Path}}
\def \maxscc {\mathit{MaxSCC}}
\def \swedge  {\hspace{0.1cm} \wedge \hspace{0.1cm}}
\def \svee    {\hspace{0.1cm} \vee   \hspace{0.1cm}}
\def \sprel  {\hspace{0.05cm} R \hspace{0.05cm}}
\newcommand{\ucl}[0]{\mathsf{UClosure}}
\newcommand{\dcl}[0]{\mathsf{DClosure}}
\newcommand{\udcl}[0]{\mathsf{UDClosure}}

\newcommand{\mychange}[1]{{#1}}

\newcommand{\subfigureautorefname}{\figureautorefname}
\let\orgautoref\autoref
\renewcommand{\autoref}
{\def\figureautorefname{Figure}%
\def\subfigureautorefname{Figure}%
\def\sectionautorefname{Section}%
\def\subsectionautorefname{Section}%
\def\subsubsectionautorefname{Section}%
\def\tableautorefname{Table}%
\def\appendixautorefname{Appendix}%
\def\propositionautorefname{Proposition}%
\def\propautorefname{Proposition}%
\def\lemmaautorefname{Lemma}%
\orgautoref
}

\theoremstyle{plain}

\begin{document}

\bibliographystyle{plain}

\title[A Static Analysis Framework for Livelock Freedom in
  CSP]{A Static Analysis Framework\\ for Livelock Freedom in
  CSP\rsuper*}
%\thanks{Research in this paper is supported by grants from EPSRC
%    and ONR.}

\author[J.~Ouaknine]{Jo\"el Ouaknine\rsuper a}				%required
\address{{\lsuper{a,c,d}}Department of Computer Science, Oxford University, UK}		%required
\email{\{joel, awr, jbw\}@cs.ox.ac.uk}  						%optional

\author[H.~Palikareva]{Hristina Palikareva\rsuper b}			%optional
\address{{\lsuper b}Department of Computing, Imperial College London, UK}	%optional
\email{h.palikareva@imperial.ac.uk}  %optional
% \thanks{thanks 2, optional.}	%optional

\author[A.~W.~Roscoe]{A. W. Roscoe\rsuper c}	%optional
%\address{Department of Computer Science, Oxford University, UK}	%optional
%\email{awr@cs.ox.ac.uk}  %optional
% \thanks{thanks 3, optional.}	%optional

\author[J.~Worrell]{James Worrell\rsuper d}	%optional
%\address{Department of Computer Science, Oxford University, UK}	%optional
%\email{jbw@cs.ox.ac.uk}  %optional
% \thanks{thanks 4, optional.}	%optional

%% required for running head on odd and even pages, use suitable
%% abbreviations in case of long titles and many authors:

%% mandatory lists of keywords and classifications:

\titlecomment{{\lsuper*}This paper is an extended version of~\cite{OPR11}. This
  work was supported by grants from EPSRC and ONR}

\ACMCCS{[{\bf Theory of computation}]: Semantics and
  reasoning---Program reasoning---Program verification\,/\,Program
  analysis; Logic---Logic and verification; }

\begin{abstract}
In a process algebra with hiding and recursion it is possible to
create processes which compute internally without ever communicating
with their environment. Such processes are said to diverge or
livelock. In this paper we show how it is possible to conservatively
classify processes as livelock-free through a static analysis of their
syntax. In particular, we present a collection of rules, based on the
inductive structure of terms, which guarantee livelock-freedom of the
denoted process. This gives rise to an algorithm which conservatively
flags processes that can potentially livelock. We illustrate our
approach by applying both BDD-based and SAT-based implementations of
our algorithm to a range of benchmarks, and show that our technique in
general substantially outperforms the model checker FDR whilst
exhibiting a low rate of inconclusive results.
\end{abstract}

\maketitle

\section{Introduction}

It is standard in process algebra to distinguish between the visible
and invisible (or silent) actions of a process.  The latter correspond
to state changes arising from internal computations such as resolving of 
nondeterminism, unfolding of a recursion, abstraction of details. 
Their occurrence is \mychange{silent and is} not detectable or controllable by the environment.  A
process is said to \emph{diverge} or \emph{livelock} if it reaches a
state from which it may forever compute internally through an infinite
sequence of invisible actions. This is usually a highly undesirable
feature of the process, described in the literature as ``even worse
than deadlock\mychange{, in that like an endless loop it may consume unbounded computing resources without achieving anything}''~\cite[page~156]{Hoa85}. Livelock invalidates certain
analysis methodologies\mychange{, \eg it signifies lack of progress}, and is often symptomatic of a bug in the
modelling. However the possibility of writing down divergent processes
arises from the presence of two crucial constructs, recursion and
hiding.  The latter converts visible actions into invisible ones and
is a key device for abstraction.

We distinguish two ways in which a process may livelock.  In the
first, a process may be able to communicate an infinite unbroken
sequence of some visible event, and this process then occurs inside
the scope of an operator which hides that event. Alternatively, a
process may livelock owing to the presence of an unguarded recursion.
Roughly speaking, the latter means that the process may recurse
without first communicating a visible action.

This paper is concerned with the problem of determining whether a
process may livelock in the context of the process algebra CSP,
although the principles upon which our analysis is based should be
transferable to other process algebras as well. While it is
straightforward to show that the problem is in general undecidable%
\footnote{For example, CSP can encode counters, and is therefore
  Turing-powerful.}, we are still able to provide a conservative
(i.e., sound but incomplete) method of checking for the possibility of
livelock: this method either correctly asserts that a given process is
livelock-free, or is inconclusive. The algorithm is based on a static
analysis\footnote{Here \emph{static analysis} is used to distinguish
  our approach from the state-space exploration methods that underlie
  model checking or refinement checking.} of the given process,
principally in terms of the interaction of hiding, renaming, and
recursion. This analysis naturally divides into two parts according to
the two sources of livelock outlined above.

The basic intuitions underlying our approach are fairly
straightforward.  In part they mirror the guardedness requirements
which ensure that well-behaved CSP process equations have unique,
livelock-free fixed points~\cite[Chap.~8]{Ros97}.  However, we
extend the treatment of~\cite{Ros97} by allowing guarded recursions to
include instances of the hiding operator. Incidentally, Milner's
notion of guarded recursions in CCS is similarly restricted by the
requirement that variables not occur inside parallel
compositions~\cite{Mil89}. Complications arise mainly because we want
to be able to fully incorporate hiding and renaming in our treatment,
both of which can have subtle indirect effects on guardedness.

We note that the idea of guarded recursions is standard in process
algebra. For instance, in Milner's framework, a variable is `strongly
guarded' in a given term if every free occurrence of the variable in
the term occurs within the scope of a prefixing
operator~\cite{Mil89}. This notion is introduced in order to justify
certain proof principles, such as that guaranteeing the uniqueness of
fixed points up to bisimilarity.  Strong guardedness has also been
extended to a calculus with hiding and action
refinement~\cite{BG02}. A key difference between our approach and
these notions is that we seek to guarantee livelock-freedom, rather
than merely the existence of unique fixed points.
 
In fact, there are few papers which deal with the problem of
guaranteeing livelock-freedom in the setting of concurrent process
calculi.\footnote{In contrast, there are numerous works treating
termination for the $\lambda$-calculus or combinatory
logic~\cite{GLT88,Mit96,Gan80}.} The existing work on livelock-freedom
has mostly been carried out in the context of mobile calculi.
\cite{San02} presents an approach for guaranteeing livelock-freedom
for a certain fragment of the $\pi$-calculus. Unlike the combinatorial
treatment presented here, this approach makes use of the rich theory
of types of the $\pi$-calculus, and in particular the technique of
logical relations. Another study of divergence-freedom in the
$\pi$-calculus appears in~\cite{YBK01}, and uses the notions of graph
types.

Note that CSP is predicated upon \emph{synchronous} 
communication. In terms of livelock analysis, different issues (and
additional difficulties) arise in an asynchronous context (assuming
unbounded communication buffers); see, e.g.,~\cite{LSW06,LSW08}.

Of course, one way to check a process for divergence is to search for
reachable cycles of silent actions in its state space, which is a
labelled transition system built from the operational
semantics. Assuming this graph is finite, this can be achieved by
calculating its strongly connected components\mychange{, using, e.g., Tarjan's algorithm~\cite{CLRS01}}. The latter can be
carried out in time linear in the size of the graph, which may however
be exponential (or worse) in the syntactic size of the term describing
the process. By circumventing the state-space exploration, we obtain a
static analysis algorithm which in practice tends to substantially
outperform state-of-the-art model-checking tools such as FDR---see \autoref{experiments} for experimental comparisons.

Naturally, there is a trade-off between the speed and accuracy of
livelock checking. It is not hard to write down processes which are
livelock-free but which our analysis indicates as potentially
divergent. However, when modelling systems in practice, it makes sense
to try to check for livelock-freedom using a simple and highly
economical static analysis before invoking computationally expensive
state-space exploration algorithms.  Indeed, as
Roscoe~\cite[page~208]{Ros97} points out, the calculations required to
determine if a process diverges are significantly more costly than
those for deciding other aspects of refinement, and it is advantageous
to avoid these calculations if at all possible.

Recent works in which CSP livelock-freedom plays a key role
include~\cite{Dim11} as well as~\cite{Sch1,Sch2}; see also references
within.

\section{CSP\@: Syntax and Conventions}\label{sec_csp_syntax} 

Let $\Sigma$ be a finite set of \emph{events}, with $\tick \notin
\Sigma$.  We write $\Sigma^{\tick\!}$ to denote $\Sigma \union
\set{\tick}$ and $\Sigma^{*\tick\!}$ to denote the set of finite
sequences of elements from $\Sigma$ which may end with $\tick$.  In
the notation below, we have $a \in \Sigma$ and $A \subseteq
\Sigma$. $R$ denotes a binary (renaming) relation on $\Sigma$; its
lifting to $\Sigma^{\tick\!}$ is understood to relate $\tick$ to
itself. The variable $X$ is drawn from a fixed infinite set of process
variables.

CSP terms are constructed according to the following grammar:
\begin{align*}
P & ::=  \STOP \; \mid \; a \then P \; \mid \; \SKIP \; \mid \; P_1
\intchoice P_2 \; \mid \; P_1 \extchoice P_2 \; \mid \; 
                          P_1 \mathrel{\parallel[A]} P_2 \; \mid \\ 
\phass P_1 \semi P_2 \; \mid \;  P \hide A \; \mid \; P[R] \; \mid \; 
       X  \; \mid \; \mu X \centerdot P  \; \mid \; \DIV \enspace .
\end{align*}
$\STOP$ is the deadlocked process. The prefixed process $a \then P$
initially offers to engage in the event $a$, and subsequently behaves
like $P$.  $\SKIP$ represents successful termination, and is willing
to communicate $\tick$ at any time.  $P \extchoice Q$ denotes the
external choice of $P$ and $Q$, whereas $P \intchoice Q$ denotes the
internal (or nondeterministic) alternative. The distinction is
orthogonal to our concerns, and indeed both choice operators behave
identically over our denotational model. The parallel composition $P_1
\parallel[A] P_2$ requires $P_1$ and $P_2$ to synchronise
\mychange{(\ie handshake)} on all events in $A$, and to behave
independently of each other with respect to all other events.  $P
\semi Q$ is the sequential composition of $P$ and $Q$: it denotes a
process which behaves like $P$ until $P$ chooses to terminate
(silently), at which point the process seamlessly starts to behave
like $Q$. $P \hide A$ is a process which behaves like $P$ but with all
communications in the set $A$ hidden.  The renamed process $P[R]$
derives its behaviours from those of $P$ in that, whenever $P$ can
perform an event $a$, $P[R]$ can engage in any event $b$ such that $a
\mathrel{R} b$. To understand the meaning of $\mu X \centerdot P$,
consider the equation $X = P$, in terms of the unknown $X$. While this
equation may have several solutions, it always has a unique
least\footnote{The relevant partial order is defined in
  \autoref{OperationalandDenotationalSemantics}.} such, written $\mu X
\centerdot P$. Moreover, as it turns out, if $\mu X \centerdot P$ is
livelock-free then the equation $X = P$ has no other
solutions. Lastly, the process $\DIV$ represents livelock, i.e., a
process caught in an infinite loop of silent events.

A CSP term is \emph{closed} if every occurrence of a variable $X$ in
it occurs within the scope of a $\mu X$ operator; we refer to such
terms as \emph{processes}. We denote by $\CSP$ the set of all CSP processes and by $\TCSP$ the set of all CSP terms, both open and closed.

Let us state a few conventions. When hiding a single event $a$, we
write $P \hide a$ rather than $P \hide \set{a}$. For $R$ a renaming relation on $\Sigma$ and $U \subseteq \Sigma$, we denote by $R(U)$ the set $\set{y \st \exists x \in U \exst x \mathrel{R} y}$. 
The binding scope of the $\mu X$ operator extends as far to the right as possible. We also
often express recursions by means of the equational notation $X =
P$, rather than the functional $\mu X \centerdot P$.

Let us also remark that CSP processes are often defined via
\emph{vectors} of mutually recursive equations. These can always be
converted to our present syntax, thanks to Beki\v{c}'s
theorem~\cite[Chap.~10]{Win93}.\footnote{Our rules for livelock
  detection require that processes be defined using the fixed-point
  operator $\mu$, as opposed to systems of mutually recursive process
  definitions. Beki\v{c}'s theorem expresses fixed points of self-maps
  on the product space $X \times Y$ in terms of fixed points of
  self-maps on the respective components $X$ and $Y$.  For example,
  consider a mutually recursive process definition of the form $P =
  f(P,Q)$, $Q=g(P,Q)$. The idea is first to define a parameterised
  fixed point of $g$ via the expression $\mu Y.g(X,Y)$, and then
  substitute in the expression for $P$, yielding $P = \mu X. f(X,\mu
  Y.g(X,Y))$.  This process can be generalised to transform mutually
  recursive definitions of arbitrary dimension into expressions using
  only the single-variable fixed-point operator $\mu$.} Accordingly,
we shall freely make use of the vectorised notation in this paper,
viewed as syntactic sugar.

\section{Operational and Denotational Semantics}
\label{OperationalandDenotationalSemantics}

We present congruent (equivalent) operational and denotational
semantics for CSP\@. For reasons of space, \mychange{some} details and clauses
are omitted. An
extensive treatment of a variety of different CSP models can also be
found in~\cite{Ros97,Ros11}. The semantics presented below only
distill those ideas from~\cite{Ros97,Ros11} that are relevant in our
setting.

\subsection{\mychange{Operational semantics}}

The operational semantics is presented as a list of inference rules in
SOS form.
% ; we only give below rules for prefixing, recursion, parallel composition, and hiding.
%(cf.~\cite{AFV01}).
In what follows, $a$ stands for a visible event, i.e., belongs to
$\Sigma^{\tick\!}$. $A \subseteq \Sigma$ and $A^{\tick\!} = A \union
\set{\tick}$. $\gamma$ can be a visible event or a silent one ($\gamma
\in \Sigma^{\tick\!} \union \set{\tau}$). $P \progress{\gamma} P'$
means that $P$ can perform an immediate and instantaneous
$\gamma$-transition, and subsequently become $P'$ (communicating
$\gamma$ in the process if $\gamma$ is a visible event). If $P$ is a
term with a single free variable $X$ and $Q$ is a process, $[Q/X]P$
represents the process $P$ with $Q$ substituted for every free
occurrence of $X$.

{\allowdisplaybreaks
\begin{gather*}
%% Note -- no rules for STOP
 \ded{}{(a \then P) \progress{a} P}{}  \qquad
 \ded{}{\SKIP \progress{\tick} \STOP}{}                        \\
 \ded{}{P_1 \intchoice P_2 \progress{\tau} P_1}{} 
        \qquad
   \ded{}{P_1 \intchoice P_2 \progress{\tau} P_2}{}                \\[0.8ex] 
\ded{P_1 \progress{\tau} P_1'}{P_1 \extchoice P_2 \progress{\tau} P_1'
\extchoice P_2}{} \qquad
\ded{P_2 \progress{\tau} P_2'}{P_1 \extchoice P_2 \progress{\tau} P_1
\extchoice P'_2}{} \\[0.3ex]
\ded{P_1 \progress{a} P_1'}{P_1 \extchoice P_2 \progress{a} P_1'}{}
 \qquad
\ded{P_2 \progress{a} P_2'}{P_1 \extchoice P_2 \progress{a} P'_2}{} \\
\ded{P_1 \progress{\gamma} P'_1}
  {P_1 \parallel[A] P_2 \progress{\gamma} P'_1 \parallel[A] P_2}
  {\gamma \notin A^{\tick\!}}  \qquad 
 \ded{P_2 \progress{\gamma} P'_2}
  {P_1 \parallel[A] P_2 \progress{\gamma} P_1 \parallel[A] P'_2}
  {\gamma \notin A^{\tick\!}}                                        \\
 \ded{P_1 \progress{a} P'_1 \quad P_2 \progress{a} P'_2}
  {P_1 \parallel[A] P_2 \progress{a} P'_1 \parallel[A] P'_2}
  {a \in A^{\tick\!}}                                                   \\
 \ded{P_1 \progress{\tick} P'_1}
  {P_1 \semi P_2 \progress{\tau} P_2}{}                      \qquad
 \ded{P_1 \progress{\gamma} P'_1}
  {P_1 \semi P_2 \progress{\gamma} P'_1 \semi P_2}{\gamma \neq \tick}   \\[0.8ex]
 \ded{P \progress{a} P'}{P \hide A \progress{\tau} 
        P' \hide A}{a \in A}                                 \qquad
 \ded{P \progress{\gamma} P'}
  {P \hide A \progress{\gamma} P' \hide A}{\gamma \notin A}             \\
 \ded{P \progress{\tau} P'}
  {P[R] \progress{\tau} P'[R]}{}                      \qquad
 \ded{P \progress{a} P'}{P[R] \progress{b} P'[R]}{a \mathrel{R} b}  \\
 \ded{}{\mu X \centerdot P \progress{\tau} 
              [(\mu X \centerdot P)/X]P}{} \qquad
 \ded{}{\DIV \progress{\tau} \DIV \makebox[0pt]{\hspace*{1.2em}.}}{}
\end{gather*}
}

These rules allow us to associate to any CSP process a labelled
transition system (LTS) representing its possible executions.  We say that a
process \emph{diverges} if it has an infinite path whose actions are
exclusively $\tau$'s. A process is \emph{livelock-free} if it never
reaches a point from which it diverges.

\subsection{\mychange{Denotational semantics}}
\label{denotationalsemantics}

% $(T,D)$, where $T \subseteq \Sigma^{*\tick\!}$ is the set of visible event traces that the process may perform, and $D \subseteq T$ is the set of traces after which it may diverge.
The denotational semantics ascribes to any CSP process a pair
$(\btraces{P}, \divergences{P})$, where $\btraces{P} = \traces{P}
\union \divergences{P} \subseteq \Sigma^{*\tick\!}$ is the set of
all finite visible-event traces that $P$ may perform, and
$\divergences{P} \subseteq \btraces{P}$ is the set of traces after
which it may diverge.\footnote{Standard models of CSP also take
  account of the liveness properties of a process by modelling its
  \emph{refusals}, i.e., the sets of events it cannot perform after a
  given trace.  However, this information is orthogonal to our
  concerns: the divergences of a process are independent of its
  refusals---see~\cite[Section~8.4]{Ros97}.}
Following~\cite{Ros11}, we write $\TD$ for the set of pairs $(T,D) \in
\mathcal{P}(\Sigma^{*\tick\!}) \times \mathcal{P}(\Sigma^{*\tick\!})$
satisfying the following axioms (where $\cat$ denotes trace
concatenation):
\begin{enumerate}[(1)]
\item $D \subseteq T$.
\item $s \cat \langle \tick \rangle \in D$ implies $s \in D$.
\item $T \subseteq \Sigma^{*\tick\!}$ is non-empty and prefix-closed.
\item $s \in D \cap \Sigma^*$ and $t \in \Sigma^{*\tick\!}$ implies
$s\cat t \in D$.
\end{enumerate}

Axiom 4 says that the set of divergences is postfix-closed. Indeed,
since we are only interested in \emph{detecting} divergence, we treat
it as catastrophic and do not attempt to record any meaningful
information past a point from which a process may diverge;
accordingly, our semantic model takes the view that a process may
perform \emph{any} sequence of events after divergence. Thus the only
reliable behaviours of a process are those in $T - D$.

Axiom 2 reflects the intuition that $\tick$ represents successful
termination.  In particular, there is no way a process may diverge
after a $\tick$ unless it is already divergent.

Given a process $P$, its denotation $\bracket{P} =
(\btraces{P},\divergences{P}) \in \TD$ is calculated by induction on
the structure of $P$; in other words, the model $\TD$ is
compositional. The complete list of clauses can be found
in~\cite[Chap.~8]{Ros97}, and moreover the traces and divergences of a
process may also be extracted from the operational semantics in
straightforward fashion. We provide the inductive rules in
Figures~\ref{fig_static_rules_traces} and \ref{fig_static_rules_divs}
to facilitate the proofs.  In the last three rules in
\autoref{fig_static_rules_divs}, $r$ ranges over $\Sigma^{*\tick\!}$,
in accordance with Axiom~4. The lifting of the renaming relation $R$
to traces is carried out element-wise. The precise definition of $s
\parallel[A] t$ in the rule for parallel composition is presented in
\autoref{fig_static_s_parallel_t}~\cite{Ros97}.
% The precise definition of $s \parallel[A] t$ in the rule for parallel composition can be found in ~\cite[page~70]{Ros97}.

\begin{figure}
\centering
\begin{boxedminipage}{10em}
{\allowdisplaybreaks
 \begin{align*}
 \traces{\STOP} \eq \set{ \tup{ } }  \\ 
 \traces{\SKIP} \eq \set{ \tup{ }, \tup{\tick}  } \\
 \traces{\DIV}  \eq \emptyset \\
 \traces{a \then P} \eq \set{ \tup{ } } \union \set{ \tup{a} \concat t   \st t \in \traces{P}  } \\
 \traces{P \extchoice Q} \eq \traces{P} \union \traces{Q} \\
 \traces{P \intchoice Q} \eq \traces{P} \union \traces{Q} \\
 \traces{P \semi Q} \eq (\traces{P} \inter \Sigma^{\ast}) \union \set{ t \concat s \st t \concat \tup{\tick} \in \traces{P}, s \in \traces{Q} } \\
 \traces{P \hide A} \eq \set{ t \restr (\Sigma \setminus A) \st t \in \traces{P} } \\
 \traces{P \renam{R} } \eq \set{ t \st \exists s \in \traces{P} \exst s \mathrel{R} t} \\
 \traces{P \parallel[A] Q} \eq \Union \, \set{ s \parallel[A] t \st s \in \traces{P}, t \in \traces{Q} } 
 \end{align*}
}
\end{boxedminipage}
\caption{The model $\TD$: inductive rules for calculating traces.}  \label{fig_static_rules_traces}
\end{figure}

\begin{figure}
\centering
\begin{boxedminipage}{14em}
{\allowdisplaybreaks
 \begin{align*}
 \divergences{\STOP} \eq \emptyset  \\ 
 \divergences{\SKIP} \eq \emptyset \\
 \divergences{\DIV}  \eq \Sigma^{\ast \tick} \\
 \divergences{a \then P} \eq  \set{ \tup{a} \concat t  \st t \in \divergences{P}  } \\
 \divergences{P \extchoice Q} \eq \divergences{P} \union \divergences{Q} \\
 \divergences{P \intchoice Q} \eq \divergences{P} \union \divergences{Q} \\
 \divergences{P \semi Q} \eq \divergences{P} \union \set{ t \concat s \st t \concat \tup{\tick} \in \btraces{P}, s \in \divergences{Q} } \\
 \divergences{P \hide A} \eq \set{ (t \restr (\Sigma \setminus A)) \concat r \st t \in \divergences{P} } \, \union \\
			   &\phantom{ \; = \;}  \set{ (u \restr (\Sigma \setminus A)) \concat r \st u \in \Sigma^{\omega}, u \restr (\Sigma \setminus A) \mbox{ finite}, \forall t < u \exst t \in \btraces{P} } \\
 \divergences{P \renam{R} } \eq \set{ t \concat r \st \exists s \in \divergences{P} \inter \Sigma^{\ast} \exst s \mathrel{R} t} \\
 \divergences{P \parallel[A] Q} \eq \set{ u \concat r \st \exists s \in \btraces{P}, \exists t \in \btraces{Q} \exst u \in ( s \parallel[A] t \inter \Sigma^{\ast}), \\
			   &\phantom{ \; = \;} (s \in \divergences{P} \mbox{ or } t \in \divergences{Q}) } 
 \end{align*}
}
\end{boxedminipage}
\caption{The model $\TD$: inductive rules for calculating divergences.}  \label{fig_static_rules_divs}
\end{figure}

\begin{figure}
\centering
\begin{boxedminipage}{14em}
{\allowdisplaybreaks
 \begin{align*} 
  s \parallel[A] t \eq t \parallel[A] s \\
  \tup{ } \parallel[A] \tup{ } \eq \set{ \tup{ } } \\
  \tup{ } \parallel[A] \tup{a} \eq \set{ \tup{ } } \\
  \tup{ } \parallel[A] \tup{b} \eq \set{ \tup{b} } \\ 
  \tup{a} \concat s \parallel[A] \tup{b} \concat t \eq \set{ \tup{b}\concat u \st u \in \tup{a}\concat s \parallel[A] t } \\
  \tup{a} \concat s \parallel[A] \tup{a} \concat t \eq \set{ \tup{a}\concat u \st u \in  s \parallel[A] t } \\  
  \tup{a} \concat s \parallel[A] \tup{a'} \concat t \eq \set{ } \ \ 
         \mbox{if $a \neq a'$} \\
  \tup{b} \concat s \parallel[A] \tup{b'} \concat t \eq \set{ b \concat u \st u \in s \parallel[A] \tup{b'} \concat t } \union      			\set{ b' \concat u \st u \in \tup{b} \concat s  \parallel[A]  t } 
 \end{align*} 
}
\end{boxedminipage}
\caption{Interleaving operator on traces (where $s,t \in \Sigma^{\ast \tick}$, $A \subseteq \Sigma^{\tick}$, $a \in A$, $b \notin A$).}  \label{fig_static_s_parallel_t}
\end{figure}

\begin{defi} 
A process $P$ is \defemph{livelock-free} if $\mathsf{divergences}(P) =
\emptyset$.
\end{defi}

\subsubsection{Reasoning About Infinite Traces}\label{sec_infinite_traces}

In general, reasoning about livelock requires reasoning about infinite
behaviours.  Hiding a set of events $A \subseteq \Sigma$ from a
process $P$ introduces divergence if $P$ is capable of performing an
infinite unbroken sequence of events from $A$.  
Although our model only records the finite traces of a process, the
finitely branching nature of our operators\footnote{All CSP operators
  are finitely branching under the assumptions that the alphabet
  $\Sigma$ is finite and that there is no unbounded
  nondeterminism~\cite{Ros97}.} entails (via K\"onig's lemma) that a
process may perform an infinite trace $u$ if and only if it can
perform all finite prefixes of $u$. In other words, the set of finite
traces of a process conveys enough information for deducing the set of
its infinite traces as well.  To keep the notation
simple, given an infinite trace $u \in \Sigma^\omega$, we will write
$$u \in \inftraces{P} \mbox{ whenever } \{ t \in \Sigma^{\ast} \st t <
u \} \subseteq \traces{P},$$ where $<$ denotes the strong prefix order
on $\Sigma^\infty = \Sigma^\ast \union \Sigma^\omega$. Furthermore, we
will write $\alltraces{P}$ to denote $\traces{P} \union
\inftraces{P}$, the set of all finite and infinite traces of $P$. We
note that traces in $\inftraces{P}$, and hence finite prefixes
thereof, cannot contain $\tick$, which denotes successful
termination. 

We now state the semantic properties we use in case of infinite traces. The proofs for all lemmas can be found in Appendix~\ref{app_infinite_traces}.

\begin{lem}\label{inft_prefix}
  Let $u \in \inftraces{a \then P}$. Then there exists $u' \in \inftraces{P}$ such that $u = \langle a \rangle \concat u'$.
\end{lem}
% Lemma~\ref{app_inft_prefix} in Appendix~\ref{app_infinite_traces}.

\begin{lem}\label{inft_choice}
  Let $u \in \inftraces{P \oplus Q}$ for $\oplus \in \{ \extchoice,
  \intchoice \}$. Then $u \in \inftraces{P}$ or $u \in \inftraces{Q}$.
\end{lem}
% Lemma~\ref{app_inft_choice} in Appendix~\ref{app_infinite_traces}.

\begin{lem}\label{inft_seq_comp}
  Let $u \in \inftraces{P \semi Q}$. Then $u \in \inftraces{P}$ or $u = t \concat u'$ with $t \concat \ticks \in \traces{P}$,  $u' \in \inftraces{Q}$.
\end{lem}
% Lemma~\ref{app_inft_seq_comp} in Appendix~\ref{app_infinite_traces}.

\begin{lem}\label{inft_hide}
  Let $u \in \inftraces{P \hide A}$ and $P \hide A$ be livelock-free. Then there exists $v \in \inftraces{P}$ such that $u = v \restriction (\Sigma \backslash A)$. 
\end{lem}
% Lemma~\ref{app_inft_hide} in Appendix~\ref{app_infinite_traces}.

\begin{lem}\label{inft_rename}
  Let $u \in \inftraces{P \renam{R}}$. Then there exists $v \in \inftraces{P}$ such that $v \mathrel{R} u$. 
\end{lem}
% Lemma~\ref{app_inft_rename} in Appendix~\ref{app_infinite_traces}.

\begin{lem}\label{inft_parallel}
  Let $u \in \inftraces{P \parallel[A] Q}$. Then there exist $u_1 \in \alltraces{P}$ and $u_2 \in \alltraces{Q}$ such that $u \in u_1 \parallel[A] u_2$, and $u_1 \in \Sigma^{\omega}$ or $u_2 \in \Sigma^{\omega}$. 
\end{lem}
% Lemma~\ref{app_inft_parallel} in Appendix~\ref{app_infinite_traces}.

\subsubsection{Handling Recursion}
% \paragraph{\textbf{Handling Recursion.}}
We interpret recursive processes in the standard way by introducing a
partial order $\sqsubseteq$ on $\TD$.  We write $(T_1,D_1) \sqsubseteq
(T_2,D_2)$ if $T_2 \subseteq T_1$ and $D_2 \subseteq D_1$.  In other
words, the order on $\TD$ is reverse inclusion on both the trace and
the divergence components. \mychange{The resulting partial order $(\TD, \sqsubseteq)$ is a complete lattice.}
The bottom element of $(\TD,\sqsubseteq)$
is $(\Sigma^{*\tick\!},\Sigma^{*\tick\!})$, i.e., the denotation of
the immediately divergent process $\DIV$.  
\mychange{The top element is $(\set{ \langle \rangle }, \emptyset)$, i.e., the denotation of the immediately deadlocking process $\STOP$.}
The least upper bound \mychange{and the greatest lower bound}
of a family $\{ (T_i,D_i) \st i \in I \}$ \mychange{are} given by
$\bigsqcup_{i \in I} (T_i,D_i) = (\bigcap_{i \in I} T_i,\bigcap_{i \in I} D_i)$ 
\mychange{ and $\bigsqcap_{i \in I} (T_i,D_i) = (\bigcup_{i \in I} T_i,\bigcup_{i \in I} D_i)$, respectively}. 

It is readily verified that each $n$-ary CSP operator other than
recursion can be interpreted as a Scott-continuous function $(\TD)^n
\rightarrow \TD$. 
The continuity of hiding rests on our assumption
that $\Sigma$ is finite (cf.~\cite[Lemma~8.3.5]{Ros97}). 
By induction we have that any CSP expression $P$ in variables
$X_1,\ldots,X_n$ is interpreted as a Scott-continuous map $(\TD)^n
\rightarrow \TD$.  Recursion is then interpreted using the least fixed
point operator $\mathrm{fix} :[\TD\rightarrow\TD]\rightarrow\TD$. For
instance $\bracket{\mu X \centerdot X}$ is the least fixed point of
the identity function on $\TD$, i.e., the immediately divergent
process.  Our analysis of livelock-freedom is based around an
alternative treatment of fixed points in terms of metric spaces.

\section{\mychange{A Family of Metrics}}\label{sec_family_metrics}

In what follows, we make repeated use of standard definitions and
facts concerning metric spaces. We refer the reader who might be
unfamiliar with this subject matter to the accessible
text~\cite{Sut75}.

Let $F(X)$ be a CSP term with a free variable $X$.
$F$ can be seen as a selfmap of $\TD$. Assume that 
there exists some metric on $\TD$ which is
complete\footnote{\mychange{A metric space $(\TD,d)$ is \emph{complete} if every Cauchy sequence converges.}}
 and under which $F$ is a contraction%
\footnote{A selfmap $F$ on a metric space $(\TD,d)$ is a
\emph{contraction} if there exists a non-negative constant $c < 1$
such that, for any $P,Q \in \TD$, $d(F(P),F(Q)) \leqslant c \cdot
d(P,Q)$. \mychange{Intuitively this means that the distance between any $P,Q \in \TD$ is strictly greater (by some factor) than the distance between their image under $F$, as depicted in \autoref{fig_contr}.}}. 
Then it follows from the Banach fixed point theorem~\cite{Sut75} that
$F$ has a unique (possibly divergent)
fixed point $\mu X \centerdot F(X)$ in $\TD$. \mychange{Furthermore, starting from \emph{any point} in  $\TD$, iterated application of $F$ is guaranteed to converge to this unique fixed point.}

\begin{figure} 
\begin{center}
 \subfigure[ A contractive map \label{fig_contr}] {
     \beginpgfgraphicnamed{contraction}
       \input{contractive.tex}      
      \endpgfgraphicnamed       
 }\hspace{0.5cm}
 \subfigure[A nonexpansive map \label{fig_non_exp}] {
    \beginpgfgraphicnamed{nonexpansion}
       \input{nonexpans.tex}    
    \endpgfgraphicnamed  
 }
\caption{Contractive and nonexpansive maps.} \label{fig_contr_nonexp_maps}
\end{center}
\end{figure}

There may be several such metrics, or none at all. Fortunately, a
class of suitable metrics can be systematically elicited from the sets
of guards of a particular recursion. Roughly speaking, the metrics
that we consider are all variants of the well-known `longest common
prefix' metric on traces\footnote{In this metric the distance between
two traces $s$ and $t$ is the infimum in $[0,1]$ of the set $\set{
2^{-k} \st \mbox{$s$ and $t$ possess a common prefix of length
$k$}}$. \mychange{I.e., the longer prefix two traces share, the closer they are, with the standard lifting to sets of traces and, therefore, to processes.}}, which were first studied by Roscoe in his doctoral
dissertation~\cite{Ros82}, and independently by de Bakker and
Zucker~\cite{dBZ82}. The reason we need to consider such variants is
that hiding fails to be nonexpansive\footnote{\mychange{A selfmap $F$ on a metric space $(\TD,d)$ is 
\emph{nonexpansive} if, for any $P,Q \in \TD$, $d(F(P),F(Q)) \leqslant d(P,Q)$, as illustrated in \autoref{fig_non_exp}.}} in the `longest common prefix'
metric.  For instance, the distance between the traces $\langle a , a,
b \rangle$ and $\langle a , a, c \rangle$ is $\frac{1}{4}$.  However,
after the event $a$ is hidden, the distance becomes $1$.  The
solution, in this particular case, is to change the definition of the
length of a trace by only counting non-$a$ events.  To formalise these
ideas let us introduce a few auxiliary definitions.  These are all
parametric in a given set of events $U \subseteq \Sigma$.

Given a trace $s \in \Sigma^{*\tick\!}$, the $U$-length of $s$,
denoted $\length_U(s)$, is defined to be the number of occurrences of
events from $U$ occurring in $s$.  Given a set of traces $T \subseteq
\Sigma^{*\tick\!}$ and $n \in \mathbb{N}$ the restriction of $T$ to
$U$-length $n$ is defined by
$T \restriction_U n \defs \{ s \in T \st \length_U(s) \leqslant n \}$.
We extend this
restriction operator to act on our semantic domain $\TD$ by defining
$(T,D) \restriction_U n \defs (T',D')$, where
\begin{enumerate}[(1)]
\item $D' = D \cup \{s \cat t \st s \in T \inter \Sigma^{*} \mbox {
  and } \length_U(s) = n \}$.
\item $T' = D' \union \{ s \in T \st \length_U(s) \leqslant n \}$.
\end{enumerate}
Thus $\semslap{P} \restriction_U n$ denotes a process which behaves
like $P$ until $n$ events from the set $U$ have occurred, after
which it diverges \mychange{unless it has already terminated}. It is the least process which agrees with $P$
on traces with $U$-length no greater than $n$.

We now define a metric $d_U$ on $\TD$ by 
\[ d_U(P,Q) \defs \inf \{ 2^{-n}
\st \semslap{P} \restriction_U n = \semslap{Q} \restriction_U n\} \enspace ,\] 
where the infimum is taken in the interval $[0,1]$.

\mychange{
\begin{prop}\label{prop_um_space}
Let $U \subseteq \Sigma$. Then $(\TD, d_U)$ is an ultrametric space. 
\end{prop}

\proof
It is easy to prove that $(\TD, d_U)$ satisfies the following laws for each $P, Q, R \in \TD$:
\begin{center}
\begin{tabular}{r c l l}
$d_U(P, Q) = 0$	& $\Leftrightarrow$ 	&  $ P = Q $ 		& diagonal law \\
$d_U(P, Q) $ 	& $=$ 			& $d_U(Q, P)$ 			& symmetry \\
$d_U(P, Q)$ 	& $\leq$ 		& $d_U(P, R) + d_U(R, Q)$ 	& triangle inequality \\ 
$d_U(P, Q)$	& $\leq$ 		& $\max(d_U(P, R), d_U(R, Q))$& ultrametric inequality  \\
\end{tabular}
\end{center}
The proofs for the first two laws are trivial. Regarding the triangle and ultrametric laws, let us suppose that $d_U(P, R) = 2^{-n}$, $d_U(R, Q) = 2^{-m}$ and $k = \min(n, m)$. Then, $\semslap{P} \restriction_U k = \semslap{R} \restriction_U k = \semslap{Q} \restriction_U k$. Therefore, 
$$ d_u(P, Q) \leq 2^{-k} = \max(d_U(P, R), d_U(R, Q)) \leq d_U(P, R) + d_U(R, Q). $$\qedhere
}

Notice that the function $U \mapsto d_U$ is antitone: if $U \subseteq
V$ then $d_U \geqslant d_V$, \mychange{\ie for any $P,Q \in \TD$, $d_U(P, Q) \geq d_V(P, Q)$.} 
%(where the order on metrics is defined pointwise). 
In particular, the greatest of all the $d_U$ is $d_\emptyset$; this is
the discrete metric on $\TD$.  Furthermore, the least of all the $d_U$
is $d_\Sigma$; this is the standard metric on $\TD$ as defined
in~\cite[Chap.~8]{Ros97}.

\begin{prop}
Let $U \subseteq \Sigma$.  Then $\TD$ equipped with the metric $d_U$
is a complete ultrametric space and the set of livelock-free processes
is a closed subset of $\TD$. Furthermore, if
${F}:{\TD}\rightarrow{\TD}$ is contractive with respect to $d_U$, then
$F$ has a unique fixed point given by $\lim_{n
\rightarrow \infty} F^n(\STOP)$. (Note that this fixed point may be
divergent.)
\end{prop}

\mychange{
\proof By Proposition \ref{prop_um_space}, $(\TD, d_U)$ is an ultrametric space. The proofs that $(\TD, d_U)$ is a complete  metric space and that the set of livelock-free processes is a closed subset of $\TD$ are presented in \autoref{app_family_metrics} (as Proposition \ref{prop_app_complete_um_space} and Proposition \ref{prop_app_set_lf_closed}, respectively).

Let ${F}:{\TD}\rightarrow{\TD}$ be contractive with respect to $d_U$. Since $(\TD, d_U)$ is a complete metric space, it follows from Banach's fixed point theorem~\cite{Sut75,Ros97} that $F$ has a unique fixed point given by $\lim_{n
\rightarrow \infty} F^n(\theta)$, where $\theta$ can be any element of $\TD$ and, in particular, the process $\STOP$. The unique fixed point may or may not be livelock free, however.
\qed
}

In the rest of this paper, the only metrics we are concerned with are
those associated with some subset of $\Sigma$; accordingly, we freely
identify metrics and sets when the context is unambiguous.

\mychange{
\subsection{Nonexpansiveness of CSP operators}\label{sec_static_nonexp}

Let us fix $U \subseteq \Sigma$. The following lemmas prove that each
CSP operator, other than recursion, is at least nonexpansive with
respect to $d_U$ in each of its arguments (for some operators we need
to impose certain conditions). Proofs can be found in
\autoref{app_static_nonexp}.

\begin{lem}\label{lemma_nonexp_in_args}
For any CSP processes $P$, $P'$, $Q$, and $Q'$ the following inequalities hold:
$$ d_{U}(P \extchoice Q, P' \extchoice Q) \leq d_{U}(P, P') \mbox{ and } d_{U}(P \extchoice  Q, P \extchoice Q') \leq d_{U}(Q, Q')$$
$$ d_{U}(P \intchoice Q, P' \intchoice Q) \leq d_{U}(P, P') \mbox{ and } d_{U}(P \intchoice  Q, P \intchoice Q') \leq d_{U}(Q, Q')$$  
$$ d_{U}(P \semi Q, P' \semi Q) \leq d_{U}(P, P') \mbox{ and } d_{U}(P \semi  Q, P \semi Q') \leq d_{U}(Q, Q')$$
$$ \phantom{.} d_{U}(P \parallel[A] Q, P' \parallel[A] Q) \leq d_{U}(P, P') \mbox{ and } d_{U}(P \parallel[A]  Q, P \parallel[A] Q') \leq d_{U}(Q, Q').$$ 

% \proof
% The proofs are given as Lemma~\ref{prop_app__arg_seq_comp}, Lemma~\ref{prop_app__arg_seq_comp_2}, Lemma~\ref{prop_app__arg_nond_ch}, Lemma~\ref{prop_app__arg_det_ch} and Lemma~\ref{prop_app__arg_par_comp} in Appendix~\ref{app_static_nonexp}.
% \qed
\end{lem} 

\begin{lem}\label{lemma_nonexp_prefix}
Let $P$ and $Q$ be CSP processes and let $a \in \Sigma$. Then:
$$ d_{U}(a \then P, a \then Q) \leq d_{U}(P, Q).$$
Furthermore, if $a \in U$, then the inequality is strict. 
% \qed
% $$ d_{U}(a \then P, a \then Q) < d_{U}(P, Q).$$ \qed
\end{lem} 

\begin{lem}\label{lemma_nonexp_hiding}
Let $P$ and $Q$ be CSP processes and let $A \subseteq \Sigma $ satisfy $A \cap U = \emptyset$. Then:
$$ d_{U}(P \hide A, Q \hide A) \leq d_{U}(P, Q).$$ 
% \proof
% The proof is given as Lemma~\ref{prop_app__arg_hiding} in Appendix~\ref{app_static_nonexp}.
% \qed 
\end{lem}

\begin{lem}\label{lemma_nonexp_renaming}
Let $P$ and $Q$ be CSP processes, $R \subseteq \Sigma \times \Sigma$ be a renaming relation on $\Sigma$ and $R(U) = \set{y \st \exists x \in U \exst x \mathrel{R} y}$. Then:
   $$ d_{R(U)}(P [R], Q [R]) \leq d_{U}(P, Q).$$
% \proof
% The proof is given as Lemma~\ref{prop_app__arg_renaming} in Appendix~\ref{app_static_nonexp}.
% \qed
\end{lem}

\begin{lem}\label{lemma_contr_seq}
 Let $P$, $Q$ and $Q'$ be CSP processes. Let $P$ always communicate an event from $U$ before it does a $\tick$. Then:
$$ d_{U}(P \semi  Q, P \semi Q') \leq \frac{1}{2} d_{U}(Q, Q').$$ 
% \proof
% The proof is given as Lemma~\ref{prop_app__arg_seq_comp_2_n_x} in Appendix~\ref{app_static_nonexp}.
% \qed
\end{lem}
}

%The proof is similar to that presented in~\cite{Ros97} and can be
%found in~\cite{slaptr}.

\section{Static Livelock Analysis}
\label{StaticLivelockAnalysis}

We present an algorithm based on a static analysis which
conservatively flags processes that may livelock. In other words, any
process classified as livelock-free really is livelock-free, although
the converse may not hold.  

Divergent behaviours originate in three different ways, two of which
are non-trivial. The first is through direct use of the process
$\DIV$; the second comes from unguarded recursions; and the third is
through hiding an event, or set of events, which the process can
perform infinitely often to the exclusion of all others.

Roscoe~\cite[Chap.~8]{Ros97} addresses the second and third points by
requiring that all recursions be \emph{guarded}, i.e., always perform
some event prior to recursing, and by banning use of the hiding
operator \mychange{under recursion}. Our idea is to extend Roscoe's
requirement that recursions should be guarded by stipulating that one
may never hide \emph{all} the guards. In addition, one may not hide a
set of events which a process is able to perform infinitely often to
the exclusion of all others. This will therefore involve a certain
amount of book-keeping.

\subsection{\mychange{Nonexpansiveness and guardedness}}
We first treat the issue of guardedness of the recursions. Our task
is complicated by the renaming operator, in that a purported guard may
become hidden only after several unwindings of a recursion.  The
following example illustrates some of the ways in which a recursion
may fail to be guarded, and thus diverge.

\begin{exa}\label{ex:unguarded}
Let $\Sigma = \set{a,b,a_0,a_1,\ldots,a_n}$ and let $R = \set{ (
 a_i,a_{i + 1}) \st 0 \leqslant i < n }$ and $S = \set{( a,b), (b,a)}$
 be renaming relations on $\Sigma$.  Consider the following processes.
\begin{enumerate}[(1)]
\item $\mu X\centerdot X$.
\item $\mu X \centerdot a \then ( X \hide a )$.
\item $\mu X \centerdot (a \then ( X \hide b )) \intchoice (b \then ( X
\hide a))$.
\item $\mu X \centerdot (a_0 \then ( X \hide a_n )) 
                               \intchoice (a_0 \then X[R])$.
\item $\mu X \centerdot \SKIP \intchoice a \then (X \semi ( X[S]\hide b) )$.
\end{enumerate}

The first recursion is trivially unguarded.  In the second recursion
the guard $a$ is hidden after the first recursive call.  In the third
process the guard in each summand is hidden in the other summand; this
process will also diverge once it has performed a single event. In
the fourth example we cannot choose a set of guards which is both
stable under the renaming operator and does not contain $a_n$.  This
process, call it $P$, makes the following sequence of visible
transitions:
\[ P \stackrel{a_0}{\longrightarrow} P \hide a_n
\stackrel{a_0}{\longrightarrow} P[R] \hide a_n
\stackrel{a_1}{\longrightarrow} P[R][R] \hide a_n
\stackrel{a_2}{\longrightarrow}
\ldots \stackrel{a_{n-1}}{\longrightarrow} P[R][R]\ldots[R] \hide
{a_n}.\]
But the last process diverges, since $P$ can make an infinite sequence
of $a_0$-transitions which get renamed to $a_n$ by successive
applications of $R$ and are then hidden at the outermost level.

A cursory glance at the last process might suggest that it is guarded
in $\{a\}$.  However, similarly to the previous example, hiding and
renaming conspire to produce divergent behaviour.  In fact the 
process, call it $P$, can make an $a$-transition to $P \semi
(P[S]\hide b)$, and thence to $(P[S]\hide b)[S]\hide b$ via two
$\tau$-transitions.  But this last process can diverge. \qed
\end{exa}

The intuitions underlying our definitions of nonexpansiveness and
guardedness are as follows. Let $U \subseteq \Sigma$ be fixed, giving
rise to a metric $d_U$ on $\TD$, and let $P = P(X)$ be a CSP term with
a single free variable $X$. Then $P$---viewed as a selfmap on $\TD$---is
by definition contractive with respect to $d_U$ (with contraction
factor $1/2$) provided that, for every $T_1, T_2 \in \TD$, it is the
case that 
\begin{equation}
\label{naive_contraction}
d_U(P(T_1), P(T_2)) \leq \frac{1}{2} d_U(T_1,T_2) \,.
\end{equation}

Now if $P$ happens to apply a one-to-one renaming operator $R$ to its
argument, say, then it becomes necessary to rephrase
Equation~\ref{naive_contraction} above as requiring that
\begin{equation}
\label{better_contraction}
d_V(P(T_1), P(T_2)) \leq \frac{1}{2} d_U(T_1,T_2) \, ,
\end{equation}
where $d_V$ is a new metric such that $R(U) = V$. Indeed, since $P$
renames events in $U$ to ones in $V$, the distance between $P(T_1)$
and $P(T_2)$ must be measured with respect to the renamed events,
rather than the original ones.

\begin{figure}
\centering 
 \beginpgfgraphicnamed{contr-UV}
 \input{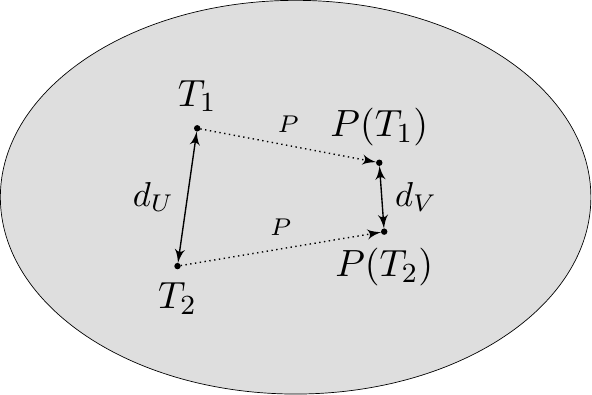}
 \endpgfgraphicnamed
\caption{$P$ is contractive in $(U,V)$, i.e., when the distance
  between inputs is measured with $d_U$ and the distance between
  outputs is measured with $d_V$.}\label{fig_NC_uv}
% \caption{Collecting sets of pairs of metrics}\label{fig_NC_uv}
\end{figure}

This leads us to the concept of a function that is contractive
\emph{with respect to two different metrics $d_U$ and $d_V$}, in which
the first metric is used to measure the distance between two inputs,
whereas the second metric measures the distance between the
corresponding two outputs of the function under
consideration---see~\autoref{fig_NC_uv}.  Following our convention of
identifying sets and metrics, we would say that $P$ is
\emph{contractive in the pair $(U,V)$}.

This reasoning needs to be slightly refined in order to handle
non-injective renamings as well as hiding. Our goal is then to define,
by induction on the structure of CSP terms, a function $\C_X : \TCSP
\longrightarrow \mathcal{P}(\mathcal{P}(\Sigma) \times
\mathcal{P}(\Sigma))$, which associates to each CSP term $P(X)$ a set
of pairs of metrics $(U,V)$ such that
Equation~\ref{better_contraction} holds. Of course, such a definition
would also need to handle terms with several free variables (in
addition to $X$), which can be done using a standard projection.

It turns out that, in order to define such a function $\C_X$, it is
first necessary to compute a function $\N_X : \TCSP \longrightarrow
\mathcal{P}(\mathcal{P}(\Sigma) \times \mathcal{P}(\Sigma))$ which
calculates, for every CSP term $P(X)$, a set of pairs of metrics
$(U,V)$ such that $P$ is nonexpansive in $(U,V)$, following the same
convention of measuring the distance between inputs via the metric
$d_U$ and the distance between outputs via the metric $d_V$.

It is also necessary to calculate an auxiliary function $\G: \TCSP
\longrightarrow \mathcal{P}(\mathcal{P}(\Sigma))$, which itself
depends on a certain function $\F: \TCSP \longrightarrow
\mathcal{P}(\mathcal{P}(\Sigma) \times \mathcal{P}(\Sigma))$. This
may seem problematic, since (as we shall see) $\F$ itself depends on
$\C_X$, but this mutual recursion is well-defined because uses of $\F$
in the definition of $\G$ only occur on subterms, and likewise for
uses of $\G$ in $\C_X$ and uses of $\C_X$ in $\F$.

We provide the intuitions underlying the definitions of $\G$ and $\F$
later on, as these functions are introduced. For now let us finally
remark that all the functions that we define are \emph{conservative
  underapproximations}, i.e., sound, but not necessarily complete. For
example, $\N_X(P)$ as defined below generates some but not necessarily
all of the pairs of metrics that witness the nonexpansiveness of $P$.

%Given a variable $X$ and a CSP term $P = P(X)$, we aim to define
%inductively a collection $\C_X(P)$ of metrics for which $P$ is
%contractive as a function of $X$ (bearing in mind that processes may
%have several free variables). It turns out that it is first necessary
%to identify those metrics in which $P$ is merely nonexpansive as a
%function of $X$, the collection of which we denote
%$\N_X(P)$. Intuitively, the role of $\N_X(P)$ is to keep track of all
%hiding and renaming in $P$. A set $U \subseteq \Sigma$ then induces a
%metric $d_U$ under which $P$ is contractive in $X$ provided $P$ is
%nonexpansive in $U$ and $\mu X \centerdot P$ always communicates an
%event from $U$ prior to recursing.

%The collections of metrics that we produce are conservative, i.e.,
%sound, but not necessarily complete. In other words, we generate some
%but not necessarily all metrics that witness the contractiveness of
%$P$. 

%As the examples above suggest, the calculation is made somewhat
%complicated by the possibility of recursing under hiding and
%renaming. As an application of hiding or renaming may require a switch
%to a different metric (see \eg Lemma \autoref{lemma_nonexp_renaming}),
%% For reasons that will soon become apparent,
%$\N_X(P)$ and $\C_X(P)$ consist of sets of \emph{pairs} of metrics, or
%in other words are identified with subsets of $\mathcal{P}(\Sigma)
%\times \mathcal{P}(\Sigma)$.  As illustrated in \autoref{fig_NC_uv},
%we employ two (possibly different) metrics $d_U$ and $d_V$ for
%measuring the distance between points, respectively, pre and post an
%application of a CSP operator $P$.  

Intuitively, the role of $\N_X(P)$ is to keep track of all hiding and
renaming in $P$. The key property of the function $\N_X$ is given by
the following \mychange{proposition}. 
%Let us recall that we denote by
%$\CSP$ the set of all CSP processes and by $\TCSP$ the set of all CSP
%terms, both open and closed.

%%%%%%%%%%%%%%%%%%%%%%%%%%%%%%%%%%%%%%%%%%%%%%%%%%%%%%%%%%%%%%%%%%%%%%%%%%%%%%5

\begin{prop}\label{prop_non_exp}
 Let $P(X, Y_{1}, \ldots, Y_{n}) = P(X, \overline{Y})$ be a CSP term
 whose free variables are contained within the set $\{ X, Y_{1},
 \ldots, Y_{n} \}$. Let $N_{X} : \TCSP \longrightarrow
 \mathcal{P}(\mathcal{P}(\Sigma) \times \mathcal{P}(\Sigma))$ be
 defined recursively on the structure of $P$ as shown in
 \autoref{fig_nonexp_0}. If $(U, V) \in N_{X}(P)$, then for all
 $T_{1}, T_{2}, \varTheta_{1}, \ldots, \varTheta_{n} \in \TD$, we have
 $d_{V}(P(T_{1}, \overline{\varTheta}),
 P(T_{2}, \overline{\varTheta})) \leq d_{U}(T_{1}, T_{2})$.
\end{prop}

\begin{figure}[h]
\centering 
\begin{boxedminipage}{0.98 \linewidth}
{%\allowdisplaybreaks
{\normalsize
 \begin{align*}
 \N_X(P) \deq \mathcal{P}(\Sigma) \times \mathcal{P}(\Sigma) \ \ 
         \mbox{\textbf{whenever $X$ is not free in $P$; otherwise:}} \\[1ex]
 \N_X(a \then P) \deq \N_X(P)  \\
 \N_X(P_1 \oplus P_2) \deq \N_X(P_1) \inter \N_X(P_2) \ \ 
      \mbox{if $\oplus \in \set{\intchoice, \extchoice, \semi, \parallel[A]}$}\\
 \N_X(P \hide A) \deq \{(U,V) \st (U,V') \in \N_X(P) \land V' \cap A = \emptyset \land V' \subseteq V \} \\
 \N_X(P[R]) \deq \{(U,V) \st (U,V') \in \N_X(P) \land R(V') \subseteq V \} \\
 \N_X(X) \deq \set{(U,V) \st U \subseteq V} \\
 \N_X(\mu Y \centerdot P) \deq \set{(U,V) \st
                (U',V') \in \N_X(P) \land (V',V') \in \N_Y(P) \land U \subseteq U' \land V' \subseteq V } \\
        \pheq \hspace{0.6 \linewidth}  \mbox{if $Y \neq X$} \enspace .
 \end{align*}
}
}
\end{boxedminipage} 
\caption{Nonexpansive sets.}  \label{fig_nonexp_0}
\end{figure}

\proof The proof proceeds by structural induction on $P$ and is presented in \autoref{app_static_analysis}. 
% Proposition~\ref{prop_app_N}

\qed

Note that, by construction, $\N_X(P)$ is always downwards-closed in
its first component and upwards-closed in its second component, which
is sound due to antitoneness (if $U \subseteq U'$ then $d_U \geqslant
d_{U'}$). Some of the rules are plainly straightforward, whereas
others (hiding and especially recursion) require careful thought. An
intuition for correctness is probably best obtained by an examination
of the proof.

%%%%%%%%%%%%%%%%%%%%%%%%%%%%%%%%%%%%%%%%%%%%%%%%%%%%%%%%%%%%%%%%%%%%%%%%%%%%%%%%%%%%%%%%%%%%%%%%%%%%%%%%%%%%%%%55

We now move to the function $\G$. Intuitively, $\G(P) \subseteq
\mathcal{P}(\Sigma)$ lists the `guards' of $\tick$ for $P$. Formally:

\begin{prop}\label{prop_guard_sets}
Let $P(X,Y_1,\ldots,Y_n) = P(X,\overline{Y})$ be a term whose free
variables are contained within the set $\{X,Y_1,\ldots,Y_n\}$. If $V
\in \G(P)$, then, with any processes---and in particular
$\DIV$---substituted for the free variables of $P$, $P$ must
communicate an event from $V$ before it can do a $\tick$.\qed
\end{prop}

The inductive clauses for $\G$ are given in \autoref{fig_guards_0}. As
mentioned earlier, note that these make use of the collection of
\emph{fair sets} $\F(P_i)$ of $P_i$, which is presented later on in
\autoref{section_fair_sets}. The definition is nonetheless
well-founded since $\F$ is here only applied to subterms. The salient
property of $\F(P_i) \neq \emptyset$ is that the process $P_i$ is
guaranteed to be livelock-free. \mychange{The proof for 
Proposition~\ref{prop_guard_sets} proceeds by structural induction
on $P$ and is presented together with Proposition~\ref{prop_app_GCF} in
\autoref{app_static_analysis}.}

\begin{figure}[h!]
\centering 
\begin{boxedminipage}{0.9 \linewidth}%\setlength{\fboxsep}{} 
{\allowdisplaybreaks
{\normalsize
\begin{align*}
\G(\STOP) \deq \mathcal{P}(\Sigma) \\ 
\G(a \then P) \deq \G(P) \union \set{V \st a \in V} \\
\G(\SKIP) \deq \emptyset \\
\G(P_1 \oplus P_2) \deq \G(P_1) \inter \G(P_2) \ \
    \mbox{if $\oplus \in \set{\extchoice, \intchoice}$}\\
\G(P_1 \semi P_2) \deq
     \left\{ \begin{array}{ll}
             \G(P_1) \union \G(P_2) &
                   \mbox{ if $P_1$ is closed and $\F(P_1) \neq \emptyset$}\\
             \G(P_1) & \mbox{ otherwise} 
     \end{array} \right.\\  
\G(P_1 \parallel[A] P_2) \deq
     \left\{ \begin{array}{ll}
             \G(P_1) \union \G(P_2) &
                   \mbox{ if, for $i=1,2$, $P_i$ is closed and 
                                            $\F(P_i) \neq \emptyset$}\\
             \G(P_1) \inter \G(P_2) & \mbox{ otherwise} 
     \end{array} \right.  \\
\G(P \hide A) \deq \left\{ \begin{array}{ll}
             \{ V \st V' \in \G(P) \land V' \inter A = \emptyset \land V' \subseteq V \} 
             & \mbox{ if $P$ is closed and}\\ 
             & \mbox{ $(\emptyset, \Sigma - A) \in \F(P)$}\\
             \emptyset & \mbox{ otherwise} 
      \end{array} \right.  \\
\G(P[R]) \deq \{V \st V' \in \G(P) \land R(V') \subseteq V \} \\
\G(X) \deq \emptyset \\ 
\G(\mu X \centerdot P) \deq \G(P) \enspace .
\end{align*}
}
}
\end{boxedminipage}
\caption{Guard sets.}  \label{fig_guards_0}
\end{figure}

\indent We are now ready to define $\C_X(P) \subseteq
\mathcal{P}(\Sigma) \times \mathcal{P}(\Sigma)$, whose central
property is given by the following proposition.

\begin{prop}\label{prop_contr_sets}
Let $P(X,Y_1,\ldots,Y_n) = P(X,\overline{Y})$ be a term whose free
variables are contained within the set $\{X,Y_1,\ldots,Y_n\}$. 
Let $\C_{X} : \TCSP \longrightarrow \mathcal{P}(\mathcal{P}(\Sigma) \times \mathcal{P}(\Sigma))$ be defined recursively on the structure of $P$ as shown in \autoref{fig_contract_0}.
If
$(U,V) \in \C_X(P)$, then for all $T_1,T_2,\theta_1, \ldots,\theta_n
\in \TD$, we have
$d_V(P(T_1,\overline{\theta}),P(T_2,\overline{\theta})) \leq
\frac{1}{2}d_U(T_1,T_2)$.
\end{prop}

\begin{figure}[h!]
\centering 
\begin{boxedminipage}{0.98 \linewidth}%\setlength{\fboxsep}{} 
{\allowdisplaybreaks
{\normalsize
\begin{align*}
\C_X(P) \deq \mathcal{P}(\Sigma) \times \mathcal{P}(\Sigma) \ \ 
\mbox{\textbf{whenever $X$ is not free in $P$; 
                                              otherwise:}} \\[1ex]
\C_X(a \then P) \deq \C_X(P) \union \set{(U,V) \in \N_X(P) 
                                             \st a \in V} \\
\C_X(P_1 \oplus P_2) \deq \C_X(P_1) \inter \C_X(P_2) \ \
    \mbox{if $\oplus \in \set{\extchoice, \intchoice, \parallel[A]}$}\\
\C_X(P_1 \semi P_2) \deq 
               \C_X(P_1) \cap (\C_X(P_2) \cup
             \set{(U,V) \in \N_X(P_2) \st V \in \G(P_1)}) \\
\C_X(P \hide A) \deq \{ (U,V) \st (U,V') \in \C_X(P) \land
                              V' \cap A = \emptyset \land V' \subseteq V \} \\ 
\C_X(P[R]) \deq \{ (U,V) \st (U,V') \in \C_X(P) \land R(V') \subseteq V \} \\
\C_X(X) \deq \emptyset \\
\C_X(\mu Y \centerdot P) \deq \set{(U,V) \st
               (U',V') \in \C_X(P) \land (V',V') \in \N_Y(P) \land U \subseteq U' \land V' \subseteq V } \\
        \pheq \hspace{0.6 \linewidth}  \mbox{if $Y \neq X$} \enspace .
\end{align*}
}
}
\end{boxedminipage}
\caption{Contractive sets.}  \label{fig_contract_0}
\end{figure}

\proof
  The proof proceeds by structural induction on $P$ and is presented together with Proposition~\ref{prop_app_GCF} in Appendix~\ref{app_static_analysis}.
\qed

\indent Note that contraction guarantees a unique fixed point, albeit
not necessarily a livelock-free one. For instance, $P(X) = (a \then X
\hide b) \extchoice (\mu Y \centerdot b \then Y)$ has a unique fixed
point which can diverge after a single event.

\subsection{Fair sets and hiding}\label{section_fair_sets}

In order to prevent livelock, we must ensure that, whenever a process
can perform an infinite%
\footnote{Recall our understanding that a process can `perform' an
  infinite trace iff it can perform all its finite prefixes.}
unbroken sequence of events from a particular set $A$, then we never
hide the whole of $A$. To this end, we now associate to each CSP term
$P$ a collection of (pairs of) \emph{fair sets} $\F(P) \subseteq
\mathcal{P}(\Sigma) \times \mathcal{P}(\Sigma)$: intuitively, this
allows us to keep track of the events which the process is guaranteed
to perform infinitely often in any infinite execution of $P$. 
As with nonexpansiveness and contractiveness, the potential presence
of renaming and hiding requires us separately to keep track of events
performed by the input processes and the output (or compound) process.

Given a set $W \subseteq \Sigma$, we say that a process is $W$-fair if
any of its infinite traces contains infinitely many events from
$W$. We now have:

\begin{prop}\label{prop_fair_sets}
Let $P(X_1,\ldots,X_n) = P(\overline{X})$ be a CSP term whose free
variables are contained within the set $\{X_1,\ldots,X_n\}$. 
Let $\F : \TCSP \longrightarrow \mathcal{P}(\mathcal{P}(\Sigma) \times \mathcal{P}(\Sigma))$ be defined recursively on the structure of $P$ as shown in \autoref{fig_fairsets_0}.
If $(U,V) \in \F(P)$, then, for any collection of livelock-free, $U$-fair
processes $\theta_1, \ldots, \theta_n \in \TD$, the process
$P(\theta_1, \ldots, \theta_n)$ is livelock-free and $V$-fair.
\end{prop}

\begin{figure}[h!]
\centering 
\begin{boxedminipage}{0.98 \linewidth}%\setlength{\fboxsep}{} 
{\allowdisplaybreaks
{\normalsize
\begin{align*}
\F(\STOP) \deq \mathcal{P}(\Sigma) \times \mathcal{P}(\Sigma) \\
\F(a \then P) \deq \F(P)  \\
\F(\SKIP) \deq \mathcal{P}(\Sigma) \times \mathcal{P}(\Sigma) \\
\F(P_1 \oplus P_2) \deq \F(P_1) \cap \F(P_2) \ \
\mbox{if $\oplus \in \set{\intchoice,\extchoice,\semi}$} \\
\F(P_1 \mathrel{\smash{\parallel[A]}} P_2) \deq 
(\F(P_1) \cap \F(P_2)) \cup \mbox{} \\
\pheq \set{(U_1 \cap U_2, V_1) \st (U_1,V_1) \in \F(P_1) \land
                             (U_2,A) \in \F(P_2)} \cup \mbox{} \\
\pheq \set{(U_1 \cap U_2, V_2) \st (U_2,V_2) \in \F(P_2) \land
                             (U_1,A) \in \F(P_1)} \\
\F(P \hide A) \deq \set{(U,V) \st (U,V') \in \F(P) \land
                               V' \cap A = \emptyset \land V' \subseteq V} \\
\F(P[R]) \deq \set{(U,V) \st (U,V') \in \F(P) \land R(V') \subseteq V }            \\
\F(X) \deq \set{(U,V) \st U \subseteq V} \\
\F(\mu X \centerdot P) \deq 
       \left\{ \begin{array}{ll}
             \set{(U,V) \st (W,W) \in \C_X(P) \cap \F(P) \land U \subseteq W \subseteq V} &
                   \mbox{ if $\mu X \centerdot P$ is open} \\
             \mathcal{P}(\Sigma) \times
             \set{V \st (W,W) \in \C_X(P) \cap \F(P) \land W \subseteq V} &
                   \mbox{ otherwise}\enspace . 
     \end{array} \right.                      
\end{align*}
}
}
\end{boxedminipage}
\caption{Fair sets.}  \label{fig_fairsets_0}
\end{figure}

\proof
   The proof proceeds by structural induction on $P$ and is presented together with Proposition~\ref{prop_app_GCF} in Appendix~\ref{app_static_analysis}.
\qed 

Note that, by construction, $\F(P)$ is
always downwards-closed in its first component and upwards-closed in
its second component; this is sound since if $U \subseteq U'$ and $P$
is $U$-fair, then $P$ is automatically $U'$-fair as well.

\indent We now obtain one of our main results as an immediate
corollary:

\begin{thm}
\label{main}
Let $P$ be a CSP process (i.e., closed term) not containing $\DIV$ in
its syntax. If $\F(P) \neq \emptyset$, then $P$ is livelock-free.
\end{thm}
\mychange{ \proof Let $\F(P) \neq \emptyset$ and $(U, V) \in \F(P)$
  for some $U, V \subseteq \Sigma$. Since $P$ is closed, $P$ has no
  free variables. Then, by Proposition~\ref{prop_fair_sets}, $P$ is
  livelock-free (and $V$-fair).  \qed }

Theorem~\ref{main} gives rise to a procedure for establishing
livelock-freedom of a given process $P$ over alphabet $\Sigma$, whose
complexity is at most quadratic in the syntactic size of $P$ and
exponential in the cardinality of $\Sigma$: indeed, for fixed
$\Sigma$, one computes $\N_X(Q)$, $\G(Q)$, $\C_X(Q)$, and $F(Q)$ for
every variable $X$ appearing in $P$ and every subterm $Q$ of
$P$. Since the number of variables and the number of subterms are both
at most linear in the size of $P$, the computation is at most
quadratic in $P$. On the other hand, each of $\N_X(Q)$, $\C_X(Q)$, and
$F(Q)$ is a collection of pairs of subsets of $\Sigma$, whereas
$\G(Q)$ is a collection of subsets of $\Sigma$. Thus for $\Sigma$ not
fixed, these pieces of data are potentially exponentially large.

In practice, applications often make use of moderately large
alphabets, making the direct set-based approach described above
prohibitively expensive. However, an inspection of the rules defining
$\N_X(Q)$, $\G(Q)$, $\C_X(Q)$, and $F(Q)$ reveals that these objects
can be represented \emph{symbolically}, either as propositional
formulas or as BDDs---further implementation details are provided in
Section~\ref{experiments}. As a result, the problem of deciding
whether $\F(P) \neq \emptyset$ can be seen to lie in NP\@.

\section{Structurally Finite-State Processes}\label{sec_static_SFS}

The techniques developed in Section~\ref{StaticLivelockAnalysis} allow
us to handle the widest range of CSP processes; among others, they
enable one to establish livelock-freedom of numerous infinite-state
processes including examples making use of infinite buffers or
unbounded counters.
%---see~\cite{slaptr} for examples.
Such processes are of
course beyond the reach of explicit-state model checkers such as
FDR\@. In order to create them in CSP, it is necessary to use devices
such as recursing under the parallel operator. In practice, however,
the vast majority of processes tend to be finite state.

Let us therefore define a CSP process to be \emph{structurally finite
  state} if it never syntactically recurses under any of parallel, the
left-hand side of a sequential composition, hiding, or renaming. 

More precisely, we first define a notion of \emph{sequential} CSP
terms: $\STOP$, $\SKIP$, and $X$ are sequential; if $P$ and $Q$ are
sequential, then so are $a \then P$, $P \intchoice Q$, $P \extchoice
Q$, and $\mu X \centerdot P$; and if in addition $P$ is closed, then
$P \semi Q$, $P \hide A$, and $P[R]$ are sequential. Observe that
sequential processes give rise to labelled transition systems of size
linear in the length of their syntax.

Now any closed sequential term is deemed to be structurally finite
state; and if $P$ and $Q$ are structurally finite state, then so are
$a \then P$, $P \intchoice Q$, $P \extchoice Q$, $P \parallel[A] Q$,
$P \semi Q$, $P \hide A$, and $P[R]$. Note that structurally
finite-state CSP terms are always closed, i.e., are processes.
Let us write $\SFS$ to denote the collection of all structurally
finite-state processes.

Whether a given process is structurally finite state can easily be
established by syntactic inspection, for example by using Beki\v{c}'s
theorem~\cite{Win93} (see \autoref{sec_csp_syntax}) and analysing the resulting $\mu$ expression. 
For such processes, it turns out that we can
substantially both simplify and sharpen our livelock analysis. More
precisely, the computation of nonexpansive and contractive data is
circumvented by instead directly examining closed sequential
components in isolation. Furthermore, the absence of free variables in
compound processes makes some of the earlier fairness calculations
unnecessary, thereby allowing more elaborate and finer data to be
computed efficiently, as we now explain.

Let $u$ be an infinite trace over $\Sigma$, and let $F,C \subseteq
\Sigma$ be two sets of events. We say that $u$ is \emph{fair in $F$}
if, for each $a \in F$, $u$ contains infinitely many occurrences of
$a$,\footnote{Note that this notion of `fairness' differs from that
  used in the previous section.} and we say that $u$ is \emph{co-fair
  in $C$} if, for each $b \in C$, $u$ contains at most finitely many
occurrences of $b$.  We lift this to sets of traces in the following
way: let $T \subseteq \Sigma^{\omega}$ be a set of infinite traces
over $\Sigma$, and let $\mathcal{F} = \set{(F_1, C_1),
  \ldots,(F_k,C_k)} \subseteq \mathcal{P}(\Sigma) \times
\mathcal{P}(\Sigma)$ be a collection of pairs of subsets of
$\Sigma$. We say that $T$ is \emph{fair/co-fair in $\mathcal{F}$}
provided that, for every infinite trace $u \in T$, there exists a pair
$(F_i,C_i) \in \mathcal{F}$ such that $u$ is both fair in $F_i$ and
co-fair in $C_i$.

Our aim is the following. Given a structurally finite-state
process $P$, we wish to compute:
\begin{iteMize}{$\bullet$}%\addtolength{\itemsep}{-0.3\baselineskip}
\item  a Boolean-valued \emph{livelock flag} $\delta(P) \in
  \set{\true,\false }$, together with
\item a collection of pairs of disjoint sets
   $\Phi(P) = \set{(F_1, C_1), \ldots, (F_k,C_k)} \subseteq
   \mathcal{P}(\Sigma) \times \mathcal{P}(\Sigma)$, 
\end{iteMize}
 such that if $\delta(P) = \false$, then (i)~$P$ is livelock-free, and 
moreover (ii)~$\inftraces{P}$ is fair/co-fair in
$\Phi(P)$.

\subsection{Handling Sequential $\SFS$ Processes}\label{section_static_atomic_sfs}

For $P$ a sequential $\SFS$ process, let us denote by $\mathit{M_P}$ its
associated labelled transition system as derived from the operational
semantics; let us assume that we construct $\mathit{M_P}$ so that all
states are reachable from the initial state. As noted earlier,
$\mathit{M_P}$ has size linear in the syntactic description of $P$. We
can then compute the livelock flag $\delta(P)$ and the set of
fair/co-fair pairs $\Phi(P) = \set{(F_1, C_1), \ldots,(F_k,C_k)}$
exactly, directly from $\mathit{M_P}$. More precisely, we set
$\delta(P)$ to $\true$ or $\false$ depending on whether or not $P$ can
eventually diverge, \ie whether $\mathit{M_P}$ contains a
$\tau$-cycle. This can be carried out using Tarjan's algorithm in time
linear in the number of states in $\mathit{M_P}$.

Assuming the livelock flag $\delta(P)$ is false, we compute the set of
fair/co-fair pairs $\Phi(P)$ as follows. We add a pair of disjoint
sets of events $(F,C)$ to $\Phi(P)$ if and only if $\mathit{M_P}$
comprises some infinite trace which is fair in $F$ and co-fair in $C$.
Note that if $P$ has no infinite trace, $\Phi(P)$ will therefore be
empty.
%Note that this will guarantee that our abstraction will be
%exact. Let us also observe that we lose nothing by assuming that for
%each pair $(F,C)$, $F \union C = \Sigma$.

It is worth pointing out how the computation of $\Phi(P)$ can be
achieved efficiently. Given a non-empty set $L \subseteq \Sigma$ of
events, we delete all ($\Sigma-L$)-labelled transitions from
$\mathit{M_P}$. If the resulting graph contains a (not necessarily
reachable) strongly connected component which comprises every event in
$L$, we include $(L,\Sigma-L)$ as a fair/co-fair pair for $P$, and
otherwise we do not.

Of course, in actual implementations, it is likely not desirable to
iterate explicitly over all possible subsets of $\Sigma$. The
computation we described can be carried out symbolically using a
Boolean circuit of size polynomial in $P$, using well-known circuit
algorithms for computing the transitive closure of
relations. Consequently, $\Phi(P)$ can be represented symbolically and
compactly either as a BDD or as a propositional formula. Further
implementation details are provided in \autoref{experiments} and
\autoref{app_encoding}.

\subsection{Compositional Rules for $\SFS$ Processes}

% Given a structurally finite-state process $P$, we compute a collection
% of \emph{fair/co-fair} pairs of disjoint sets $\Phi(P) \subseteq
% \mathcal{P}(\Sigma) \times \mathcal{P}(\Sigma)$, together with a
% Boolean-valued \emph{livelock flag} $\delta(P) \in
% \{\mathrm{true},\mathrm{false}\}$, giving rise to our second main
% result:

\begin{thm}
\label{main2}
Let $P$ be a structurally finite-state process. Let $\Phi: \SFS
\longrightarrow \mathcal{P}(\mathcal{P}(\Sigma) \times
\mathcal{P}(\Sigma))$ and $\delta: \SFS \longrightarrow \{ \true,
\false \}$ be defined recursively on the structure of $P$ as shown in
Figures~\ref{fig_fair_cofair_sets} and \ref{fig_fair_cofair_div},
respectively. Then, if $\delta(P) = \false$, $P$ is
livelock-free. Moreover, if in addition $\Phi(P) = \{ (F_1, C_1),
\ldots, (F_k, C_k) \}$, then, for every infinite trace $u$ of $P$,
there exists $1 \leq i \leq k$, such that $u$ is both fair in $F_i$
and co-fair in $C_i$.

%  Write 
% $\Phi(P)=$\linebreak 
% $\{(F_1,C_1),\ldots,(F_k,C_k)\}$. If $\delta(P) = \mathrm{false}$,
% then $P$ is livelock-free, and moreover, for every infinite trace $u$
% of $P$, there exists $1 \leq i \leq k$ such that $u$ is both fair in
% $F_i$ and co-fair in $C_i$.
\end{thm}
% \proof
%    The proof proceeds by structural induction on $P$ and is presented as Proposition~\ref{app_static_SFS} in Appendix~\ref{app_static_analysis}.
% \qed

% The calculation of $\Phi(P)$ proceeds inductively as follows. For $P$
% a closed sequential process, $\Phi(P)$ is computed directly from the
% labelled transition system associated with $P$.%
% \footnote{It is worth pointing out how this can be achieved
%   efficiently. Given a set $L \subseteq \Sigma$ of events, delete all
%   ($\Sigma-L$)-labelled transitions from $P$'s labelled transition
%   system. If the resulting graph contains a (not necessarily
%   reachable) strongly connected component which comprises every single
%   event in $L$, include $(L,\Sigma-L)$ as a fair/co-fair pair for $P$.
% 
% Of course, in actual implementations, it is not necessary to iterate
% explicitly over all possible subsets of $\Sigma$. The computation we
% described can be carried out symbolically using a Boolean circuit of
% size polynomial in $|\Sigma|$, using well-known circuit algorithms for
% computing the transitive closure of relations. Consequently, $\Phi(P)$
% can be represented symbolically and compactly either as a BDD or a
% propositional formula. Further implementation details are provided in
% Sect.~\ref{experiments}.}
% The rules for compound $\SFS$ processes are given in \autoref{fig_fair_cofair_sets}.

\begin{figure}[h!]
\centering 
\begin{boxedminipage}{0.95 \linewidth}%\setlength{\fboxsep}{} 
{\allowdisplaybreaks
{\normalsize
\begin{align*}
\Phi(P) \deq \mbox{computed from $P$'s LTS (see \autoref{section_static_atomic_sfs})} \\
\pheq \mbox{\textbf{whenever $P$ is a sequential $\SFS$ process; 
                                              otherwise:}} \\[1ex]
\Phi(a \then P) \deq \Phi(P) \\
\Phi(P_1 \oplus P_2) \deq
\Phi(P_1) \union \Phi(P_2) \ \ \mbox{if $\oplus \in
  \set{\intchoice,\extchoice,\semi}$} \\
\Phi(P_1 \mathrel{\smash{\parallel[A]}} P_2) \deq 
 \{(F,C) \st F \inter C = \emptyset \land (F_i, C_i) \in
  \Phi(P_i) \mbox{ for } i = 1,2 \land \\ 
  \pheq \phantom{\{(F,C) \st} F = F_1 \union F_2 \land \\ 
  \pheq \phantom{\{(F,C) \st} C = (C_1 \inter A) \union (C_2 \inter A) \union
  ((C_1 - A) \inter (C_2 - A)) \} \union \mbox{} \\
  \pheq \{(F, C)
  \st (F, C) \in \Phi(P_1) \land F \inter A = \emptyset\} \union \mbox{} \\
  \pheq \{(F, C) \st (F, C) \in \Phi(P_2) \land F \inter A =
  \emptyset\} \\
\Phi(P \hide A) \deq \{(F - A, C \union A) \st (F,C) \in \Phi(P)\} \\
\Phi(P[R])  \deq \set{ (F,C) \st (F',C') \in \Phi(P) \land F' \subseteq R^{-1}(F) \land F \subseteq R(F') \land \\
   \pheq \phantom{\{(F,C) \st}  C = \set{b \in \Sigma \st R^{-1}(b) \subseteq C'} }  \enspace .
\end{align*}
}
}
\end{boxedminipage}
\caption{Fair/co-fair sets.}\label{fig_fair_cofair_sets}
\end{figure}

% $\Phi(P \renam{R})  = \set{ (F,C) \st (F',C') \in \Phi(P) \land F' \subseteq R^{-1}(F) \land F \subseteq R(F') \land \\
% \phantom{\Phi(P \renam{R}) = \{(F,C) \st} C = \set{b \in \Sigma \st R^{-1}(b) \subseteq C'}}$

% \indent Note that by construction, all fair/co-fair pairs of sets thus
% generated remain disjoint; this is key in the rule for parallel
% composition, where the fair/co-fair data of individual sub-components
% enables one to rule out certain pairs for the resulting parallel
% process.
% 
% The calculation of $\delta(P)$ similarly proceeds inductively,
% making use of the fair/co-fair data, as follows. If $P$ is a closed
% sequential process, then $\delta(P)$ is determined directly from the
% labelled transition system associated with $P$, according to whether
% the latter contains a $\tau$-cycle or not (using, e.g., Tarjan's
% algorithm \cite{CLRS01}). Otherwise (see \autoref{fig_fair_cofair_div}):
%

\begin{figure}[h!]
\centering 
\begin{boxedminipage}{0.95 \linewidth}%\setlength{\fboxsep}{} 
{\allowdisplaybreaks
{\normalsize
\begin{align*}
\delta(P) \deq \mbox{computed from $P$'s LTS (see \autoref{section_static_atomic_sfs})} \\ %\ \  
\pheq \mbox{\textbf{whenever $P$ is a sequential $\SFS$ process; 
                                              otherwise:}} \\[1ex]
\delta(a \then P) \deq \delta(P) \\
\delta(P_1 \oplus P_2) \deq \delta(P_1) \lor \delta(P_2) \ \
\mbox{if $\oplus \in \set{\intchoice,\extchoice,\parallel[A],\semi}$} \\
\delta(P \hide A) \deq \left\{ \begin{array}{ll} \mathrm{false} & 
   \mbox{ if $\delta(P) = \mathrm{false}$ and, for each 
    $(F,C) \in \Phi(P)$, $F - A \neq \emptyset$}\\ 
    \mathrm{true} & \mbox{ otherwise}
     \end{array} \right.\\           
\delta(P[R])  \deq \delta(P) \enspace .
\end{align*}
}
}
\end{boxedminipage}
\caption{$\delta$-bit.}  \label{fig_fair_cofair_div}
\end{figure}

% \footnotetext{Let us remark that the clause for the hiding operator is
% phrased here so as to make the rule as intuitively clear as
% possible. In practice, one however need not iterate over all possible
% pairs $(F,C) \in \Phi(P)$: it is simpler instead to evaluate the
% negation, an existential calculation which is easily integrated within
% either a SAT or BDD implementation.}

The proof of Theorem~\ref{main2} proceeds by structural induction on
$P$ and is presented 
% as Proposition~\ref{prop_fair_cofair}
in Appendix~\ref{app_static_SFS}.

\indent Note that by construction, all fair/co-fair pairs of sets thus
generated remain disjoint; this is key in the rule for parallel
composition, where the fair/co-fair data of individual sub-components
enables one to rule out certain pairs for the resulting parallel
process. Also, as shown in the proof of Theorem~\ref{main2}, whenever
$(F,P)$ appears as a fair/co-fair pair in some $\Phi(P)$, $F$ is never
empty.

Let us also remark that the $\delta$ clause for the hiding operator is
here phrased in a way so as to make the rule as intuitively clear as
possible. In practice, one however need not iterate over all possible
pairs $(F,C) \in \Phi(P)$: it is simpler instead to evaluate the
negation, an existential calculation which is easily integrated within
either a SAT or BDD implementation.

\subsection{Static Livelock Analysis Algorithm}

Theorems~\ref{main} and \ref{main2} yield a conservative algorithm for
livelock-freedom: given a CSP process $P$ (which we will assume does
not contain $\DIV$ in its syntax), determine first whether $P$ is
structurally finite state. If so, assert that $P$ is livelock-free if
$\delta(P) = \mathrm{false}$, and otherwise report an inconclusive
result. If $P$ is not structurally finite state, assert that $P$ is
livelock-free if $\F(P) \neq \emptyset$, and otherwise report an
inconclusive result. 

The complexity of this procedure is in the worst case quadratic in the
syntactic size of $P$ and exponential in the cardinality of $\Sigma$,
by invoking a similar line of reasoning as that presented following
Theorem~\ref{main}. Likewise, determining for an $\SFS$ process $P$
whether $\delta(P)$ is true is easily seen to lie in NP\@.

It is perhaps useful to illustrate how the inherent incompleteness of
our procedure can manifest itself in very simple ways. For example,
let $P = a \then Q$ and $Q = (a \then P) \extchoice (b \then Q)$, and
let $R = (P \parallel[\{a,b\}] Q) \hide b$. Using Beki\v{c}'s
procedure, $R$ is readily seen to be (equivalent to) a structurally
finite-state process. Moreover, $R$ is clearly livelock-free, yet
$\delta(R) = \mathrm{true}$ and $\F(R) = \emptyset$. Intuitively,
establishing livelock-freedom here requires some form of state-space
exploration, to see that the `divergent' state $(Q \parallel[\{a,b\}]
Q) \hide b$ of $R$ is in fact unreachable, but that is precisely the
sort of reasoning that our static analysis algorithm is not geared to
do.

Nonetheless, we have found in practice that our approach succeeded in
establishing livelock-freedom for a wide range of existing benchmarks;
we report on some of our experiments in \autoref{experiments}, and
also present in~\autoref{app_abp} a small case study illustrating the
intuitions underlying the rules given in
Figures~\ref{fig_fair_cofair_sets} and \ref{fig_fair_cofair_div}.

We conclude by noting that, for structurally finite-state processes,
Theorem~\ref{main2} is stronger than Theorem~\ref{main}, i.e., it
correctly classifies a larger class of processes as being
livelock-free, as stated in the following proposition. Empirically,
algorithms based on Theorem~\ref{main2} have also been found to run
considerably faster in practice.

\begin{prop}
\label{stronger}
For any structurally finite-state process $P$, if $\F(P) \neq
\emptyset$ then $\delta(P) = \false$. 
\end{prop}
\proof
A proof sketch is given in \autoref{app_static_SFS}.
\qed

\section{Implementation and Experimental Results}
\label{experiments}

We have implemented both the general framework and the framework for structurally finite-state processes in a tool called \textsc{slap}, which is an acronym for \textsc{Static Livelock Analyser of Processes}. 
Computationally, the crux of our algorithms revolves around the generation and manipulation of sets. The algorithms fit very naturally into a symbolic paradigm; hence \textsc{slap} is fully symbolic. The choice of an underlying symbolic engine is configurable, with support for using a SAT engine (based on MiniSAT 2.0), a BDD engine (based on CUDD 2.4.2), or running
a SAT and a BDD analyser in parallel and reporting the results of the first one to finish. Some details regarding the symbolic part of our frameworks and algorithms are presented in \autoref{app_encoding}.

We have also integrated the framework for analysing structurally
finite-state processes directly into FDR~\cite{AGLOPRW12}, where it now
constitutes an alternative back-end for establishing livelock
freedom.  The binaries for the latter can
be downloaded from the following location:
\begin{center}
 \url{http://www.cs.ox.ac.uk/projects/concurrency-tools/slap/}
\end{center}

% Computationally, the crux of our algorithm revolves around the
% manipulation of sets. We have built both BDD-based and
% propositional-formula-based implementations, using respectively
% CUDD~2.4.2 and MiniSat~2.0 for computations. Our resulting tool was
% christened \textsc{slap}, for \textsc{Static Livelock Analyser of Processes}.

We experimented with a wide range of benchmarks, including
parameterised, parallelised, and piped versions of Milner's Scheduler,
the Alternating Bit Protocol, the Sliding Window Protocol, the Dining
Philosophers, Yantchev's Mad Postman Algorithm~\cite{YJ89}, as well as a
Distributed Database algorithm.\footnote{Scripts and descriptions for
  all benchmarks are available from the website associated
  with~\cite{Ros11}.} In all our examples,
internal communications were hidden, so that livelock-freedom can be
viewed as a progress or liveness property. All benchmarks were
livelock-free, although the reader familiar with the above examples
will be aware that manually establishing livelock-freedom for several
of these can be a subtle exercise.

In all cases apart from the Distributed Database algorithm, \textsc{slap} was
indeed correctly able to assert livelock-freedom (save for rare
instances of timing out). (Livelock-freedom for the Distributed
Database algorithm turns out to be remarkably complex;
see~\cite{Ros97} for details.) In almost all instances, both BDD-based
and SAT-based implementations of \textsc{slap} substantially outperformed the
state-of-the-art CSP model checker FDR, often completing orders of
magnitude faster. On the whole, BDD-based and SAT-based
implementations performed comparably, with occasional
discrepancies. All experiments were carried out on a 3.07GHz Intel
Xeon processor running under Ubuntu with 8~GB of RAM\@. Times in
seconds are given in Table~\ref{fig_results}, with * indicating a
30-minute timeout. %Further details of the experiments are provided in~\cite{slaptr}.

\begin{table}[ht] \setlength{\tabcolsep}{2pt}
\centering
\hspace{-2ex}
\begin{minipage}[b]{0.50\linewidth} \centering
\begin{tabular}{c  c  c  c }
\toprule
\textbf{Benchmark}  &  \textbf{FDR}  	&  \textbf{Static}  &  \textbf{Static}  \\  
		    &			&  (BDD) 	    &  (SAT) \\	

\midrule
\midrule

Milner-10 & 0	& 0.06	&  0.05  \\

Milner-15 & 0 	& 0.19 & 0.14 \\ 

Milner-20 & 409 & 0.63 & 0.28 \\ 

Milner-21 & 948& 0.73 & 0.23 \\ 

Milner-22 & * & 0.93 & 0.25 \\ 

Milner-25 & * & 1.63 & 0.41 \\ 

Milner-30 & * & 7.56 & 0.8 \\[0.6ex] 

\midrule 

ABP-0 & 0 & 0.03 & 0.11 \\ 

ABP-0-inter-2 & 0 & 0.03 & 0.23 \\ 

ABP-0-inter-3 & 23 & 0.06 & 0.35 \\ 

ABP-0-inter-4 & * & 0.08 & 0.47 \\ 

ABP-0-inter-5 & * & 0.09 & 0.63 \\ 

ABP-0-pipe-2 & 0 & 0.04 & 0.35 \\ 

ABP-0-pipe-3 & 2 & 0.06 & 0.75 \\ 

ABP-0-pipe-4 & 175 & 0.08 & 1.27 \\ 

ABP-0-pipe-5 & * & 0.10 & 1.85 \\ 

ABP-0-pipe-6 & * & 0.11 & 2.91 \\ [0.5ex]  

ABP-4 & 0 & 0.11 & * \\ 

ABP-4-inter-2 & 39 & 0.16 & * \\ 

ABP-4-inter-3 & * & 0.22 & * \\ 

ABP-4-inter-7 \ & * & 0.39 & * \\ 

ABP-4-pipe-2 & 12 & 0.38 & * \\ 

ABP-4-pipe-3 & * & 0.38 & * \\ 

ABP-4-pipe-7 & * & 0.39 & * \\ 

\bottomrule
\end{tabular}
\end{minipage}%
\begin{minipage}[b]{0.50\linewidth} \centering

\begin{tabular}{c  c  c  c }
\toprule

\textbf{Benchmark}  &  \textbf{FDR}  	&  \textbf{Static}  &  \textbf{Static}  \\  
		    &			&  (BDD) 	    &  (SAT) \\	

\midrule
\midrule

SWP-1 & 0 & 0.03 & 7.06 \\ 

SWP-2 & 0 &  0.46& * \\ 

SWP-3 & 0 & 46.81 & * \\ 

SWP-1-inter-2 & 0 & 0.04 & 14.84 \\ 

SWP-1-inter-3 & 31 & 0.06 & 24.02 \\ 

SWP-1-inter-4 & * & 0.08 & 29.44 \\ 

SWP-1-inter-7 & * & 0.13 & 58.82 \\ 

SWP-2-inter-2 & 170 & 0.71 & * \\ 

SWP-2-inter-3 & * & 0.94 & * \\ 

SWP-1-pipe-2 & 0 & 0.04 & 28.09 \\ 

SWP-1-pipe-3 & 0 & 0.07 & 66.71 \\ 

SWP-1-pipe-4 & 3 & 0.09 & 121.09 \\ 

SWP-1-pipe-5 & 246 & 0.10 & 192.39 \\

SWP-1-pipe-7 & * & 0.14 &  399.55\\  

\midrule
Philosophers-5 & 0 & 0.30 & 0.10 \\ 

Philosophers-7 & 2 & 1.62 & 0.21 \\ 

Philosophers-8 & 20 & 2.51 & 0.35 \\ 

Philosophers-9 & 140 & 3.98 & 0.50 \\ 

Philosophers-10 \ & 960 & 7.49 & 0.72 \\

\midrule

Mad Postman-2 & 0 & 0.06 & 0.03 \\ 

Mad Postman-3 & 6 & * & 0.20 \\ 

Mad Postman-4 & * & * & 0.89 \\ 

Mad Postman-5 & * & * & 4.21 \\ 

Mad Postman-6 & * & * & 20.75 \\  

\bottomrule

\end{tabular}
\end{minipage}
\vspace{2ex}

\caption{Times reported are in seconds, with * denoting a 30-minute
timeout.}\label{fig_results}
\end{table}

% \textsc{slap} 

% FDR is largely written in C++ and runs on Linux, Mac OS X and Solaris on
% SPARC. The binaries, as well as a user manual, are available for download from:
% http://www.cs.ox.ac.uk/projects/concurrency-tools/.
% There are two ways of using FDR: either through its own GUI or through a
% command-line interface that is primarily used by other veriﬁcation tools which
% use FDR as a back end. Details can be found in the user manual. Collections of
% CSP scripts can be downloaded from http://www.cs.ox.ac.uk/ucs/CSPM.

\section{Future Work}

An interesting property of our approach is the possibility for our
algorithm to produce a \emph{certificate} of livelock-freedom,
consisting among others in the various sets supporting the final
judgement. Such a certificate could then be checked by an independent tool.
%in polynomial time

Other directions for future work include improving the efficiency of
\textsc{slap} by incorporating various abstractions (such as collapsing all
events on a given channel, or placing \emph{a priori} bounds on the
size of sets), or conversely increasing accuracy at modest
computational cost, for example by making use of algebraic laws at the
syntactic level, such as bounded unfoldings of parallel compositions.

\clearpage

% \bibliography{digit3}
% \input{slap_journal5.bbl}

\appendix

\section{Proofs for Section~\ref{sec_infinite_traces}}\label{app_infinite_traces}

Throughout the section we will use the following notation. For every
$u \in \Sigma^{\omega}$ and $i \in \mathbb{N}$ we will denote by $u_i$
the prefix of $u$ of length $i$. 
Then, as explained in \autoref{sec_infinite_traces}, $u \in \inftraces{P}$ if and only if for each $i \in \mathbb{N}$, $u_i \in  \traces{P} \inter \Sigma^{\ast}$. Let us recall that for every $i \in \mathbb{N}$, $u_i$ cannot contain $\tick$ and is therefore an element of $\Sigma^{\ast}$. 
We will frequently make use of the following observation which relies on the set $\traces{P}$ being prefix-closed. If $u_i \in  \traces{P}$ for infinitely many $i \in \mathbb{N}$, then $u_i \in  \traces{P}$ for all $i \in \mathbb{N}$ and, therefore, $u \in \inftraces{P}$. Most proofs will be based on K\"{o}nig's Lemma, which we now recall.

\begin{thm}[K\"{o}nig's Lemma] Suppose that for each $i \in \mathbb{N}$, $X_i$ is a non-empty finite set and $f_i: X_{i+1} \rightarrow X_i$ is a total function. Then there is a sequence $\langle x_i \st i \in \mathbb{N} \rangle$, such that $x_i \in X_i$ and $f_i(x_{i+1}) = x_i$.\qed
\end{thm}

In our proofs we will define the sets $X_i$ as specific subsets of $ \traces{P} \inter \Sigma^{i}$. For each $i \in \mathbb{N}$, $x_i \in X_i$ and  $x_{i+1} \in X_{i+1}$, $f_i(x_{i+1}) = x_i$ will imply that $x_i < x_{i+1}$, where $<$ denotes the strict prefix order of traces on $\Sigma^{\ast}$. For a given $x_{i+1} \in X_{i+1}$, the choice for $f_i(x_{i+1})$ might not be unique, but we can take an arbitrary prefix $x_i$ of $x_{i+1}$ from $X_i$ satisfying certain properties. Then the sequence $\langle x_i \st i \in \mathbb{N} \rangle$ will form an infinite chain $x_0 < x_1 < x_2 < \ldots x_n \ldots$ under prefix and $x = \lim_{i = 0}^{\infty} x_{i} \in \inftraces{P}$.

\vspace{0.2cm}\noindent
\textbf{Lemma~\ref{inft_prefix}.} \textit{Let $u \in \inftraces{a \then P}$. Then there exists $u' \in \inftraces{P}$, such that $u = \langle a \rangle \concat u'$.}

\proof
   Let $u \in \inftraces{a \then P}$. By definition, for each $t < u$, $t \in \traces{a \then P}$. Then, for each $t < u$, $t = \langle a \rangle \concat t'$ for some $t' \in \traces{P}$. Let $u = \langle a \rangle \concat u'$ for some $u' \in \Sigma^{\omega}$. Then, for each $t' < u'$, $t' \in \traces{P}$. Therefore, by definition, $u' \in \inftraces{P}$.
\qed                                                                                                                                                                                                                                                                                                                                                                                                                                                                                                                                                                                                                                                                                                                                                                                                                                                                                                                                                                                                                                                                                                                                                                                                                                                                                                                                                                                                                                                                                                                                                                                                                                                                                                                                                                                                                                                                                                                                                                                                                                                                                                                                                                                                                                                                                                                                                                                                                                                                                                                                                                                                                                                                                                                                                                                                                                                                                                                                                                                                                                                                                                                                                                                                                                                                                                                                                                                                                                                                                                                                                                                                                                                                                                                                                                                                                                                                                                                                                                                                                                                                                                                                                                                                                                                                   
                                                                                                                                                                                                                                                                                                                                                                                                                                                                                                                                                                                                                                                                                                                                                                                                                                                                                                                                                                                                                                                                                                                                                                                                                                                                                                                                                                                                                                                                                                                                                                                                                                                                                                                                                                                                                                                                                                                                                                                                                                                                                                                                                                                                                                                                                            
% \begin{lem}\label{app_inft_choice}
%   Let $u \in \inftraces{P \oplus Q}$ for $\oplus \in \{ \extchoice, \intchoice \}$. Then $u \in \inftraces{P}$ or $u \in \inftraces{Q}$.	
% \end{lem}

\vspace{0.2cm}\noindent
\textbf{Lemma~\ref{inft_choice}.} \textit{Let $u \in \inftraces{P \oplus Q}$ for $\oplus \in \{ \extchoice, \intchoice \}$. Then $u \in \inftraces{P}$ or $u \in \inftraces{Q}$.}

\proof
  Let $u \in \inftraces{P \oplus Q}$. By definition, for each $i \in \mathbb{N}$, $u_i \in  \traces{P \oplus Q}$. Therefore, for each $i \in \mathbb{N}$, $u_i \in \traces{P}$ or $u_i \in \traces{Q}$. Then, due to the pigeonhole principle, $u_i \in \traces{P}$ for infinitely many $i \in \mathbb{N}$ or $u_i \in \traces{Q}$ for infinitely many $i \in \mathbb{N}$. Let without loss of generality the former holds. Then, $u_i \in \traces{P}$ for all $i \in \mathbb{N}$ and, hence,  $u \in \inftraces{P}$.
\qed

% \begin{lem}\label{app_inft_seq_comp}
%    Let $u \in \inftraces{P \semi Q}$. Then $u \in \inftraces{P}$ or $u = t \concat u'$ with $t \concat \ticks \in \traces{P}$,  $u' \in \inftraces{Q}$.
% \end{lem}

\vspace{0.2cm}\noindent
\textbf{Lemma~\ref{inft_seq_comp}.} \textit{Let $u \in \inftraces{P \semi Q}$. Then $u \in \inftraces{P}$ or $u = t \concat u'$ with $t \concat \ticks \in \traces{P}$,  $u' \in \inftraces{Q}$.}

\proof
Let $u \in \inftraces{P \semi Q}$.
By definition, for each $i \in \mathbb{N}$, \mbox{$u_i \in \traces{P \semi Q}$}. Therefore, for each $i \in \mathbb{N}$, $u_i \in \traces{P}$ or $u_i = t_{1} \concat t_{2}$ with $t_{1} \concat \ticks \in \traces{P} \inter \Sigma^{\ast \checkmark}$,  $t_{2} \in \traces{Q} \inter \Sigma^{\ast}$. If for each $i \in \mathbb{N}$, $u_i \in \traces{P}$, then, by definition, $u \in \inftraces{P}$.
Otherwise, there exists $N \in \mathbb{N}$, such that $u_0, u_1, \ldots, u_N \in \traces{P}$, but $u_{N+1} \notin \traces{P}$. Therefore, for $j \geq 1$, $u_{N+j} \notin \traces{P}$. By assumption, for every $i \in \mathbb{N}$, $u_i \in \traces{P \semi Q}$. Therefore, for $j \geq 1$, $u_{N + j} = t_{j} \concat v_{j}$ where $t_j \concat \ticks \in \traces{P}$ (and therefore $t_j \leq u_N$) and $v_j \in \traces{Q}$. Then, there must be some $t \leq u_N$, such that $t_j = t$ for infinitely many $u_{N + j}$'s. Let us write $u_j = t \concat w_j$ for $j \geq |t|$. We have that $t \concat \ticks \in \traces{P}$ and infinitely often $w_j \in \traces{Q}$. Since for $j < j'$, $w_j < w_{j'}$, and the set of traces is prefix-closed, $w_j \in \traces{Q}$ for each $j \geq |t|$. Then, by definition, $u' = \lim_{j = |t|}^{\infty} w_j \in \inftraces{Q}$.
\qed
% 

% \begin{lem}\label{app_inft_hide}
%    Let $u \in \inftraces{P \hide A}$ and $P \hide A$ be livelock-free. Then there exists $v \in \inftraces{P}$, such that $u = v \restriction (\Sigma \backslash A)$. 
% \end{lem}

\vspace{0.2cm}\noindent
\textbf{Lemma~\ref{inft_hide}.} \textit{Let $u \in \inftraces{P \hide A}$ and $P \hide A$ be livelock-free. Then there exists $v \in \inftraces{P}$, such that $u = v \restriction (\Sigma \backslash A)$.}

\proof
   Let $u \in \inftraces{P \hide A}$. By definition, for each $i \in \mathbb{N}$, $u_i \in \traces{P \hide A}$, \ie there exists $v_{j_i} \in \traces{P}$, such that $u_i = v_{j_i} \restriction (\Sigma \backslash A)$.

   Let, for $i \in \mathbb{N}$, $\backslash^{-1}(u_i) = \{  v \in \alltraces{P} \st  v \restriction (\Sigma \backslash A) = u_i\}$. We claim that, for each  $i \in \mathbb{N}$, $\backslash^{-1}(u_i)$ is finite. Suppose, for the sake of the argument, that $\backslash^{-1}(u_k)$ is infinite. We will prove that $P \hide A$ is divergent, which will be a contradiction with $P \hide A$ being livelock-free. Let $u_k = \langle a_1, a_2, \ldots a_k\rangle$. It is clear that $\{ a_1, a_2, \ldots, a_k \} \inter A = \emptyset$. Then, $\backslash^{-1}(u_k) = ( (A^{\ast} \union A^{\omega}) \; a_1 \; (A^{\ast} \union A^{\omega}) \; a_2 \; (A^{\ast} \union A^{\omega}) \ldots (A^{\ast} \union A^{\omega}) \; a_k \; (A^{\ast} \union A^{\omega}) ) \inter \alltraces{P} $. Let for $i \in \{ 0, \ldots, k-1\}$, $n_i$ be the maximum number of occurrences of consecutive events from $A$ before the occurrence of $a_{i+1}$ and let $n_k$ be the maximum number of consecutive events from $A$ after $a_k$. Then, for $i = \{0, \ldots, k\}$, $n_i \in \mathbb{N} \union \{ \omega\}$. Since $\backslash^{-1}(u_k)$ is infinite, there exists $j \in \{0, \ldots, k\}$, such that $n_j = \omega$. Let $j_{\min}$ be the minimal $j$ with this property. Then, for $i < j_{\min}$, $n_i \in \mathbb{N}$. Let $v \in ( A^{\ast}  a_1 A^{\ast} a_2 A^{\ast} \ldots A^{\ast}  a_{j_{\min}} A^{\omega})  \inter \alltraces{P} $. Therefore, $v \restriction (\Sigma \backslash A) = \langle a_1, a_2, \ldots a_{j_{\min}}\rangle  = u_{j_{\min}}  \in \divergences{P \hide A}$ which is a contradiction with $P \hide A$ being livelock-free. Hence, for $i = \{0, \ldots, k\}$, $n_i \in \mathbb{N}$ and therefore, $\backslash^{-1}(u_k)$ is finite.
   Therefore,  for each  $i \in \mathbb{N}$, we have: 
 
 		\begin{enumerate}[(1)]
 		 \item $\backslash^{-1}(u_i) \neq \emptyset$ because $u_i \in \traces{P \hide A}$ \label{item_H_notempty}
 		 \item $\backslash^{-1}(u_i)$ is finite 
 		 \item For each $j > i$, for each $w \in \backslash^{-1}(u_j)$, there exists $v \in \backslash^{-1}(u_i)$, such that $v < w$. The trace $v$ can be defined as  an arbitrary prefix of $w$ of $(\Sigma \backslash A)$-length $i$.  \label{item_H_prefixing}
 		\end{enumerate}
 
\noindent Therefore, by K\"{o}nig's Lemma, there exists an infinite sequence $v_{j_1} < v_{j_2} < \ldots < v_{j_n} < \ldots$, such that for $i \in \mathbb{N}$, $v_{j_i} \in \backslash^{-1}(u_i)$, i.e., $v_{j_i} \in \traces{P}$ and  $u_i = v_{j_i} \restriction (\Sigma \backslash A)$. Therefore, $v = \lim_{i = 0}^{\infty} v_{j_i} \in \inftraces{P}$ and $u = v \restriction (\Sigma \backslash A)$. 
\qed
 
% \begin{lem}\label{app_inft_rename}
%    Let $u \in \inftraces{P \renam{R}}$. Then there exists $v \in \inftraces{P}$, such that $v \renamomega u$. 
% \end{lem}

\vspace{0.2cm}\noindent
\textbf{Lemma~\ref{inft_rename}.} \textit{Let $u \in \inftraces{P \renam{R}}$. Then there exists $v \in \inftraces{P}$, such that $v \renamomega u$.}

\proof
  	Let $u \in \inftraces{P \renam{R}}$. By definition, for each $i \in \mathbb{N}$, $u_i \in \traces{P \renam{R}} \inter \Sigma^{\ast}$.  	
 	Therefore, for each $i \in \mathbb{N}$, there exists $v_{j_i} \in \traces{P} \inter \Sigma^{\ast}$, such that $v_{j_i} \mathrel{R} u_i$, i.e., $\length(u_i) = \length(v_{j_i}) = i$ and for each $0 \leq k \leq i$, $v_{j_i}(k) 	 \mathrel{R}  u_i(k)$. Let, for $i \in \mathbb{N}$, $R^{-1}(u_i) = \{  v \in \traces{P} \st  v \mathrel{R} u_i \}$. Then, for $i \in \mathbb{N}$:
 		\begin{enumerate}[(1)]
 		 \item $R^{-1}(u_i) \neq \emptyset$ because $u_i \in \traces{P \renam{R}}$ \label{item_R_notempty}
 		 \item $R^{-1}(u_i)$ is finite because $\Sigma$, and therefore $R$, are finite
 		 \item For each $j > i$ and each $w \in R^{-1}(u_j)$, there exists $v \in R^{-1}(u_i)$, such that $v < w$. The trace $v$ can be constructed as the prefix of $w$ of length $i$.  \label{item_R_prefixing}
 		\end{enumerate}
 
\noindent  Therefore, by K\"{o}nig's Lemma, there exists an infinite sequence $v_{j_1} < v_{j_2} < \ldots < v_{j_n} < \ldots$, such that for $i \in \mathbb{N}$, $v_{j_i} \in R^{-1}(u_i)$, \ie $v_{j_i} \in \traces{P}$ and  $v_{j_i} \mathrel{R} u_i $. Therefore, $v = \lim_{i = 0}^{\infty} v_{j_i} \in \inftraces{P}$ and $v \renamomega u$.
\qed
% 

% \begin{lem}\label{app_inft_parallel}
%    Let $u \in \inftraces{P \parallel[A] Q}$. Then there exist $u_1 \in \alltraces{P}$, $u_2 \in \alltraces{Q}$, such that $u \in u_1 \parallel[A] u_2$ and, $u_1 \in \Sigma^{\omega}$ or $u_2 \in \Sigma^{\omega}$. 
% \end{lem}

\vspace{0.2cm}\noindent
\textbf{Lemma~\ref{inft_parallel}.} \textit{Let $u \in \inftraces{P \parallel[A] Q}$. Then there exist $u_1 \in \alltraces{P}$, $u_2 \in \alltraces{Q}$, such that $u \in u_1 \parallel[A] u_2$ and, $u_1 \in \Sigma^{\omega}$ or $u_2 \in \Sigma^{\omega}$.}

\proof
	Let $u \in \inftraces{P \parallel[A] Q}$. Then, by definition, for each $n \in \mathbb{N}$, there exist $v_{i_n} \in \traces{P} \inter \Sigma^{\ast} $ and $w_{j_n} \in \traces{Q} \inter \Sigma^{\ast}$, such that $u_n \in v_{i_n}  \parallel[A] w_{j_n}$ and $ n \leq |v_{i_n}| + |w_{j_n}| \leq 2n$. Therefore, for each such triple $(u_n, v_{i_n}, w_{j_n})$ there exists a function $f^n:\{1, \ldots, n  \} \mapsto \{ 0, 1, 2\}$ specifying a possible interleaving of $v_{i_n}$ and $w_{j_n}$ for obtaining $u_n$. More specifically, $f^n(i)$ indicates which process contributes for communicating the $i$-th event of $u_n$, with $0$ denoting both $P$ and $Q$ (for events in $A$), $1$ denoting only $P$, $2$ denoting only $Q$. Given $u_n = \langle a_1, \ldots a_n \rangle$ and $f^n$, $v_{i_n}$ and $w_{j_n}$ are identified uniquely as $v_{i_n} = \langle a_i   \st 1 \leq i \leq n, f^n(a_i) \subseteq \{0, 1\} \rangle $, $w_{j_n} = \langle a_j   \st 1 \leq j \leq n, f^n(a_j) \subseteq \{0, 2\} \rangle $.

 	Let us define a partially ordered set $((\Sigma^{\ast \checkmark})^{2}, \leq)$ with $(v, w) \leq (v', w')$ iff $v \leq v'$ and $w \leq w'$, where $\leq$ denotes a non-strict prefix on traces. We will prove that there exists an infinite chain $(v_{i_1}, w_{j_1}) \leq \ldots \leq (v_{i_n}, w_{j_n}) \leq \ldots$, such that for each $n \in \mathbb{N}$, $v_{i_n} \in \traces{P}$, $w_{j_n} \in \traces{Q}$, $u_n \in v_{i_n}  \parallel[A] w_{j_n}$.

 	Let for $k \in \mathbb{N}$, $\parallel[A]^{-1}(u_k) = \{ (v_{i_k}, w_{j_k})  \st v_{i_k} \in \traces{P}, w_{j_k} \in \traces{Q}, u_k \in v_{i_k}  \parallel[A] w_{j_k}  \}$. Then:
 
 	\begin{enumerate}[(1)]
 		 \item $\parallel[A]^{-1}(u_k) \neq \emptyset$ because $u_i \in \traces{P \parallel[A] Q}$. 
 		 \item $\parallel[A]^{-1}(u_k)$ is finite because $\Sigma$ is finite.
 		 \item For each $k > l$ and each $(v_{i_k}, w_{j_k}) \in \parallel[A]^{-1}(u_k)$, there exists $(v_{i_l}, w_{j_l}) \in \parallel[A]^{-1}(u_l)$, such that $(v_{i_l}, w_{j_l}) < (v_{i_k}, w_{j_k})$. The pair of traces $(v_{i_l}, w_{j_l})$ can be constructed as follows. Let for the triple $(u_k, v_{i_k}, w_{j_k})$ the function $f^k:\{1, \ldots, n  \} \mapsto \{ 0, 1, 2\}$ specifies a possible interleaving of $v_{i_k}$ and $w_{j_k}$ for obtaining $u_k$. We define $f^{l}(i) = f^k(i)$ for $1 \leq i \leq l$. Then, $(v_{i_l}, w_{j_l})$ is the pair that is uniquely identified by $f^{l}$ and $u_l$.
 	\end{enumerate}

\noindent Therefore, by K\"{o}nig's Lemma, for each $n \in \mathbb{N}$, there exist $(v_{i_1}, w_{j_1}) \leq (v_{i_2}, w_{j_2})  \leq \ldots \leq  (v_{i_n}, w_{j_n})$, such that for each $1 \leq k \leq n$,  $v_{i_k} \in \traces{P}$, $w_{j_k} \in \traces{Q}$, $u_k \in v_{i_k}  \parallel[A] w_{j_k}$, $ k \leq |v_{i_k}| + |w_{j_k}| \leq 2k$. Let $v = \lim_{k = 1}^{\infty} v_{i_k}$, $w = \lim_{k = 1}^{\infty} w_{j_k}$. Then clearly, $v \in \alltraces{P}$, $w \in \alltraces{Q}$ and $u \in v  \parallel[A] w$. Let us assume that both $v$ and $w$ are finite, i.e., $|v| = l_v$, $|w| = l_w$ for some $l_v, l_w \in \mathbb{N}$. Then, each prefix of $u$ will be of length at most $l_v + l_w \in \mathbb{N}$, which is a contradiction with $u$ being infinite. Therefore, at least one of $v$ and $w$ is infinite.
\qed

\section{Proofs for Section~\ref{sec_family_metrics}}\label{app_family_metrics}

\begin{lem}[\mbox{\cite[Lemma~9.2.5]{Sut75}}]\label{lemma_cauchy_subseq}
 In any metric space, if $s$ is a Cauchy sequence that has a subsequence that converges to a point $x$, then $s$ also converges to $x$.\qed
\end{lem}

\begin{prop}\label{prop_app_complete_um_space}
Let $U \subseteq \Sigma$.  Then $\TD$ equipped with the metric $d_U$ is a complete metric space.
\end{prop}

\proof
	We will prove that every Cauchy sequence converges.

	Let $\langle P_i \st i \in \mathbb{N} \rangle $ be a Cauchy sequence in $(\TD, d_U)$. By definition, for every $\varepsilon > 0$, there exists $N_\varepsilon \in \mathbb{N}$ such that, for every $n, m \geq N_\varepsilon$, $d_U(P_n, P_m) < \varepsilon$.
	Therefore, for every $r \in \mathbb{N}$ and $\varepsilon = 2^{-r}$, there exists $N_r \in \mathbb{N}$ such that, for every $n, m \geq N_r$, $d_U(P_n, P_m) < 2^{-r}$, \ie $P_n \restriction_U r = P_m \restriction_U r$.  Then, for every $r, m \in \mathbb{N}$, $d_U(P_{N_{r}}, P_{N_{r+m}}) < 2^{-r}$.
	Therefore, the subsequence $\langle P_{N_{r}} \st r \in \mathbb{N} \rangle $ of $\langle P_i \st i \in \mathbb{N} \rangle $ is itself a Cauchy sequence.

	Let us define $P = \bigsqcap_{q \in \mathbb{N}} \bigsqcup_{r \geq q} P_{N_{r}}$. $P \in \TD$ because $(\TD, \sqsubseteq)$ is a complete lattice. 	We will prove that the subsequence $\langle P_{N_{r}} \st r \in \mathbb{N} \rangle $ converges to $P$, i.e., that for every $r \in \mathbb{N}$,  $d_U(P_{N_{r}}, P) < 2^{-r}$.

	Let us fix $r$. Suppose, for the sake of the argument, that $d_U(P_{N_{r}}, P) \geq 2^{-r}$ and let, without loss of generality, $P_{N_{r}}$ and $P$ disagree on the sets of their divergences. Therefore, there exists $t \in \Sigma^{\ast \checkmark}$ such that $\length_U(t) < r$ and, either $t \in \divergences{P_{N_{r}}} \backslash \divergences{P}$ or $t \in \divergences{P} \backslash \divergences{P_{N_{r}}}$. To remind, by construction we have $\divergences{P} = \bigcup_{q \in \mathbb{N}} \bigcap_{r \geq q} \divergences{P_{N_{r}}}$. We explore both alternatives.

	\begin{iteMize}{$\bullet$}
	 \item 	Suppose $t \in \divergences{P_{N_{r}}} \backslash \divergences{P}$. Since $t \not \in \divergences{P}$, for every $q \in \mathbb{N}$ there exists $s_q \geq q$ such that $t \not \in \divergences{P_{N_{s_q}}}$. Therefore, for $q = r$ there exists $s_r \geq r$ such that $t \not \in \divergences{P_{N_{s_r}}}$. Hence, since $t \in \divergences{P_{N_{r}}}$ and $\length_U(t) < r$, $d_U(P_{N_r}, P_{N_{s_r}}) \geq 2^{-r}$ which is a contradiction with $d_U(P_{N_{r}}, P_{N_{r+m}}) < 2^{-r}$ for $m \geq 0$.

	 \item 	Therefore, $t \in \divergences{P} \backslash \divergences{P_{N_{r}}}$. Since $t \in \divergences{P}$, there exists $q \in \mathbb{N}$ such that for every $s \geq q$, $t \in \divergences{P_{N_s}}$. However, as $t \not \in \divergences{P_{N_{r}}}$ and $\length_U(t) < r$, for every $s \geq r$, $t \not \in \divergences{P_{N_{s}}}$ which again leads to a contradiction.
	\end{iteMize}

\noindent Therefore, for every $r \in \mathbb{N}$,  $d_U(P_{N_{r}}, P) < 2^{-r}$ and, hence, the subsequence $\langle P_{N_{r}} \st r \in \mathbb{N} \rangle $ converges to $P$. Therefore, from Lemma~\ref{lemma_cauchy_subseq},  $\langle P_i \st i \in \mathbb{N} \rangle $ also converges to $P$ and, hence, $(\TD,d_U)$ is a complete metric space.\qed

\begin{prop}\label{prop_app_set_lf_closed}
Let $U \subseteq \Sigma$. Then the set of livelock-free processes is a closed subset of $(\TD, d_U)$.
\end{prop}

\proof
Let $\langle P_i \st i \in \mathbb{N} \rangle $ be a sequence of livelock-free elements of $\TD$ converging to a process $Q \in \TD$. Therefore, by definition, for every $\varepsilon > 0$, there exists $N \in \mathbb{N}$ such that, for every $n \geq N$, $d_U(P_n, Q) < \varepsilon$. We will prove that $Q$ is also livelock-free.

Suppose for the sake of the argument that $Q$ can diverge. Let $t \in \divergences{Q}$ and $\length_U(t) = k$. If we take $\varepsilon = 2^{-k}$, since $\langle P_i \st i \in \mathbb{N} \rangle $ converges to $Q$, there exists $N_t \in \mathbb{N}$ such that, for every $n \geq N_t$, $d_U(P_n, Q) < 2^{-k}$ and, therefore,  $P_n \restriction_U k = Q \restriction_U k$. Therefore, for every $n \geq N_t$, $t \in \divergences{P_n}$, which is a contradiction with $\langle P_i \st i \in \mathbb{N} \rangle $ being all livelock-free.

Therefore, $Q$ is livelock-free and, hence, the set of livelock-free processes is closed.
\qed

\section{Proofs for Section~\ref{sec_static_nonexp}}\label{app_static_nonexp}

Throughout this section let us fix a set of events $U \subseteq \Sigma$.

% \begin{lem}\label{prop_app__arg_seq_comp}
%  For any CSP processes $P$, $P'$ and $Q$:
% $$ d_{U}(P \semi Q, P' \semi Q) \leq d_{U}(P, P').$$
% \end{lem}

\vspace{0.2cm}\noindent
\textbf{Lemma~\ref{lemma_nonexp_in_args} ($\semi_1$).} \textit{For any CSP processes $P$, $P'$, and $Q$:
 $$ d_{U}(P \semi Q, P' \semi Q) \leq d_{U}(P, P').$$}
\vspace{-0.5cm}

\proof
 Suppose $(T_{P}, D_{P}) \restriction_U k = (T_{P'}, D_{P'}) \restriction_U k$.  We will prove that $(T_{P \semi Q}, D_{P \semi Q}) \restriction_U k = (T_{P'  \semi Q}, D_{P'  \semi Q}) \restriction_U k$, from which we can conclude that $ d_U(P \semi Q, P' \semi Q) \leq d_U(P, P')$.\\

Let $t \in \divergences{P \semi Q}$ and $\length_{U}(t) \leq k$. We will prove that $t \in \divergences{P' \semi Q}$ and therefore, $D_{P \semi Q} \restriction_U k \subseteq D_{P' \semi Q} \restriction_U k$. The reverse containment is established similarly by symmetry.

Since $t \in \divergences{P \semi Q}$, by definition, $t \in \divergences{P}$ or $t=t_{1} \concat t_{2}$ with $t_{1} \concat \langle \checkmark \rangle \in \btraces{P}$, $t_{2} \in \divergences{Q}$. We consider both cases.

\begin{iteMize}{$\bullet$}
 \item Suppose $t \in \divergences{P}$. Since $\length_U(t) \leq k$ and $(T_{P}, D_{P}) \restriction_U k = (T_{P'}, D_{P'}) \restriction_U k$, $t \in \divergences{P'}$. Therefore, by definition, $t \in \divergences{P' \semi Q}$.

% \item Suppose $t=t_{1} \concat t_{2}$ with $t_{1} \concat \langle
%   \checkmark \rangle \in \btraces{P}$, $t_{2} \in
%   \divergences{Q}$. Observe that $\length_U(t_{1} \concat \langle
%   \checkmark \rangle) = \length_U(t_{1}) \leq \length_U(t) \leq k
%   $. Then, since $(T_{P}, D_{P}) \restriction_U k = (T_{P'}, D_{P'})
%   \restriction_U k$, $t_{1} \concat \langle \checkmark \rangle \in
%   \btraces{P'}$. Hence, by definition, $t_{1} \concat t_{2} = t \in
%   \divergences{P' \semi Q}$.
 \item Suppose $t=t_{1} \concat t_{2}$ with $t_{1} \concat \langle
   \checkmark \rangle \in \btraces{P}$, $t_{2} \in
   \divergences{Q}$. Observe that $\length_U(t_{1} \concat \langle
   \checkmark \rangle) = \length_U(t_{1}) \leq \length_U(t) \leq k
   $. Then $t_{1} \concat \langle \checkmark \rangle \in
   \btraces{P'}$, since $(T_{P}, D_{P}) \restriction_U k = (T_{P'}, D_{P'})
   \restriction_U k$. Hence, by definition, $t_{1} \concat t_{2} = t \in
   \divergences{P' \semi Q}$.
\end{iteMize}

\noindent
 Now let $t \in \btraces{P \semi Q}$ and $\length_U(t) \leq k$.
 We will prove that $t \in \btraces{P'  \semi  Q}$ and therefore, $T_{P \semi Q} \restriction_U k \subseteq T_{P' \semi Q} \restriction_U k$. The reverse containment is established similarly by symmetry.
 Since $t \in \btraces{P \semi Q}$, $t \in \divergences{P  \semi  Q}$ or $t \in \traces{P  \semi  Q}$. The latter reduces to  $t \in \traces{P} \inter \Sigma^{\ast}$ or $t=t_{1} \concat t_{2}$ with $t_{1} \concat \langle \checkmark \rangle \in \traces{P}$, $t_{2} \in \traces{Q}$. We consider all three alternatives.
\begin{iteMize}{$\bullet$}
 \item Suppose first that $t \in \divergences{P  \semi  Q}$. We already proved that $t \in \divergences{P'  \semi  Q}$ and therefore, $t \in \btraces{P'  \semi  Q}$.
 \item Suppose now what $t \in \traces{P} \inter \Sigma^{\ast}$. Therefore, $t \in \btraces{P} \inter \Sigma^{\ast}$. Then, since $(T_{P}, D_{P}) \restriction_U k = (T_{P'}, D_{P'}) \restriction_U k$, $t \in \btraces{P'} \inter \Sigma^{\ast}$. 
 \begin{iteMize}{$-$}
 	\item If $t \in \traces{P'} \inter \Sigma^{\ast}$, then by definition, $t \in \traces{P' \semi Q} \subseteq \btraces{P' \semi Q}$.
 	\item If $t \in \divergences{P'} \inter \Sigma^{\ast}$, then by definition, $t \in \divergences{P' \semi Q} \subseteq \btraces{P' \semi Q}$. 
 \end{iteMize}

 \item Suppose finally that $t=t_{1} \concat t_{2}$ with $t_{1} \concat \langle \checkmark \rangle \in \traces{P}$, $t_{2} \in \traces{Q}$. We note that $\length_U(t_{1} \concat \langle \checkmark \rangle) = \length_U(t_{1}) \leq \length_U(t) \leq k $. Then, since $t_{1} \concat \langle \checkmark \rangle \in \traces{P}$  and $(T_{P}, D_{P}) \restriction_U k = (T_{P'}, D_{P'}) \restriction_U k$,  $t_{1} \concat \langle \checkmark \rangle \in \btraces{P'}$.
 \begin{iteMize}{$-$}
 	\item Let $t_{1} \concat \langle \checkmark \rangle \in \traces{P'}$. By definition, $t \in \traces{P' \semi Q} \subseteq \btraces{P' \semi Q}$.
 	\item Let $t_{1} \concat \langle \checkmark \rangle \in \divergences{P'}$. By Axiom 2 of $\TD$, $t_{1}  \in \divergences{P'}$. Since $t_{1} \in \Sigma^{\ast}$, by Axiom 4 of $\TD$, $t=t_{1} \concat t_{2} \in \divergences{P'}$. Then by definition,  $t \in \divergences{P' \semi Q} \subseteq \btraces{P' \semi Q}$.
 \end{iteMize}
\end{iteMize}

\noindent Therefore, $(T_{P \semi Q}, D_{P \semi Q}) \restriction_U k = (T_{P'  \semi Q}, D_{P'  \semi Q}) \restriction_U k$ and, hence, $ d_U(P \semi Q, P' \semi Q) \leq d_U(P, P')$. 
\qed
% \end{lem}

%%%%%%%%%%%%%%%%%%%%%%%%%%%%%%%%%%%%%%%%%%%%%%%%%%%%%%%%%%%%%%%%%%%%%%%%%%%%%%%%%%%%%%%%%%%%%%%%%%%%%%%%%%%%5

% \begin{lem}\label{prop_app__arg_seq_comp_2}
%  For any CSP processes $P, Q$ and $Q'$:
% $$ d_U(P \semi  Q, P \semi Q') \leq d_U(Q, Q').$$
% \end{lem}

\vspace{0.2cm}\noindent
\textbf{Lemma~\ref{lemma_nonexp_in_args} ($\semi_2$).} \textit{For any CSP processes $P, Q$, and $Q'$:
 $$ d_U(P \semi  Q, P \semi Q') \leq d_U(Q, Q').$$}
\vspace{-0.5cm}

\proof
	Suppose $(T_{Q}, D_{Q}) \restriction_U k = (T_{Q'}, D_{Q'}) \restriction_U k$. We will prove that $(T_{P \semi Q}, D_{P \semi Q}) \restriction_U k = (T_{P  \semi Q'}, D_{P  \semi Q'}) \restriction_U k$, from which $ d_U(P \semi Q, P \semi Q') \leq d_U(Q, Q')$ follows immediately. \\

\noindent
	Let $t \in \divergences{P \semi Q}$ and $\length_U(t) \leq k$.
	\begin{iteMize}{$\bullet$}
	 \item Suppose $t \in \divergences{P}$. By definition, $t \in \divergences{P  \semi  Q'}$.
	 \item  Suppose $t=t_{1} \concat t_{2}$ with $t_{1} \concat \langle \checkmark \rangle \in \btraces{P}$, $t_{2} \in \divergences{Q}$ and $\length_U(t_{2}) \leq \length_U(t)  \leq k$.
	 Since by assumption $ D_{Q} \restriction_U k =  D_{Q'} \restriction_U k$, $t_{2} \in \divergences{Q'}$. Then by definition, $t \in \divergences{P \semi Q'}$.
	\end{iteMize}

\noindent	
	Let $t \in \btraces{P \semi Q}$ and $\length_U(t) \leq k$.
	\begin{iteMize}{$\bullet$}
	 \item Let first $t \in \divergences{P  \semi  Q}$. We already proved that $t \in \divergences{P  \semi  Q'}$ and therefore, $t \in \btraces{P  \semi  Q'}$.
	 \item Let now $t \in \traces{P} \inter \Sigma^{\ast}$. Then by definition, $t \in \traces{P  \semi  Q'} \subseteq \btraces{P  \semi  Q'}$.
	 \item Let finally $t=t_{1} \concat t_{2}$ with $t_{1} \concat \langle \checkmark \rangle \in \traces{P} \subseteq \btraces{P}$, $t_{2} \in \traces{Q} \subseteq \btraces{Q}$. Since $\length_U(t) \leq k$, $\length_U(t_{2}) \leq k$. Then, by assumption, $t_{2} \in \btraces{Q'}$.
	\begin{iteMize}{$-$}
		\item If $t_{2} \in \traces{Q'}$, by definition, $t=t_{1} \concat t_{2} \in \traces{P \semi Q'} \subseteq \btraces{P \semi Q'}$.
		\item Let $t_{2} \in \divergences{Q'}$. Since $t_{1} \concat \langle \checkmark \rangle \in  \btraces{P}$, by definition, $t=t_{1} \concat t_{2} \in \divergences{P \semi Q'} \subseteq \btraces{P \semi Q'}$.
	\end{iteMize} 
\end{iteMize}

\noindent
Therefore, $D_{P \semi Q} \restriction_U k \subseteq D_{P \semi Q'} \restriction_U k$ and $T_{P \semi Q} \restriction_U k \subseteq T_{P \semi Q'} \restriction_U k$. The reverse containments are established similarly by symmetry. Therefore, $(T_{P \semi Q}, D_{P \semi Q}) \restriction_U k = (T_{P  \semi Q'}, D_{P  \semi Q'}) \restriction_U k$ and, hence, $ d_U(P \semi Q, P \semi Q') \leq d_U(Q, Q')$.  
\qed

% \begin{lem}\label{prop_app__arg_seq_comp_2_n_x} 
% Let $P, Q$ and $Q'$ be CSP processes. Let $P$ always communicate an event from $U \subseteq \Sigma$ before it does a $\tick$. Then: $$ d_U(P \semi Q, P \semi Q') \leq \frac{1}{2} d_U(Q, Q').$$
% \end{lem}

\vspace{0.2cm}\noindent
\textbf{Lemma~\ref{lemma_contr_seq}.} \textit{Let $P, Q$, and $Q'$ be CSP processes. Let $P$ always communicate an event from $U \subseteq \Sigma$ before it does a $\tick$. Then: $$ d_U(P \semi Q, P \semi Q') \leq \frac{1}{2} d_U(Q, Q').$$}
\vspace{-0.5cm}

\proof
	Suppose $(T_{Q}, D_{Q}) \restriction_U k = (T_{Q'}, D_{Q'}) \restriction_U k$. We will prove that  $(T_{P \semi Q}, D_{P \semi Q}) \restriction_U k+1 = $ $ (T_{P \semi Q'}, D_{P \semi Q'}) \restriction_U k+1$, which implies $ d_U(P \semi Q, P \semi Q') \leq \frac{1}{2} d_U(Q, Q').$\\

	\noindent	
	Let $t \in \btraces{P \semi Q}$ and $\length_U(t) \leq k+1$. 

	\begin{iteMize}	{$\bullet$}
	 \item Suppose $t \in \divergences{P \semi Q}$. 
		\begin{iteMize}{$-$}
			\item If $t \in \divergences{P}$, by definition, $t \in \divergences{P \semi Q'} \subseteq \btraces{P  \semi  Q'}$.
			\item Let $t=t_{1} \concat t_{2}$ with $t_{1} \concat \langle \checkmark \rangle \in \btraces{P}$, $t_{2} \in \divergences{Q} \subseteq \btraces{Q}$.  
			%Since $V \subseteq \G(P)$
			Since $P$ always communicates an event from $U \subseteq \Sigma$ before it can do a $\tick$, $t_{1}$ contains an event from $U$. Therefore, $\length_U(t_{2}) \leq k$. Then, since by assumption $(T_{Q}, D_{Q}) \restriction_U k = (T_{Q'}, D_{Q'}) \restriction_U k$, $t_{2} \in \divergences{Q'}$. Therefore, by definition, $t=t_{1} \concat t_{2} \in \divergences{P \semi Q'} \subseteq \btraces{P  \semi  Q'}$.
		\end{iteMize}

	 \item Suppose $t \in \traces{P \semi Q}$.
		\begin{iteMize}{$-$}
			\item If $t \in \traces{P} \inter \Sigma^{\ast}$, then by definition, $t \in \traces{P \semi Q'} \subseteq \btraces{P  \semi  Q'}$.
			\item Let $t=t_{1} \concat t_{2}$ with $t_{1} \concat \ticks \in \traces{P}$, $t_{2} \in \traces{Q}$. 
			%Since $V \subseteq \G(P)$
			Since $P$ always communicates an event from $U \subseteq \Sigma$ before it does a $\tick$, $t_{1}$ contains an event from $U$. Therefore,  $\length_U(t_{2}) \leq k$. Then, by assumption, $t_{2} \in \btraces{Q'}$.
			\begin{iteMize}{$*$}
				\item If $t_{2} \in \traces{Q'}$, by definition, $t=t_{1} \concat t_{2} \in \traces{P \semi Q'} \subseteq \btraces{P \semi Q'}$.
				\item Let $t_{2} \in \divergences{Q'}$. Since $t_{1} \concat \langle \checkmark \rangle \in  \traces{P}$, by definition, $t=t_{1} \concat t_{2} \in \divergences{P \semi Q'} \subseteq \btraces{P \semi Q'}$.
			\end{iteMize} 
		\end{iteMize}
		
	\end{iteMize}

\noindent Therefore, $D_{P \semi Q} \restriction_U k+1 \subseteq D_{P
  \semi Q'} \restriction_U k+1$ and $T_{P \semi Q} \restriction_U k+1
\subseteq T_{P \semi Q'} \restriction_U k+1$. The reverse containments
are established similarly by symmetry. Therefore, $(T_{P \semi Q},
D_{P \semi Q}) \restriction_U k+1 = (T_{P \semi Q'}, D_{P \semi Q'})
\restriction_U k+1$ and, hence, $ d_U(P \semi Q, P \semi Q') \leq
\frac{1}{2} d_U(Q, Q')$.  \qed

% \begin{lem}\label{prop_app__arg_nond_ch}
% For any CSP processes $P, P'$ and $Q$: 
% $$ d_U(P \intchoice Q, P' \intchoice Q) \leq d_U(P, P').$$
% \end{lem}

\vspace{0.2cm}\noindent
\textbf{Lemma~\ref{lemma_nonexp_in_args} ($\intchoice$).} \textit{For any CSP processes $P, P'$, and $Q$: 
 $$ d_U(P \intchoice Q, P' \intchoice Q) \leq d_U(P, P').$$}
\vspace{-0.5cm}

\proof
 Suppose $(T_{P}, D_{P}) \restriction_U k = (T_{P'}, D_{P'}) \restriction_U k$. We will prove that $(T_{P \intchoice Q}, D_{P \intchoice Q}) \restriction_U k = (T_{P'  \intchoice Q}, D_{P'  \intchoice Q}) \restriction_U k$, which directly implies $ d_U(P \intchoice Q, P' \intchoice Q) \leq d_U(P, P')$.
\\

\noindent
 Let $t \in \divergences{P \intchoice Q}$ and $\length_U(t) \leq k$.
\begin{iteMize}{$\bullet$}
 \item Suppose $t \in \divergences{P}$. By assumption, $ D_{P} \restriction_U k =  D_{P'} \restriction_U k$. Therefore,  $t \in \divergences{P'}$ $\subseteq \divergences{P' \intchoice Q}$.
 \item Suppose $t \in \divergences{Q}$. By definition, $t \in  \divergences{P' \intchoice Q}$.
\end{iteMize}

\noindent
 Let $t \in \btraces{P \intchoice Q}$ and $\length_U(t) \leq k$. We have that $\btraces{P \intchoice Q} = \traces{P \intchoice Q} \union \divergences{P \intchoice Q} = \traces{P} \union \divergences{P} \union \traces{Q} \union \divergences{Q} = \btraces{P} \union \btraces{Q}$.
\begin{iteMize}{$\bullet$}
 \item Let $t \in \btraces{P}$. By assumption, $ T_{P} \restriction_U k =  T_{P'} \restriction_U k$. Therefore,  $t \in \btraces{P'} \subseteq  \btraces{P' \intchoice Q}$.
 \item Let $t \in \btraces{Q}$. By definition, $t \in \btraces{P' \intchoice Q}$.
\end{iteMize}

\noindent Therefore, $D_{P \intchoice Q} \restriction_U k \subseteq D_{P'  \intchoice Q} \restriction_U k$ and $T_{P  \intchoice Q} \restriction_U k \subseteq T_{P'  \intchoice Q} \restriction_U k$. The reverse containments are established similarly by symmetry. Therefore, $(T_{P \intchoice Q}, D_{P \intchoice Q}) \restriction_U k = (T_{P'  \intchoice Q}, D_{P'  \intchoice Q}) \restriction_U k$ and, hence, $ d_U(P \intchoice Q, P' \intchoice Q) \leq d_U(P, P')$. 
\qed

% \begin{lem}\label{prop_app__arg_det_ch}
%  For any CSP processes $P, P'$ and $Q$:
% $$ d_U(P \extchoice Q, P' \extchoice Q) \leq d_U(P, P').$$
% \end{lem}

\vspace{0.2cm}\noindent
\textbf{Lemma~\ref{lemma_nonexp_in_args} ($\extchoice$).} \textit{For any CSP processes $P, P'$, and $Q$:
 $$ d_U(P \extchoice Q, P' \extchoice Q) \leq d_U(P, P').$$}
\vspace{-0.5cm}

\proof
Same as for $\intchoice$.
\qed

% \begin{lem}\label{prop_app__arg_par_comp}
%  For any CSP processes $P, P'$ and $Q$ and any $A \subseteq \Sigma$:
% $$ d_U(P \parallel[A] Q, P' \parallel[A] Q) \leq d_U(P, P').$$
% \end{lem}

\vspace{0.2cm}\noindent
\textbf{Lemma~\ref{lemma_nonexp_in_args} ($\parallel[A]$).} \textit{For any CSP processes $P, P'$, and $Q$ and any $A \subseteq \Sigma$:
 $$ d_U(P \parallel[A] Q, P' \parallel[A] Q) \leq d_U(P, P').$$}
\vspace{-0.3cm}

\proof
Suppose $(T_{P}, D_{P}) \restriction_U k = (T_{P'}, D_{P'}) \restriction_U k$. We will prove that $(T_{P \parallel[A] Q}, D_{P \parallel[A] Q}) \restriction_U k = (T_{P'  \parallel[A] Q}, D_{P'  \parallel[A] Q}) \restriction_U k$, which directly implies $ d_U(P \parallel[A] Q, P' \parallel[A] Q) \leq d_U(P, P')$.\\

\noindent
Let $t \in \divergences{P \parallel[A] Q}$ and $\length_U(t) \leq k$. Therefore, $t = u \concat   v$ with \mbox{$u \in (s \parallel[A] r  \inter \Sigma^{\ast})$}, $s \in \btraces{P}$, $r \in \btraces{Q}$ and, $s \in \divergences{P}$ or $r \in \divergences{Q}$. Let us recall that $v$ ranges over
$\Sigma^{*\tick\!}$, in accordance with Axiom 4. Let us observe that $\length_U(s) \leq \length_U(u) \leq \length_U(t) \leq k$. Therefore, by assumption, $s \in \btraces{P'}$.
\begin{iteMize}{$\bullet$}
 \item Let $s \in \divergences{P}$. By assumption, $s \in \divergences{P'}$. Therefore by definition, $t \in \divergences{P' \parallel[A] Q}$.

 \item Let $r \in \divergences{Q}$. Since $s \in \btraces{P'}$, by definition, $t \in \divergences{P' \parallel[A] Q}$.
\end{iteMize}

\noindent
 Let $t \in \btraces{P \parallel[A] Q}$ and $\length_U(t) \leq k$.

\begin{iteMize}{$\bullet$}
 \item Suppose $t \in \divergences{P \parallel[A] Q}$. We already proved that $t \in \divergences{P \parallel[A] Q'}$  $\subseteq$  \mbox{ $\btraces{P' \parallel[A] Q}$.}
 \item Suppose $t \in \traces{P \parallel[A] Q}$. Therefore, there exist $s \in \traces{P} \subseteq \btraces{P}$, $r \in \traces{Q} \subseteq \btraces{Q}$, such that $t \in s \parallel[A] r$. By assumption, $s \in \btraces{P'}$.
\begin{iteMize}{$-$}
	\item If $s \in \traces{P'}$, by definition, $t \in \traces{P' \parallel[A] Q} \subseteq \btraces{P' \parallel[A] Q}$.
	\item If $s \in \divergences{P'}$, by definition, $t \in \divergences{P' \parallel[A] Q} \subseteq \btraces{P' \parallel[A] Q}$.   
\end{iteMize} 
\end{iteMize}

\noindent Therefore, $D_{P \parallel[A] Q} \restriction_U k \subseteq D_{P'  \parallel[A] Q} \restriction_U k$ and $T_{P  \parallel[A] Q} \restriction_U k \subseteq T_{P'  \parallel[A] Q} \restriction_U k$. The reverse containments are established similarly by symmetry. Therefore, $(T_{P \parallel[A] Q}, D_{P \parallel[A] Q}) \restriction_U k = (T_{P'  \parallel[A] Q}, D_{P'  \parallel[A] Q}) \restriction_U k$ and, hence, $ d_U(P \parallel[A] Q, P' \parallel[A] Q) \leq d_U(P, P')$.  
\qed

% \begin{lem}\label{prop_app__arg_hiding}
% Let $P$ and $Q$ be CSP processes and let $A \subseteq \Sigma $ satisfy $A \inter U = \emptyset$. Then:
% $$ d_{U}(P \hide A, Q \hide A) \leq d_{U}(P, Q).$$
% \end{lem}

\vspace{0.2cm}\noindent
\textbf{Lemma~\ref{lemma_nonexp_hiding}.} \textit{Let $P$ and $Q$ be CSP processes and let $A \subseteq \Sigma $ satisfy $A \inter U = \emptyset$. Then:
$$ d_{U}(P \hide A, Q \hide A) \leq d_{U}(P, Q).$$}
\vspace{-0.5cm}

\proof
Suppose $(T_{P}, D_{P}) \restriction_U k = (T_{Q}, D_{Q}) \restriction_U k$. We will prove that $(T_{P \hide A}, D_{P \hide A}) \restriction_U k = (T_{Q \hide A}, D_{Q \hide A}) \restriction_U k$, which implies $ d_U(P \hide A, Q \hide A) \leq d_U(P, Q)$.\\

\noindent
Let $t \in \divergences{P \hide A}$ and $\length_U(t) \leq k$. We consider the possible alternatives for $t$.
\begin{iteMize}{$\bullet$}
 \item Suppose that there exists $s \in \divergences{P}$, such that $t = (s \restriction (\Sigma \backslash A) ) \concat  r$. Since $A \inter U = \emptyset$, $\length_{U}(s) = \length_{U}(s  \restriction (\Sigma \backslash A)) \leq \length_{U}(t) \leq k$. Then, by assumption, $ s \in \divergences{Q}$. Therefore, by definition, $t \in \divergences{Q \hide A}$. 
 \item Now suppose that there exists $u \in \Sigma^{\omega}$, such that $u \restriction (\Sigma \backslash A) $ is finite, for each $s < u$, $s \in \btraces{P} $, and $t = u \restriction (\Sigma \backslash A) \hspace{0.1cm} \concat \hspace{0.1cm} r$. Since $A \inter U = \emptyset$, $\length_{U}(u) = \length_{U}(u  \restriction (\Sigma \backslash A)) \leq \length_{V}(t) \leq k$. Then, by assumption,  for each $s < u$, $s \in \btraces{Q} $. Hence $t \in \divergences{Q \hide A}$ follows by definition.
\end{iteMize}

\noindent
Let $t \in \btraces{P \hide A}$ and $\length_U(t) \leq k$.

\begin{iteMize}{$\bullet$}
 \item Let first $t \in \divergences{P \hide A}$. We already proved that $t \in \divergences{Q \hide A} \subseteq$  \mbox{$\btraces{Q \hide A}$}.
 \item Let now $t \in \traces{P \hide A}$. Therefore, there exists $s \in \traces{P} \subseteq \btraces{P}$, such that $t = s  \restriction (\Sigma \backslash A)$. From $A \inter U = \emptyset$, $\length_{U}(s) = \length_{U}(s   \restriction (\Sigma \backslash A)) = \length_{U}(t) \leq k$. Then, by assumption, $s \in \btraces{Q}$.
	\begin{iteMize}{$-$}
		\item If $s \in \divergences{Q}$, by definition, $t \in \divergences{Q \hide A} \subseteq$  \mbox{$\btraces{Q \hide A}$}.	
		\item If $s \in \traces{Q}$, by definition, $t \in \traces{Q \hide A} \subseteq \btraces{Q \hide A}$.	
	\end{iteMize}
\end{iteMize}

\noindent Therefore, $D_{P \hide A} \restriction_U k \subseteq D_{Q \hide A} \restriction_U k$ and $T_{P \hide A} \restriction_U k \subseteq T_{Q \hide A} \restriction_U k$. The reverse containments are established similarly by symmetry.
 Therefore, $(T_{P \hide A}, D_{P \hide A}) \restriction_U k = (T_{Q \hide A}, D_{Q \hide A}) \restriction_U k$ and, hence, $ d_U(P \hide A, Q \hide A) \leq d_U(P, Q)$.
\qed

% \begin{lem}\label{prop_app__arg_renaming}
%  Let $P$ and $Q$ be CSP processes, $R \subseteq \Sigma \times \Sigma$ be a renaming relation on $\Sigma$ and $R(U) = \set{y \st \exists x \in U \exst x \mathrel{R} y}$. Then:
%    $$ d_{R(U)}(P\renam{R}, Q\renam{R}) \leq d_{U}(P, Q).$$
% \end{lem}

\vspace{0.2cm}\noindent
\textbf{Lemma~\ref{lemma_nonexp_renaming}.} \textit{Let $P$ and $Q$ be CSP processes, $R \subseteq \Sigma \times \Sigma$ be a renaming relation on $\Sigma$ and $R(U) = \set{y \st \exists x \in U \exst x \mathrel{R} y}$. Then:
    $$ d_{R(U)}(P\renam{R}, Q\renam{R}) \leq d_{U}(P, Q).$$}
\vspace{-0.5cm}

\proof

Suppose $(T_{P}, D_{P}) \restriction_U k = (T_{Q}, D_{Q}) \restriction_U k$.  We will prove that $(T_{P \renam{R}}, D_{P \renam{R}}) \restriction_{R(U)} k = (T_{Q \renam{R}}, D_{Q \renam{R}}) \restriction_{R(U)} k$. \\

\noindent
Let $t \in \divergences{P \renam{R}}$ and $\length_{R(U)}(t) \leq k$. Then there exist $s_{1}, t_{1} \in \Sigma^{\ast}, r \in \Sigma^{\ast \checkmark}$, such that $ s_{1} \in \divergences{P} \inter \Sigma^{\ast}$,  $ s_{1} \mathrel{R} t_{1}$ and $t = t_{1} \concat r$.  Then, $\length(s_{1}) = \length(t_{1})$ and for $1 \leq i \leq \length(s_1)$, $s_{1_i} \mathrel{R} t_{1_i}$. Therefore, $ \length_U(s_1) = \length_{R(U)}(t_1) \leq \length_{R(U)}(t) \leq k$ and, by assumption, $s_1 \in \divergences{Q} \inter \Sigma^{\ast}$. Hence, by definition, $t \in \divergences{Q \renam{R}}$. \\

\noindent
Let $t \in \btraces{P \renam{R}}$ and $\length_{R(U)}(t) \leq k$.
\begin{iteMize}{$\bullet$}
 \item If $t \in \divergences{P \renam{R}}$, we already proved that $t \in \divergences{Q \renam{R}} \subseteq \btraces{Q \renam{R}}$. 

 \item Let $t \in \traces{P \renam{R}}$. Then there exists $s \in \traces{P}$, such that  $ s \mathrel{R} t$. Therefore, $ \length_U(s) = \length_{R(U)}(t) \leq k$ and, by assumption, $s \in \btraces{Q}$.
	\begin{iteMize}{$-$}
		\item If $s \in \traces{Q}$, by definition, $t \in \traces{Q \renam{R}} \subseteq \btraces{Q \renam{R}}$.
		\item If $s \in \divergences{Q}$, by definition, $t \in \divergences{Q \renam{R}} \subseteq \btraces{Q \renam{R}}$.
	\end{iteMize}	
\end{iteMize}

\noindent Therefore, $D_{P \renam{R}} \restriction_{R(U)} k \subseteq D_{Q \renam{R}} \restriction_{R(U)} k$ and $T_{P \renam{R}} \restriction_{R(U)} k \subseteq T_{Q \renam{R}} \restriction_{R(U)} k$. The reverse containments are established similarly by symmetry.
 Therefore, $(T_{P \renam{R}}, D_{P \renam{R}}) \restriction_{R(U)} k = (T_{Q \renam{R}}, D_{Q \renam{R}}) \restriction_{R(U)} k$ and, hence, $ d_{R(U)}(P \renam{R}, Q \renam{R}) \leq d_U(P, Q)$.
\qed

%%%%%%%%%%%%%%%%%%%%%%%%%%%%%%%%%%%%%%%%%%%%%%%%%%%%%%%%%%%%%%

\section{Proofs for Section~\ref{StaticLivelockAnalysis}}\label{app_static_analysis}

\vspace{0.2cm}\noindent
\textbf{Proposition~\ref{prop_non_exp}.} \textit{
 Let $P(X, Y_{1}, \ldots, Y_{n}) = P(X, \overline{Y})$ be a CSP term whose free variables are contained within the set $\{ X, Y_{1}, \ldots, Y_{n} \}$. Let $N_{X} : \TCSP \longrightarrow \mathcal{P}(\mathcal{P}(\Sigma) \times \mathcal{P}(\Sigma))$ be defined recursively on the structure of $P$ as shown in \autoref{fig_nonexp_0}. If $(U, V)  \in N_{X}(P)$, then for all $T_{1}, T_{2}, \varTheta_{1}, \ldots, \varTheta_{n}  \in \TD$, $  d_{V}(P(T_{1}, \overline{\varTheta}), P(T_{2}, \overline{\varTheta})) \leq d_{U}(T_{1}, T_{2})$.
}
\vspace{0.2cm}

% \begin{prop}\label{prop_app_N}
%  Let $P(X, Y_{1}, \ldots, Y_{n}) = P(X, \overline{Y})$ be a CSP term whose free variables are contained within the set $\{ X, Y_{1}, \ldots, Y_{n} \}$. Let $N_{X} : \TCSP \longrightarrow \mathcal{P}(\mathcal{P}(\Sigma) \times \mathcal{P}(\Sigma))$ be defined recursively on the structure of $P$ as shown in \autoref{fig_nonexp_0}. If $(U, V)  \in N_{X}(P)$, then for all $T_{1}, T_{2}, \varTheta_{1}, \ldots, \varTheta_{n}  \in \TD$, $  d_{V}(P(T_{1}, \overline{\varTheta}), P(T_{2}, \overline{\varTheta})) \leq d_{U}(T_{1}, T_{2})$. 
% \end{prop}

\proof

Structural induction on $P$. Let us take arbitrary $T_{1}, T_{2}, \varTheta_{1}, \ldots, \varTheta_{n}  \in \TD$.

\begin{iteMize}{$\bullet$}\setlength{\leftmargin}{-3cm}

 \item $\N_{X}( P ) \widehat{=} \mathcal{P}(\Sigma) \times \mathcal{P}(\Sigma)$ whenever $X$ is not free in $P$.
      {
	\proof
	 Let $(U, V) \in \N_{X}(P)$. 	 
	 Then 
\[d_{U}(T_{1}, T_{2})  \geq 0  =  d_{V}( P(T_{1}, \overline{\varTheta}), P(T_{2}, \overline{\varTheta})) = d_{V}( P(\overline{\varTheta}), P(\overline{\varTheta})).\eqno{\qEd}
\]
      }
 \item $\N_{X}(a \then P ) \widehat{=} \N_{X}(P)$.
      {
	\proof
		Suppose $(U, V) \in \N_{X}(a \then P)$. By construction, $(U, V) \in \N_{X}(P)$.  Then:\vspace{0.1cm}
		
		\noindent
		\begin{tabular}{@{\hspace{-0.00cm}} l l l}
	 		$d_{U}(T_{1}, T_{2})$ 	& $\geq d_{V}( P(T_{1}, \overline{\varTheta}), P(T_{2}, \overline{\varTheta})) $ 	&  // induction hypothesis  \\
	 					& $\geq d_{V}( a \then P(T_{1}, \overline{\varTheta}), a \then  P(T_{2}, \overline{\varTheta})) $ &  // Lemma~\autoref{lemma_nonexp_prefix}, $>$ if $a \in V$  \\
						& $ = d_{V}( (a \then P)(T_{1}, \overline{\varTheta}), (a \then  P)(T_{2}, \overline{\varTheta})) $\rlap{\hbox to 167 pt{\hfill\qEd}} &  
	 			
		\end{tabular}
      }
 \item $\N_{X}(P \hide A) \widehat{=} \{(U,V) \st (U,V') \in \N_X(P) \land V' \cap A = \emptyset \land V' \subseteq V \}$.
      {
	\proof
	 Suppose $(U, V) \in \N_{X}(P \hide A)$. 
	 By construction, there exists $V'$, such that $(U, V') \in  \N_{X}(P)$,   $V' \subseteq V $ and $V' \cap A = \emptyset $.
	 We will prove that for any $T_{1}, T_{2} \in \TD$, $d_{U}(T_{1}, T_{2}) \geq d_{V}(P(T_{1}, \overline{\varTheta}) \hide A, P(T_{2}, \overline{\varTheta}) \hide A)$. \vspace{0.1cm}

	\noindent
	\begin{tabular}{@{\hspace{-0.00cm}} l l l}
	 $d_{U}(T_{1}, T_{2})$ 	& $\geq d_{V'}(P(T_{1}, \overline{\varTheta}), P(T_{2}, \overline{\varTheta})) $ 				&  // induction hypothesis  \\
	 			& $\geq d_{V'}(P(T_{1}, \overline{\varTheta}) \hide A, P(T_{2}, \overline{\varTheta}) \hide A ) $ &  // $V' \cap A = \emptyset $, Lemma~\ref{lemma_nonexp_hiding} \\
	 			& $\geq d_{V}(P(T_{1}, \overline{\varTheta}) \hide A, P(T_{2}, \overline{\varTheta}) \hide A ) $  &  // $V' \subseteq V $, $U \mapsto d_{U}$ antitone\rlap{\hbox to 52 pt{\hfill\qEd}}
	\end{tabular}
       }

\item $\N_{X}( P_{1} \oplus P_{2} ) \widehat{=} \N_{X}(P_{1}) \cap \N_{X}(P_{2}) = \{ (U_{1} \cap U_{2} , V_{1} \cup V_{2}) \st (U_{i}, V_{i}) \in \N_{X}(P_{i})  \}$ for every \mbox{$\oplus \in \{ \intchoice, \extchoice, \semi, \parallel[A] \}$}.

{
\proof

 Suppose $(U, V) \in \N_{X}(P_{1} \oplus  P_{2})$.
 By construction, there exist $(U_{1}, V_{1}) \in \N_{X}(P_{1})$ and $(U_{2}, V_{2}) \in \N_{X}(P_{2})$, such that $U = U_{1} \cap U_{2}$ and $V = V_{1} \cup V_{2}$. Therefore, $(U, V) \in \N_{X}(P_{1}$), $(U, V) \in \N_{X}(P_{2})$ (antitoneness).\vspace{0.1cm}

\noindent
 $d_{V}((P_{1} \oplus P_{2}) (T_{1}, \overline{\varTheta}), (P_{1} \oplus P_{2})(T_{2}, \overline{\varTheta})) $ 

\noindent 
 $ =  d_{V}(P_{1}(T_{1}, \overline{\varTheta})  \oplus P_{2}(T_{1}, \overline{\varTheta}), P_{1}(T_{2}, \overline{\varTheta}) \oplus P_{2}(T_{2}, \overline{\varTheta}))  $  

\noindent
// ultrametric inequality

\noindent
\begin{tabular}{@{\hspace{-0.00cm}} l l}
$ \leq max \{$ & $ d_{V}(P_{1}(T_{1}, \overline{\varTheta})  \oplus P_{2}(T_{1}, \overline{\varTheta}), P_{1}(T_{2}, \overline{\varTheta}) \oplus P_{2}(T_{1}, \overline{\varTheta}) ),$   \\
	       & $ d_{V}(P_{1}(T_{2}, \overline{\varTheta})   \oplus P_{2}(T_{1}, \overline{\varTheta}), P_{1}(T_{2}, \overline{\varTheta}) \oplus P_{2}(T_{2}, \overline{\varTheta}) )  \}$ \\
\end{tabular}

\noindent
// Lemma~\ref{lemma_nonexp_in_args}

\noindent
\begin{tabular}{@{\hspace{-0.00cm}} l l}
$ \leq max \{$ & $d_{V}(P_{1}(T_{1}, \overline{\varTheta}), P_{1}(T_{2}, \overline{\varTheta}))$     \hspace{0.2cm}//  $\leq d_{U}(T_{1}, T_{2})$ by induction hypothesis for $P_{1}$  \\	
	    & $d_{V}(P_{2}(T_{1}, \overline{\varTheta}), P_{2}(T_{2}, \overline{\varTheta}))  \}$ //  $\leq d_{U}(T_{1}, T_{2})$ by induction hypothesis for $P_{2}$ \\
\end{tabular}\\
$ \leq d_{U}(T_{1}, T_{2})$ 
\qed
}

\item $\N_{X}(P [R]) \widehat{=} \{(U,V) \st (U,V') \in \N_X(P) \land R(V') \subseteq V \}$.
 {   
	\proof
	  Suppose $(U, V) \in \N_{X}(P [R])$. 
	 By construction, there exists $V'$, such that $(U, V') \in  \N_{X}(P)$ and   $R(V') \subseteq V $.\vspace{0.1cm}
	 
	\noindent
	\begin{tabular}{@{\hspace{-0.00cm}}l l l}
	 $d_{U}(T_{1}, T_{2})$ 	& $\geq d_{V'}(P(T_{1}, \overline{\varTheta}), P(T_{2}, \overline{\varTheta})) $ 				&  // induction hypothesis  \\
	 			& $\geq d_{R(V')}(P(T_{1}, \overline{\varTheta}) [R], P(T_{2}, \overline{\varTheta}) [R]) $ &  //  Lemma~\ref{lemma_nonexp_renaming} \\
	 			& $\geq d_{V}(P(T_{1}, \overline{\varTheta}) [R], P(T_{2}, \overline{\varTheta})[R]) $  &  // $R(V') \subseteq V $, $U \mapsto d_{U}$ antitone \\
	\end{tabular}	\qed

}

\item $\N_{X}(X) \widehat{=} \{  (U, V) \st U \subseteq V  \}$.
{
	\proof\hfill

		\noindent
	 	\begin{tabular}{@{\hspace{-0.00cm}}l l l}
		$d_{U}(T_{1}, T_{2})$ 	& $\geq d_{V}(T_{1}, T_{2})$ 	   &  // $U \subseteq V $, $U  \mapsto d_{U}$ antitone \\
					& $= d_{V}((X)(T_{1}, \overline{\varTheta}) , (X)(T_{2}, \overline{\varTheta}))$   &  \\
					
		\end{tabular}						
	\qed
}

\item $\N_{X}(\mu Y \centerdot P) \widehat{=} \set{(U,V) \st
                (U',V') \in \N_X(P) \land (V',V') \in \N_Y(P) \land U \subseteq U' \land V' \subseteq V }$ if $Y \neq X$.
{
	\proof
	
	Suppose $(U, V) \in \N_{X}(\mu Y \centerdot P)$ for $X \neq Y$ and $X, Y$ free in $P(X, Y, Z_{1}, \ldots, Z_{n})$.	
	By construction, there exist $U_{X}, V_{X} \subseteq \Sigma$  such that:
	\begin{enumerate}[(1)]
		\item  $(U_{X}, V_{X}) \in \N_{X}(P)$
		\item  $U \subseteq U_{X} $, $V_{X} \subseteq V $ \label{rec_x}		
		\item  $(V_{X}, V_{X}) \in \N_{Y}(P)$ 			
	\end{enumerate}
	Therefore, by induction hypothesis, for all $T_1, T_2, \xi, \overline{\varTheta} \in \TD$, we have:
	\begin{equation}\label{ih_n_x}
	  d_{U_X}(T_{1}, T_{2}) \geq d_{V_{X}}(P(T_{1}, \xi, \overline{\varTheta}), P(T_{2}, \xi, \overline{\varTheta}))
	\end{equation}
	\begin{equation}\label{ih_n_y}
	 d_{V_X}(T_{1}, T_{2}) \geq d_{V_{X}}(P(\xi, T_1, \overline{\varTheta}), P(\xi, T_2, \overline{\varTheta}))
	\end{equation}

	\vspace{0.1cm}
        \noindent
	\begin{tabular}{@{\hspace{-0.00cm}}l l l}
		$d_{U}(T_{1}, T_{2})$ 	& $\geq d_{U_{X}}(T_{1}, T_{2})$   &  // $U \subseteq U_{X}$, antitoneness \\
					& $\geq d_{V_{X}}(P(T_{1}, \xi, \overline{\varTheta}), P(T_{2}, \xi, \overline{\varTheta}))$   &  //  from \ref{ih_n_x} \\ 					
	\end{tabular}	

	\vspace{0.1cm}
	Let $P_{1}(Y) = P(T_{1}, Y, \overline{\varTheta})$, $P_{2}(Y) = P(T_{2}, Y, \overline{\varTheta})$. $P_{1}(Y)$ and $P_{2}(Y)$ are continuous over $\sqsubseteq$. Therefore, there exist $\mu Y \centerdot P_{1}(Y) = \bigcap_{n = 0}^{\infty} P_{1}^{n} = P_{1}^{\ast}$ and $\mu Y \centerdot P_{2}(Y) = \bigcap_{n = 0}^{\infty} P_{2}^{n} = P_{2}^{\ast}$, where for $i=1,2$, $P_{i}^{0} = \bot = \DIV$, $P_{i}^{n+1} = P_i(P_{i}^{n})$. 

	We will prove by induction that
	\begin{equation}\label{iiih}
	  d_{V_{X}}(P_{1}^{n}, P_{2}^{n}) \leq d_{U_{X}}(T_{1}, T_{2}) \mbox{ for } n \geq 1.
	\end{equation}

	\begin{iteMize}{$-$}
	
	\item Let $n = 1$. \\
	\begin{tabular}{@{\hspace{-0.00cm}} l l l}	\vspace{0.1cm}

		\noindent
		$d_{U_X}(T_{1}, T_{2})$			& $\geq d_{V_{X}}(P(T_{1}, \DIV, \overline{\varTheta}), P(T_{2}, \DIV, \overline{\varTheta}))$   &  //  from \ref{ih_n_x} \\		
					& $ = d_{V_{X}}(P_{1}^{1}, P_{2}^{1})$   &  \\						
	\end{tabular}	

	\item Suppose $ d_{V_{X}}(P_{1}^{n}, P_{2}^{n}) \leq d_{U_X}(T_{1}, T_{2})$. \vspace{0.1cm}

	\noindent
	\begin{tabular}{@{\hspace{-0.00cm}} l l l}
		$d_{V_{X}}(P_{1}^{n+1}, P_{2}^{n+1})$ 	& $ = d_{V_{X}}(  P(T_{1}, P_{1}^{n}, \overline{\varTheta}),
							                  P(T_{2}, P_{2}^{n}, \overline{\varTheta})  )   $   \\
					& // ultrametric inequality & \\
					& $\leq max \{  d_{V_{X}}(  P(T_{1}, P_{1}^{n}, \overline{\varTheta}),
							                  P(T_{2}, P_{1}^{n}, \overline{\varTheta})  ),$   &    \\
					& $ \hspace{1.2cm}	d_{V_{X}}(  P(T_{2}, P_{1}^{n}, \overline{\varTheta}),
							                  P(T_{2}, P_{2}^{n}, \overline{\varTheta})  ) \} $   & \\

					& $\leq max \{  d_{U_{X}}(  T_{1}, T_{2} ),$   & \hspace{-1.5cm} // from \ref{ih_n_x} \\ 
					& $ \hspace{1.2cm} d_{V_{X}}(  P_{1}^{n},  P_{2}^{n}  ) \} $   & \hspace{-1.5cm} // from \ref{ih_n_y}  \\
					& $ \leq max \{  d_{U_{X}}(  T_{1}, T_{2} ),$   &     \\ 
					& $ \hspace{1.2cm} d_{U_{X}}(  T_{1}, T_{2}  \} $   & \hspace{-1.5cm} // from \ref{iiih}, local i.h.   \\
					& $ \leq  d_{U_{X}}(  T_{1}, T_{2} )$   &     \\ 					
					
	\end{tabular}	
	\end{iteMize}

\noindent Let $d_{U_{X}}(T_{1}, T_{2}) = 2^{-k}$ for some $k \in \mathbb{N}$. Now suppose for the sake of contradiction that $d_{V_{X}}(P_{1}^{\ast}, P_{2}^{\ast}) > d_{U_{X}}(T_{1}, T_{2}) = 2^{-k}$ and let, without loss of generality, $P_{1}^{\ast}$ and $P_{2}^{\ast}$ differ on the sets of their divergences. Therefore, again without loss of generality, there exists $s \in \divergences{P_{1}^{\ast}}$ such that $s \not \in \divergences{P_{2}^{\ast}}$ and $  \length_{V_{X}}(s) \leq k$. Then, since $P_{i}^{\ast} = \bigcap_{n = 0}^{\infty} P_{i}^{n}$, $s \in P_{1}^{n}$ for all $n \in \mathbb{N}$, but there exists $l \in \mathbb{N}$ such that $s \not \in P_{2}^{l}$. But then $d_{V_{X}}(P_{1}^{l}, P_{2}^{l}) > 2^{-k} = d_{U_{X}}(T_{1}, T_{2})$, which is a contradiction with \ref{iiih}. Therefore, $d_{V_{X}}(P_{1}^{\ast}, P_{2}^{\ast}) \leq d_{U_{X}}(T_{1}, T_{2})$. Then, since $U \subseteq U_{X} $ and  $V_{X} \subseteq V $, by antitoneness, $d_{V}((\mu Y \centerdot P)(T_{1}, \overline{\varTheta}), (\mu Y \centerdot P)(T_{2}, \overline{\varTheta})) = d_{V}(P_{1}^{\ast}, P_{2}^{\ast}) \leq d_{U}(T_{1}, T_{2})$.
	\qed
}
\qed
\end{iteMize}

\begin{prop}\label{prop_app_GCF}
 Let $P(X, Y_{1}, \ldots, Y_{n}) = P(X, \overline{Y})$ be a CSP term whose free variables are contained within the set $\{ X, Y_{1}, \ldots, Y_{n} \}$. Let $\G : \TCSP \longrightarrow \mathcal{P}(\mathcal{P}(\Sigma))$, $\C_{X} : \TCSP \longrightarrow \mathcal{P}(\mathcal{P}(\Sigma) \times \mathcal{P}(\Sigma))$ and $\F : \TCSP \longrightarrow \mathcal{P}(\mathcal{P}(\Sigma) \times \mathcal{P}(\Sigma))$ be defined recursively on the structure of $P$ as shown in Figures \ref{fig_guards_0}, \ref{fig_contract_0} and \ref{fig_fairsets_0}, respectively. Then:
  
\begin{enumerate}[\em(1)]\setlength{\itemsep}{0.15cm}\setlength{\headsep}{0.15cm}
  \item If $V  \in \G(P)$, then, with any processes substituted for the free variables of $P$ (and in particular $\DIV$),  $P$ must communicate an event from $V$ before it can do a $\tick$. \label{hypo_G}

  \item If $(U, V)  \in \C_{X}(P)$, then for all processes $T_{1}, T_{2}, \varTheta_{1}, \ldots, \varTheta_{n}  \in \TD$, \\ $ d_{V}(P(T_{1}, \overline{\varTheta}), P(T_{2}, \overline{\varTheta})) \leq \frac{1}{2} d_{U}(T_{1}, T_{2})$. \label{hypo_Cx}

  \item If $(U, V) \in \F(P)$, then, for any collection of $U$-fair livelock-free processes $\theta_{0}, \ldots, \theta_{n} \in \TD$, the process $P(\theta_{0}, \ldots, \theta_{n})$ is livelock-free and $V$-fair. \label{hypo_F}   
\end{enumerate} 
\end{prop}

\proof
 We carry out the proof by induction on the structure of $P$. For clarity, we prove (\ref{hypo_G}), (\ref{hypo_Cx}) and (\ref{hypo_F}) one by one, in Propositions \ref{prop_guard_sets}, \ref{prop_contr_sets} and \ref{prop_fair_sets}, respectively. In each of these propositions, our induction hypothesis is that at any point all (\ref{hypo_G}), (\ref{hypo_Cx}) and (\ref{hypo_F}) hold for any subterm of $P$.
\qed

% \begin{prop}\label{prop_app_G}
%  Let $P(X, Y_{1}, \ldots, Y_{n}) = P(X, \overline{Y})$ be a CSP term whose free variables are contained within the set $\{ X, Y_{1}, \ldots, Y_{n} \}$. Let $\G : \TCSP \longrightarrow \mathcal{P}(\mathcal{P}(\Sigma))$ be defined recursively on the structure of $P$ as shown in \autoref{fig_guards_0}. If $V  \in \G(P)$, then, with any processes substituted for the free variables of $P$ (and in particular $\DIV$),  $P$ must communicate an event from $V$ before it can do a $\tick$.
% \end{prop}

\vspace{0.2cm}\noindent
\textbf{Proposition~\ref{prop_guard_sets}.} \textit{
 Let $P(X, Y_{1}, \ldots, Y_{n}) = P(X, \overline{Y})$ be a CSP term whose free variables are contained within the set $\{ X, Y_{1}, \ldots, Y_{n} \}$. Let $\G : \TCSP \longrightarrow \mathcal{P}(\mathcal{P}(\Sigma))$ be defined recursively on the structure of $P$ as shown in \autoref{fig_guards_0}. If $V  \in \G(P)$, then, with any processes substituted for the free variables of $P$ (and in particular $\DIV$),  $P$ must communicate an event from $V$ before it can do a $\tick$.
}
\vspace{0.0cm}

\proof

Structural induction on $P$. We will write $\widehat{P}$ to denote the result of substituting all free variables in $P$ with the most general  process $ \bot = \DIV$. For each process $\xi$, $\DIV \sqsubseteq \xi$. Therefore, by monotonicity of CSP operators~\cite{Ros97}, for any process term $C(X)$, $C(\DIV) \sqsubseteq C(\xi)$.

\begin{iteMize}{$\bullet$}
  \item $\G(\STOP) = \mathcal{P}(\Sigma)$.
	{
	\proof 
	    $\STOP$  cannot terminate and, therefore, the property holds vacuously.
	\qed
	}

 \item $\G(a \then P) = \G(P) \union \{ V \st a \in V \}$. 
	{
	\proof
	Let $V \in \G(a \then P)$, $t = s \concat \ticks  \in \btraces{\widehat{a \then P}} = \btraces{a \then \widehat{P}}$.
	Therefore, $t = \langle a \rangle \concat r \concat \ticks$ for some $r \in \Sigma^{\ast}$ such that $r \concat \ticks \in \btraces{\widehat{P}}$ and $s = \langle a \rangle \concat r$. Since $V \in \G(a \then P)$, by construction, $V \in \G( P)$ or $a \in V$.
		\begin{iteMize}{$-$}
			\item Suppose $V \in \G( P)$.  Then, by induction hypothesis, $r \concat \ticks$ contains an event from $V$ and, therefore, so do $s$ and $t$.
			\item Suppose $a \in V$. Then $t = \langle a \rangle \concat r \concat \ticks$ contains the event $a \in V$  before $\checkmark$.
		\end{iteMize}\qed
	}
 
 \item $\G(\SKIP) = \emptyset$. \qed

\item $\G(P_{1} \oplus P_{2}) = \G(P_{1}) \inter \G(P_{2})$ for $\oplus \in \{ \intchoice, \extchoice \}$.
	{
	\proof
	Let $V \in \G(P_{1} \oplus P_{2})$ and $t = s \concat \ticks  \in \btraces{\widehat{P_{1} \oplus P_{2}}} = \btraces{\widehat{P_{1}}} \union \btraces{\widehat{P_{2}}}$. Therefore, $t \in \btraces{\widehat{P_{1}}}$ or  $t \in \btraces{\widehat{P_{2}}}$. Let, without loss of generality, $t \in \btraces{\widehat{P_{1}}}$. By construction, $V \in \G(P_{1})$. Then, by induction hypothesis, $s$ contains an event from $V$.
	\qed
	}
\item $\G(P_1 \semi P_2) = $ 
         $\left\{ \begin{array}{ll}
                \G(P_1) \union \G(P_2) &
                     \mbox{ if $P_{1}$ is closed and $\F(P_{1}) \neq \emptyset$} \\
                \G(P_{1}) &
                      \mbox{otherwise} \enspace \\ 
       		\end{array} \right.$ \\

	{
	\proof

	Let  $V \in \G(P_{1} \semi P_{2})$ and $t = s \concat \ticks  \in \btraces{\widehat{P_{1} \semi P_{2}}}$.

	Let first $P_1$ be closed and $\F(P_{1}) \neq \emptyset$. Then, by Proposition \ref{prop_app_GCF} (\ref{hypo_F}), $P_1$ is livelock-free and, therefore, $\divergences{P_{1}} = \emptyset$. Therefore, $t = t_{1} \concat t_{2} \concat \ticks$ with  $t_{1} \concat \ticks \in \mathsf{traces}(\widehat{P_{1}})$ and  $t_{2} \concat \ticks \in \btraces{\widehat{P_{2}}}$. In this case by construction $V \in \G(P_1) \union \G(P_2)$. Let without loss of generality $V \in \G(P_{1})$.  Then, by induction hypothesis, $t_{1}$ contains an event from $V$ and therefore so does $t$.

	Let now $P_1$ be open or $\F(P_{1}) = \emptyset$. Then by construction $V \in \G(P_1)$. We consider the two possibilities for $t$.

	\begin{iteMize}{$-$}
	  \item $t = t_{1} \concat t_{2} \concat \ticks$ with  $t_{1} \concat \ticks \in \mathit{traces}(\widehat{P_{1}})$ and  $t_{2} \concat \ticks \in \btraces{\widehat{P_{2}}}$. Since $V \in  \G(P_{1})$, by induction hypothesis, $t_{1}$ contains an event from $V$ and therefore so does $t$.
	  \item $t \in \divergences{P_{1}}$ and, therefore, $t \in \btraces{P_{1}}$. Then again, by induction hypothesis, $t_{1}$ contains an event from $V$ and therefore so does $t$.\qed
	\end{iteMize}
	}

\item $\G(P_1 \parallel[A] P_2) =$ 
       $\left\{ \begin{array}{ll}
              \G(P_1) \union \G(P_2) &
                   \mbox{ if, for $i = 1,2$, $P_{i}$ is closed and $\F(P_{i}) \neq \emptyset$ } \\
              \G(P_{1}) \inter \G(P_2) &
                   \mbox{ otherwise}\enspace \\ 
     \end{array} \right.$                     \\

	{
	\proof
  
	Let  $V \in \G(P_{1} \parallel[A] P_{2})$ and $t = s \concat \ticks  \in \btraces{\widehat{P_{1} \parallel[A] P_{2}}}$.
  
	Let first both $P_1$ and $P_2$ be closed, $\F(P_{1}) \neq \emptyset$ and $\F(P_{2}) \neq \emptyset$. Then, by Proposition~\ref{prop_app_GCF} (\ref{hypo_F}), $P_1$ and $P_2$ are livelock-free and, therefore,  $P_1 \parallel[A] P_2$ is livelock-free. Therefore, $\divergences{\widehat{P_{1} \parallel[A] P_{2}}} = \emptyset$ and $\btraces{\widehat{P_{1} \parallel[A] P_{2}}} = \traces{\widehat{P_{1} \parallel[A] P_{2}}}$. By construction, $V \in  \G(P_1) \union \G(P_2)$. Let without loss of generality $V \in \G(P_{2})$. Since $t \in \traces{\widehat{P_{1} \parallel[A] P_{2}}}$, then, due to distributed termination, there exist $t_{1}$, $t_{2}$, such that $t_{1} \concat \ticks \in \mathsf{traces}(\widehat{P_{1}})$, $t_{2} \concat \ticks \in \mathsf{traces}(\widehat{P_{2}})$ and $t \in t_{1} \parallel[A] t_{1}$. By induction hypothesis,  $t_{2}$ contains an event from $V$ and therefore so does $t$. 

	Otherwise, $t \in \traces{\widehat{P_{1} \parallel[A] P_{2}}}$ or $t \in \divergences{\widehat{P_{1} \parallel[A] P_{2}}}$. 
	We consider both alternatives. By construction, $V \in \G(P_{1}) \inter \G(P_2)$, \ie $V \in \G(P_{1})$ and $V \in \G(P_{2})$.

	\begin{iteMize}{$-$}
		\item $t = s \concat \ticks \in \traces{\widehat{P_{1} \parallel[A] P_{2}}}$. Then, due to distributed termination, there exist $t_{1}$, $t_{2}$, such that $t_{1} \concat \ticks \in \mathit{traces}(\widehat{P_{1}})$, $t_{2} \concat \ticks \in \mathit{traces}(\widehat{P_{2}})$ and $t \in t_{1} \parallel[A] t_{1}$.  
		By induction hypothesis, both $t_{1}$ and $t_{2}$ contain an event from $V$ and therefore so does $t$. 

		\item $t = s \concat \ticks \in \divergences{\widehat{P_{1} \parallel[A] P_{2}}}$. Therefore, there exist $s_{1}, s_{2}, t_{1}, t_{2}$, such that $t_{1} \in \btraces{\widehat{P_{1}}}$, $t_{2} \in \btraces{\widehat{P_{2}}}$, $s_{1} \in (t_{1} \parallel[A] t_{2}) \inter \Sigma^{\ast}$, $t = s_{1} \concat s_{2} \concat \ticks$ and, $t_{1} \in \divergences{\widehat{P_{1}}}$ or $t_{2} \in \divergences{\widehat{P_{2}}}$. Let without loss of generality $t_{1} \in \divergences{\widehat{P_{1}}}$. Then $t_{1} \in \Sigma^{\ast}$ and therefore, $t_{1} \concat \ticks \in \divergences{\widehat{P_{1}}}$. Since $V \in \G(P_{1})$, by induction hypothesis $t_{1}$ contains an event from $V$ and therefore so does $t$.\qedhere

	\end{iteMize}
	}

\item $\G(P [R]) = \{V \st V' \in \G(P) \land R(V') \subseteq V \}$.

	{
	\proof

	Let $V \in \G(P\renam{R})$. Then, by construction, there exists $V' \in \G(P)$ with $R(V') \subseteq V$. Let  $t = s \concat \ticks  \in \btraces{\widehat{P [R]}}$. Then, $t \in \divergences{\widehat{P \renam{R}}}$ or $t \in \traces{\widehat{P \renam{R}}}$. We consider both alternatives.

	\begin{iteMize}{$-$}
	\item Suppose $t = s \concat \ticks \in \divergences{\widehat{P \renam{R}}}$.
	 Therefore, there exist $s_1, s_2, r_1 \in \Sigma^{\ast}$, such that $r_1 \in \divergences{P}$, $r_1 \mathrel{R} s_1$ and $t = s_1 \concat s_2 \concat \ticks$. As $r_1 \in \divergences{P}$, by Axiom 4, $r_1 \concat \ticks \in \divergences{P}$.   Then, by induction hypothesis for $P$, $r_1$ contains an event from $V'$. Since $r_1 \mathrel{R} s_1$, $s_1$ contains an event from $R(V') \subseteq V$. Therefore,  since $t = s_{1} \concat s_2 \concat \ticks $, $t$ contains an event from $V$.

	\item Suppose $t = s \concat \ticks  \in \traces{\widehat{P \renam{R}}}$.  Therefore, there exist $t', s' \in \traces{\widehat{P}}$, such that $t' = s' \concat \ticks$ and $s' \mathrel{R} s$. By induction hypothesis for $P$ and $t'$, $s'$ contains an event from $V'$. Since $s' \mathrel{R} s$, $s$ contains an event from $R(V') \subseteq V$ and, hence, $t$ contains an event from $V$.
	\end{iteMize}
	\qed
	}

\item $\G(P \hide A) = $
      $ \left\{ \begin{array}{ll}
              \{ V \st V' \in \G(P) \land V' \inter A = \emptyset \land V' \subseteq V \} &
                   \mbox{if $P$ is closed and} \\
		   & (\emptyset,\Sigma - A) \in \F(P)  \\
% 		    & \mbox{$(U,W) \in \F(P)$ with $W \cap A = \emptyset$ } \\		    
              \emptyset &
                   \mbox{otherwise}\enspace \\  
		\end{array} \right.$                     \\

	{
	\proof

\noindent Let $V \in  \G(P \hide A)$ and let, furthermore, $P$ be closed and $(\emptyset,\Sigma - A) \in \F(P)$. Then $P$ does not have free process variables and by Proposition~ \ref{prop_app_GCF} (\ref{hypo_F}) we can conclude the following:
	\begin{enumerate}[(1)]
	    \item $P$ is livelock-free, \ie $\divergences{P} = \emptyset$ and $\btraces{P} = \traces{P}$. \label{enum_G_hide_lf}
	    \item Any infinite trace $u$ of $P$ contains infinitely many events from $\Sigma \backslash A$ and therefore $u \upharpoonright (\Sigma \backslash A)$ is infinite.\label{enum_G_hide_inf_tr}	
	\end{enumerate}

\noindent Let $t = s \concat \ticks  \in \btraces{\widehat{P} \hide A} = \btraces{P \hide A}$. Then $t \in \divergences{P \hide A}$ or $t \in \traces{P \hide A}$. We consider both alternatives.

	\begin{iteMize}{$-$}
	      \item Let $t = s \concat \ticks  \in \divergences{P \hide A}$. As from (\ref{enum_G_hide_lf}) $\divergences{P} = \emptyset$ (i.e., $t$ cannot arise from a divergence of $P$), by definition there exists $u \in \inftraces{P}$ such that $s_1 = u \upharpoonright (\Sigma \backslash A)$ is finite and $t = s_1 \concat s_2 \concat \ticks$. However, by (\ref{enum_G_hide_inf_tr}), $u \upharpoonright (\Sigma \backslash A)$ cannot be finite for any infinite trace $u$ of $P$. Due to the contradiction, this case is not possible.

 	      \item Therefore, $t = s \concat \ticks  \in \traces{P \hide A}$. Therefore, there exists $t' = s' \concat \ticks \in \traces{P}$, such that $s = s' \upharpoonright (\Sigma \backslash A)$. 	
	      Since $V \in G (P \hide A)$, by construction there exists $V' \in \G(P)$ with $V' \subseteq V$ and $V' \cap A = \emptyset$. By induction hypothesis for $t'$ and $P$, $s'$ contains an event from $V' \subseteq V$. But $V' \cap A = \emptyset$. Hence, $s = s' \upharpoonright (\Sigma \backslash A)$ contains an event from $V'$ and therefore from $V$.\qed
	\end{iteMize}
	}

\item $\G(X) = \emptyset$. \qed

\item $\G(\mu X \centerdot P) = \G(P)$.

	{
	\proof

	Let $V \in \G(\mu X \centerdot P)$ and $t = s \concat \ticks  \in \btraces{\widehat{\mu X \centerdot P}}$. $\widehat{\mu X \centerdot P} = (\mu X \centerdot P)(\overline{\mathit{\DIV}}) = P^{\ast}\! = \bigcap_{i=0}^{\infty} P^{n}$, where  $P^{0\!} = \DIV$, $P^{n+1}\! = P(P^{n}, \overline{\DIV})$. Since $t \in \btraces{P^{\ast}}$, $t \in \btraces{P^{n}}$, for each $n \in \mathbb{N}$. Therefore, 
\[t = s \concat \ticks \in \btraces{P^{1}} = \btraces{P(\DIV, \overline{\DIV})} = \btraces{\widehat{P}}.
\]
 By construction, $V \in \G(P)$. Therefore, by induction hypothesis for $P$ and $t$, $s$ contains an event from $V$. 
	\qed

	}
\end{iteMize} 

\qed

% \begin{prop}\label{prop_app_C_X}
%  Let $P(X, Y_{1}, \ldots, Y_{n}) = P(X, \overline{Y})$ be a CSP term whose free variables are contained within the set $\{ X, Y_{1}, \ldots, Y_{n} \}$. Let $\C_{X} : \TCSP \longrightarrow \mathcal{P}(\mathcal{P}(\Sigma) \times \mathcal{P}(\Sigma))$ be defined recursively on the structure of $P$ as shown in \autoref{fig_contract_0}. If $(U, V)  \in \C_{X}(P)$, then for all processes $T_{1}, T_{2}, \varTheta_{1}, \ldots, \varTheta_{n}  \in \TD$, $ d_{V}(P(T_{1}, \overline{\varTheta}), P(T_{2}, \overline{\varTheta})) \leq \frac{1}{2} d_{U}(T_{1}, T_{2})$.
% \end{prop}

\vspace{0.2cm}\noindent
\textbf{Proposition~\ref{prop_contr_sets}.} \textit{
 Let $P(X, Y_{1}, \ldots, Y_{n}) = P(X, \overline{Y})$ be a CSP term whose free variables are contained within the set $\{ X, Y_{1}, \ldots, Y_{n} \}$. Let $\C_{X} : \TCSP \longrightarrow \mathcal{P}(\mathcal{P}(\Sigma) \times \mathcal{P}(\Sigma))$ be defined recursively on the structure of $P$ as shown in \autoref{fig_contract_0}. If $(U, V)  \in \C_{X}(P)$, then for all processes $T_{1}, T_{2}, \varTheta_{1}, \ldots, \varTheta_{n}  \in \TD$, $ d_{V}(P(T_{1}, \overline{\varTheta}), P(T_{2}, \overline{\varTheta})) \leq \frac{1}{2} d_{U}(T_{1}, T_{2})$.
}
\vspace{0.2cm}

\proof
 Structural induction on $P$. Let us take arbitrary $T_{1}, T_{2}, \varTheta_{1}, \ldots, \varTheta_{n}  \in \TD$.
 \begin{iteMize}{$\bullet$}
	\item $\C_{X}( P ) \widehat{=} \mathcal{P}(\Sigma) \times \mathcal{P}(\Sigma)$ whenever $X$ is not free in $P$.
	{
	\proof
	 Let $(U, V) \in \C_{X}( P ) $.  Then
	 \[\frac{1}{2} d_{U}(T_{1}, T_{2})  \geq 0  =  d_{V}( P(T_{1}, \overline{\varTheta}), P(T_{2}, \overline{\varTheta})) = d_{V}( P(\overline{\varTheta}), P(\overline{\varTheta})).\eqno{\qEd}
\]
	}

 \item $\C_{X}(a \then P ) \widehat{=} \C_{X}( P ) \cup \{ (U, V) \in \N_X(P) \st a \in V \}$.
	{
	\proof
	Suppose $(U, V) \in \C_{X}(a \then P)$.
	By construction, $(U, V) \in \C_{X}(P)$ or,  $(U, V) \in \N_{X}(P)$ and $a \in V$. We consider both cases.
	\begin{iteMize}{$-$}
		\item Suppose $(U, V) \in \C_{X}(P)$. Then: \vspace{0.1cm}

		    \noindent
		    \begin{tabular}{@{\hspace{-0.00cm}} l l l}
			    $\frac{1}{2} d_{U}(T_{1}, T_{2})$ 	& $\geq d_{V}( P(T_{1}, \overline{\varTheta}), P(T_{2}, \overline{\varTheta})) $ 	&  // induction hypothesis  \\
						    & $\geq d_{V}( a \then P(T_{1}, \overline{\varTheta}), a \then  P(T_{2}, \overline{\varTheta})) $ &  // Lemma~\autoref{lemma_nonexp_prefix}  \\
				    
		    \end{tabular}
		\item Suppose $(U, V) \in \N_{X}(P)$ and $a \in V$. \vspace{0.1cm}

		     \noindent
		     \begin{tabular}{@{\hspace{-0.00cm}}l l l}
	 			$\frac{1}{2} d_{U}(T_{1}, T_{2})$ 	& $\geq \frac{1}{2} d_{V}( P(T_{1}, \overline{\varTheta}), P(T_{2}, \overline{\varTheta})) $ 	&  // $(U, V) \in \N_X(P) $, \\
									& &   // Proposition \ref{prop_non_exp} \\
	 					& $ = d_{V}( a \then P(T_{1}, \overline{\varTheta}), a \then  P(T_{2}, \overline{\varTheta})) $ &  // $a \in V$\rlap{\hbox to 111pt{\hfill\qEd}}  \\
	 			
		     \end{tabular}
	\end{iteMize}\bigskip
	}

 \item $\C_{X}( P_{1} \oplus P_{2} ) \widehat{=} \C_{X}(P_{1}) \cap \C_{X}(P_{2}) = \{ (U_{1} \cap U_{2} , V_{1} \cup V_{2}) \st (U_{i}, V_{i}) \in \C_{X}(P_{i})  \}$ for \mbox{$\oplus \in \{ \intchoice, \extchoice,  \parallel[A] \}$}.	
	{
	\proof

 		Suppose $(U, V) \in \C_{X}(P_{1} \oplus  P_{2})$.
		By construction, there exist $(U_{1}, V_{1}) \in \C_{X}(P_{1})$ and $(U_{2}, V_{2}) \in \C_{X}(P_{2})$, such that $U = U_{1} \cap U_{2}$ and $V = V_{1} \cup V_{2}$. Therefore, $(U, V) \in \C_{X}(P_{1}$), $(U, V) \in \C_{X}(P_{2})$ (antitoneness). Then: \vspace{0.1cm}

		\noindent
		$d_{V}((P_{1} \oplus P_{2}) (T_{1}, \overline{\varTheta}), (P_{1} \oplus P_{2})(T_{2}, \overline{\varTheta})) $

		\noindent 
		$ =  d_{V}(P_{1}(T_{1}, \overline{\varTheta})  \oplus P_{2}(T_{1}, \overline{\varTheta}), P_{1}(T_{2}, \overline{\varTheta}) \oplus P_{2}(T_{2}, \overline{\varTheta}))  $

		\noindent 
		// ultrametric inequality
		
		\noindent
		\begin{tabular}{@{\hspace{-0.00cm}} l l}
		$ \leq max \{$ & $ d_{V}(P_{1}(T_{1}, \overline{\varTheta})  \oplus P_{2}(T_{1}, \overline{\varTheta}), P_{1}(T_{2}, \overline{\varTheta}) \oplus P_{2}(T_{1}, \overline{\varTheta}) ),$   \\
	       			& $ d_{V}(P_{1}(T_{2}, \overline{\varTheta})   \oplus P_{2}(T_{1}, \overline{\varTheta}), P_{1}(T_{2}, \overline{\varTheta}) \oplus P_{2}(T_{2}, \overline{\varTheta}) )  \}$ \\
		\end{tabular}

		\noindent 
		// Lemma~\ref{lemma_nonexp_in_args}	 

		\noindent
		\begin{tabular}{@{\hspace{-0.00cm}}l l}
		$ \leq max \{$ & $d_{V}(P_{1}(T_{1}, \overline{\varTheta}), P_{1}(T_{2}, \overline{\varTheta}))$     \hspace{0.2cm}// $\leq \frac{1}{2} d_{U}(T_{1}, T_{2})$, induction hypothesis for $P_{1}$ \\	
	    			& $d_{V}(P_{2}(T_{1}, \overline{\varTheta}), P_{2}(T_{2}, \overline{\varTheta}))  \}$ //  $\leq \frac{1}{2} d_{U}(T_{1}, T_{2})$, induction hypothesis for $P_{2}$ \\		 
		\end{tabular}\\	    	
		$ \leq \frac{1}{2} d_{U}(T_{1}, T_{2})$
	\qed
	}

\item $\C_{X}( P_{1}  \semi  P_{2} ) \widehat{=} \C_{X}(P_{1}) \inter ( \C_{X}(P_{2}) \union \{ (U, V) \in \N_X(P_{2}) \st V \in \G(P_{1})  \}$.
	{
	\proof
		Suppose $(U, V) \in \C_{X}(P_{1}  \semi   P_{2})$.
		By construction, $(U, V) \in \C_{X}(P_{1}  \semi   P_{2})$ yields 2 possibilities:
		\begin{iteMize}{$-$}
			\item $(U, V) \in \C_{X}(P_{1}) \inter  \C_{X}(P_{2})$. The proof is the same as the proof for $\intchoice, \extchoice$ and $\parallel[A]$.
			\item $(U, V) \in \C_{X}(P_{1}) \inter   \{ (U, V) \in \N_X(P_{2}) \st V \in \G(P_{1})  \}$. Again, using the ultrametric inequality:\vspace{0.1cm}

			\noindent
			$ d_{V}(P_{1}(T_{1}, \overline{\varTheta})   \semi  P_{2}(T_{1}, \overline{\varTheta}), P_{1}(T_{2}, \overline{\varTheta})  \semi  P_{2}(T_{2}, \overline{\varTheta}))  $

			\noindent
			\begin{tabular}{@{\hspace{-0.00cm}} l l l}
			$ \leq max \{$ & $ d_{V}(P_{1}(T_{1}, \overline{\varTheta})   \semi  P_{2}(T_{1}, \overline{\varTheta}), P_{1}(T_{2}, \overline{\varTheta})  \semi  P_{2}(T_{1}, \overline{\varTheta}) ),$    & // Lemma~\ref{lemma_nonexp_in_args} \\
				& $ d_{V}(P_{1}(T_{2}, \overline{\varTheta})    \semi  P_{2}(T_{1}, \overline{\varTheta}), P_{1}(T_{2}, \overline{\varTheta})  \semi  P_{2}(T_{2}, \overline{\varTheta}) )  \}$ & \\ 
			\end{tabular}

			\noindent
			\begin{tabular}{@{\hspace{-0.00cm}} l l l}    
			$ \leq max \{$ & $ d_{V}(P_{1}(T_{1}, \overline{\varTheta}), P_{1}(T_{2}, \overline{\varTheta})  ),$  & // $\leq \frac{1}{2} d_U(T_{1}, T_{2})$, induction hypothesis \\
					& $ d_{V}(P_{2}(T_{1}, \overline{\varTheta}),  P_{2}(T_{2}, \overline{\varTheta}) )  \}$ & // $\leq \frac{1}{2} d_U(T_{1}, T_{2})$, Prop.~\ref{prop_app_GCF} (\ref{hypo_G}) and \ref{lemma_contr_seq} \\ 
			\end{tabular}

			$ \leq \frac{1}{2} d_{U}(T_{1}, T_{2})$ 
		\end{iteMize}
	\qed
	}

\item $\C_{X}(P \hide A) \widehat{=} \{ (U,V) \st (U,V') \in \C_X(P) \land V' \cap A = \emptyset \land V' \subseteq V \}$.
	{
	\proof
	 Suppose $(U, V) \in \C_{X}(P \backslash A)$.	
	 By construction, there exists $V'$, such that $(U, V') \in  \C_{X}(P)$,   $V' \subseteq V $ and $V' \cap A = \emptyset $.\vspace{0.1cm}

	 \noindent
	\begin{tabular}{@{\hspace{-0.00cm}}l l l}
	 $\frac{1}{2} d_{U}(T_{1}, T_{2})$ 	& $\geq d_{V'}(P(T_{1}, \overline{\varTheta}), P(T_{2}, \overline{\varTheta})) $ 				&  // induction hypothesis  \\
	 			& $\geq d_{V'}(P(T_{1}, \overline{\varTheta}) \hide A, P(T_{2}, \overline{\varTheta}) \hide A ) $ &  // $V' \cap A = \emptyset $, Lemma~\ref{lemma_nonexp_hiding} \\
	 			& $\geq d_{V}(P(T_{1}, \overline{\varTheta}) \hide A, P(T_{2}, \overline{\varTheta}) \hide A ) $  &  // $V' \subseteq V $, $U \mapsto d_{U}$ antitone \\
	\end{tabular}

	\qed
	}
\item $\C_{X}(P \renam{R}) \widehat{=} \{ (U,V) \st (U,V') \in \C_X(P) \land R(V') \subseteq V \}$.
	{
	\proof
	 Suppose $(U, V) \in \C_{X}(P \renam{R})$.	
	 By construction, there exists $V'$, such that $(U, V') \in  \C_{X}(P)$ and   $R(V') \subseteq V $. \vspace{0.1cm}

	\noindent
	\begin{tabular}{@{\hspace{-0.00cm}} l l l}
	 $\frac{1}{2} d_{U}(T_{1}, T_{2})$ 	& $\geq d_{V'}(P(T_{1}, \overline{\varTheta}), P(T_{2}, \overline{\varTheta})) $ 				&  // induction hypothesis  \\
	 			& $\geq d_{R(V')}(P(T_{1}, \overline{\varTheta}) \renam{R}, P(T_{2}, \overline{\varTheta}) \renam{R}) $ &  //  Lemma~\ref{lemma_nonexp_renaming} \\
	 			& $\geq d_{V}(P(T_{1}, \overline{\varTheta}) \renam{R}, P(T_{2}, \overline{\varTheta}) \renam{R}) $  &  // $R(V') \subseteq V $, $U \mapsto d_{U}$ antitone \\
	\end{tabular}
	\qed
	}
\item $\C_{X}(X) \widehat{=} \emptyset$. \qed

\item $\C_{X}(\mu Y \centerdot P) \widehat{=} \set{(U,V) \st
               (U',V') \in \C_X(P) \land (V',V') \in \N_Y(P) \land U \subseteq U' \land V' \subseteq V }$ if $Y \neq X$
{
\proof
	Suppose $(U, V) \in \C_{X}(\mu Y \centerdot P)$ for $X \neq Y$ and $X, Y$ free in $P(X, Y, Z_{1}, \ldots, Z_{n})$.
	
	\noindent
	Then, by construction, there exist $U', V' \subseteq \Sigma$  such that:
	\begin{enumerate}
		\item  $(U', V') \in \C_{X}(P)$
		\item  $U \subseteq U' $, $V' \subseteq V $ 		
		\item  $(V', V') \in \N_{Y}(P)$ 			
	\end{enumerate}

	\noindent
	Since $(U', V') \in \C_{X}(P)$, by induction hypothesis, for all $T_1, T_2, \xi, \overline{\varTheta} \in \TD$ we have:
	\begin{equation}\label{ih_c_x}
	    \frac{1}{2} d_{U'}(T_{1}, T_{2}) \geq d_{V'}(P(T_{1}, \xi, \overline{\varTheta}), P(T_{2}, \xi, \overline{\varTheta}))
	\end{equation}
  
	\noindent
	Since $(V', V') \in \N_{Y}(P)$, from Proposition~\ref{prop_non_exp}:
	\begin{equation}\label{ih_cn_y}
	 d_{V'}(T_{1}, T_{2}) \geq d_{V'}(P(\xi, T_1, \overline{\varTheta}), P(\xi, T_2, \overline{\varTheta}))
	\end{equation}

	\noindent
	Let  $P_{1}(Y) = P(T_{1}, Y, \overline{\varTheta})$, $P_{2}(Y) = P(T_{2}, Y, \overline{\varTheta})$. Then, $(\mu Y \centerdot P)(T_{1}, \overline{\theta}) = P_{1}^{\ast} = \bigcap_{i = 0}^{\infty} P_{1}^{n}$ and $(\mu Y \centerdot P)(T_{2}, \overline{\theta}) = P_{2}^{\ast} = \bigcap_{i = 0}^{\infty} P_{2}^{n}$, where for $i=1,2$, $P_{i}^{0} = \bot = \DIV$, $P_{i}^{n+1} = P(T_{i}, P_{i}^{n}, \overline{\theta}) = P_i(P_{i}^{n})$.
	
	\noindent
	We will prove by induction that for $n \geq 1$:
	 \begin{equation} \label{local_ih_c}
	 d_{V'}(P_{1}^{n}, P_{2}^{n}) \leq \frac{1}{2} d_{U'}(T_{1}, T_{2}) 
	 \end{equation}	 
	
	\begin{iteMize}{$-$}
	
	\item n=1.

	\noindent
	\begin{tabular}{@{\hspace{-0.00cm}} l l l}
		 	
		$\frac{1}{2} d_{U'}(T_{1}, T_{2})$ & 	$\geq  d_{V'}(P(T_{1}, \DIV, \overline{\varTheta}), P(T_{2}, 						\DIV, \overline{\varTheta}))$   &  //  from (\ref{ih_c_x}) \\		
						   & $ =  d_{V'}(P_{1}^{1}, P_{2}^{1})$   &  \\						
	\end{tabular}	

	\item Suppose $ d_{V'}(P_{1}^{n}, P_{2}^{n}) \leq \frac{1}{2} d_{U'}(T_{1}, T_{2})$.

	\noindent
	\begin{tabular}{@{\hspace{-0.00cm}}l l l}
		$d_{V'}(P_{1}^{n+1}, P_{2}^{n+1})$ 	& $ = d_{V'}(  P(T_{1}, P_{1}^{n}, \overline{\varTheta}),
							               P(T_{2}, P_{2}^{n}, \overline{\varTheta})  )   $   \\
					& $\leq max \{  d_{V'}(  P(T_{1}, P_{1}^{n}, \overline{\varTheta}),
							         P(T_{2}, P_{1}^{n}, \overline{\varTheta})  ),$   &    \\
					& $ \hspace{1.2cm}	d_{V'}(  P(T_{2}, P_{1}^{n}, \overline{\varTheta}),
							                 P(T_{2}, P_{2}^{n}, \overline{\varTheta})  ) \} $   & \\

					& $\leq max \{ \frac{1}{2} d_{U'}(  T_{1}, T_{2} ),$   &    \hspace{-1.5cm}  // from (\ref{ih_c_x}) \\ 
					& $ \hspace{1.2cm} d_{V'}(  P_{1}^{n},  P_{2}^{n}  ) \} $   & \hspace{-1.5cm}  // from (\ref{ih_cn_y}) \\ 

					& $ \leq max \{ \frac{1}{2} d_{U'}(  T_{1}, T_{2} ),$   &     \\ 
					& $ \hspace{1.2cm} \frac{1}{2} d_{U'}(  T_{1}, T_{2}  \} $   & \hspace{-1.5cm} // from (\ref{local_ih_c})  \\
					& $ \leq  \frac{1}{2} d_{U'}(  T_{1}, T_{2} )$   &     \\ 					
					
	\end{tabular}\bigskip	
	\end{iteMize}

	\noindent
	Now suppose that $d_{V'}(P_{1}^{\ast}, P_{2}^{\ast}) > \frac{1}{2} d_{U'}(T_{1}, T_{2})$ and let $d_{U'}(T_{1}, T_{2}) = 2^{-k}$. Let, without loss of generality, $P_{1}^{\ast}$ and $P_{2}^{\ast}$ differ on the sets of their divergences and let there exist $s \in \divergences{P_{1}^{\ast}}$ such that $s \not \in \divergences{P_{2}^{\ast}}$ and $\length_{V'}(s) < k+1$. Then, since $P_{i}^{\ast} = \bigcap_{n = 0}^{\infty} P_{i}^{n}$, there exists $l \in \mathbb{N}$ such that $s \notin P_{2}^{l}$, but for all $n \in \mathbb{N}$, $s \in P_{1}^{n}$. But then $d_{V'}(P_{1}^{l}, P_{2}^{l}) > 2^{-(k+1)} = \frac{1}{2} d_{U'}(T_{1}, T_{2})$, which is a contradiction with (\ref{local_ih_c}). Therefore, $d_{V'}(P_{1}^{\ast}, P_{2}^{\ast}) \leq \frac{1}{2} d_{U'}(T_{1}, T_{2})$. Then, since $U \subseteq U' $ and  $V' \subseteq V $, by antitoneness, $d_{V}((\mu Y \centerdot P)(T_{1}, \overline{\varTheta}), (\mu Y \centerdot P)(T_{2}, \overline{\varTheta})) = d_{V}(P_{1}^{\ast}, P_{2}^{\ast}) \leq \frac{1}{2} d_{U}(T_{1}, T_{2})$.
	\qed
	}

 \end{iteMize}
\qed

%%%%%%%%%%%%%%%%%%%%%%%%%%%%%%%%%%%%%%%%%%%%%5

% \begin{prop}{\label{prop_app_F}}
%  Let $P( X_{1}, \ldots, X_{n}) = P( \overline{X})$ be a CSP term whose free variables are contained within the set $\{ X_{1}, \ldots, X_{n} \}$. Let $\F : \TCSP \longrightarrow \mathcal{P}(\mathcal{P}(\Sigma) \times \mathcal{P}(\Sigma))$ be defined recursively on the structure of $P$ as shown in \autoref{fig_fairsets_0}.  If $(U, V) \in \F(P)$, then, for any collection of $U$-fair livelock-free processes $\theta_{1}, \ldots, \theta_{n} \in \TD$, the process $P(\theta_{1}, \ldots, \theta_{n})$ is livelock-free and $V$-fair.    
% \end{prop}

\vspace{0.2cm}\noindent
\textbf{Proposition~\ref{prop_fair_sets}.} \textit{
 Let $P( X_{1}, \ldots, X_{n}) = P( \overline{X})$ be a CSP term whose free variables are contained within the set $\{ X_{1}, \ldots, X_{n} \}$. Let $\F : \TCSP \longrightarrow \mathcal{P}(\mathcal{P}(\Sigma) \times \mathcal{P}(\Sigma))$ be defined recursively on the structure of $P$ as shown in \autoref{fig_fairsets_0}.  If $(U, V) \in \F(P)$, then, for any collection of $U$-fair livelock-free processes $\theta_{1}, \ldots, \theta_{n} \in \TD$, the process $P(\theta_{1}, \ldots, \theta_{n})$ is livelock-free and $V$-fair.
}
\vspace{0.2cm}

\proof

 Structural induction on $P$.

\begin{iteMize}{$\bullet$}
 \item $\F(\STOP) = \F(\SKIP) = \mathcal{P}(\Sigma) \times \mathcal{P}(\Sigma)$.
	{\proof
		$\STOP$ and $\SKIP$ are livelock-free and do not contain infinite traces.
	\qed}
		
 \item $\F(a \then P) = \F(P)$
	{\proof
	Let $(U,V) \in \F(a \then P)$, $\theta_{1}, \ldots, \theta_{n}$ be a collection of livelock-free $U$-fair processes.

	Since $(U,V) \in \F(a \then P)$, by construction, $(U,V) \in \F(P)$. Therefore, by induction hypothesis, $P(\overline{\varTheta})$ is livelock-free and $V$-fair. Therefore, $a \then P(\overline{\varTheta})$ is  livelock-free.

	We will prove that $a \then P(\overline{\varTheta})$ is $V$-fair. Let $u \in \inftraces{a \then P(\overline{\varTheta})}$. Therefore, by Lemma~\ref{inft_prefix}, there exists $u' \in  \inftraces{P(\overline{\varTheta})}$, such that $u = \langle a \rangle \concat u'$. 
	Since $P(\overline{\varTheta})$ is  $V$-fair, $u'$ contains infinitely many events from $V$, and so does therefore $u$. Hence, $a \then P(\overline{\varTheta})$ is $V$-fair.	
	\qed}

\item $\F(P_{1} \oplus P_{2}) = \F(P_{1}) \cap \F(P_{2})$ for $\{ \intchoice , \extchoice \}$
	{\proof
	Let $(U,V) \in \F(P_{1} \oplus P_{2})$, $\theta_{1}, \ldots, \theta_{n}$ be a collection of livelock-free $U$-fair processes.

	Since $(U,V) \in \F(P_{1} \oplus P_{2})$, by construction, $(U,V) \in \F(P_{1})$ and $(U,V) \in \F(P_{2})$. Therefore, by induction hypothesis, $P_{1}(\overline{\varTheta})$ and $P_{2}(\overline{\varTheta})$ are livelock-free and $V$-fair. Therefore, $P_{1}(\overline{\varTheta}) \oplus P_{2}(\overline{\varTheta})$ is livelock-free.

	Let $u \in \inftraces{P_{1}(\overline{\varTheta}) \oplus P_{2}(\overline{\varTheta})}$. Then, by Lemma \ref{inft_choice}, $u \in \inftraces{P_{1}(\overline{\varTheta})}$ or $u \in \inftraces{P_{2}(\overline{\varTheta})}$. Let without loss of generality the former holds. Then, since $P_{1}(\overline{\varTheta})$ is $V$-fair, $u$ contains infinitely many events from $V$. Therefore, $P_{1}(\overline{\varTheta}) \oplus P_{2}(\overline{\varTheta})$ is $V$-fair.	
	\qed}

\item $\F(P_{1}  \semi  P_{2}) = \F(P_{1}) \cap \F(P_{2})$ 
	{\proof
	Let $(U,V) \in \F(P_{1}  \semi  P_{2})$, $\theta_{1}, \ldots, \theta_{n}$ be a collection of livelock-free $U$-fair processes.

	Since $(U,V) \in \F(P_{1}  \semi  P_{2})$, by construction, $(U,V) \in \F(P_{1})$ and $(U,V) \in \F(P_{2})$. Therefore, by induction hypothesis, $P_{1}(\overline{\varTheta})$ and $P_{2}(\overline{\varTheta})$ are livelock-free and $V$-fair. Therefore, $P_{1}(\overline{\varTheta})  \semi  P_{2}(\overline{\varTheta})$ is livelock-free.

	Let $u \in \inftraces{P_{1}(\overline{\varTheta})  \semi  P_{2}(\overline{\varTheta})}$. Then, by Lemma~\ref{inft_seq_comp}, $u \in \inftraces{P_{1}(\overline{\varTheta})}$ or $u = t_{1} \concat u_{2}$ with $t_{1} \concat \ticks \in \traces{P_{1}(\overline{\varTheta})} \inter \Sigma^{\ast \checkmark}$,  $u_{2} \in \inftraces{ P_{2}(\overline{\varTheta})}$.

	\begin{iteMize}{$-$}
		\item Suppose $u \in \inftraces{P_{1}(\overline{\varTheta})}$. Since $P_{1}(\overline{\varTheta})$ is $V$-fair, $u$ contains infinitely many events from $V$.	
		\item Suppose $u = t_{1} \concat u_{2}$ with $t_{1} \concat \langle \checkmark \rangle \in \traces{P_{1}(\overline{\varTheta})} \inter \Sigma^{\ast \checkmark}$,  $u_{2} \in \inftraces{ P_{2}(\overline{\varTheta})}$. Since $P_{2}(\overline{\varTheta})$ is $V$-fair, $u_{2}$ contains infinitely many events from $V$ and so does therefore $u$.	
	\end{iteMize}
	Therefore, $P_{1}(\overline{\varTheta})  \semi  P_{2}(\overline{\varTheta})$ is $V$-fair. 
	\qed}

\item 
	\begin{tabular}[t]{l l}
       		$\F(P_{1} \parallel[A] P_{2}) =$ & $ (\F(P_{1}) \cap \F(P_{1})) \hspace{0.2cm} \union$ \\[-1.5 ex]
			&	$\{ (U_{1} \cap U_{2}, V_{1}  )  \st (U_{1}, V_{1}) \in \F(P_{1}) \land (U_{2}, A) \in \F(P_{2}) \} \hspace{0.2cm} \union$ \\
			&	$\{ (U_{1} \cap U_{2}, V_{2}  )  \st (U_{2}, V_{2}) \in \F(P_{2}) \land (U_{1}, A) \in \F(P_{1}) \} $ \\

      	\end{tabular}

	{\proof
	Let $(U,V) \in \F(P_{1}  \parallel[A]  P_{2})$, $\theta_{1}, \ldots, \theta_{n}$ be a collection of livelock-free $U$-fair processes.

	Since $(U,V) \in \F(P_{1} \parallel[A] P_{2})$, by construction, $\F(P_{1}) \neq \emptyset$, $\F(P_{2}) \neq \emptyset$  and, by induction hypothesis, $P_{1}(\overline{\varTheta})$ and $P_{2}(\overline{\varTheta})$ are livelock-free. Therefore, $P_{1}(\overline{\varTheta}) \parallel[A] P_{2}(\overline{\varTheta})$ is livelock-free.

	Let $u \in \inftraces{P_{1}(\overline{\varTheta})\! \parallel[A]\! P_{2}(\overline{\varTheta})}$. Hence, by Lemma~\ref{inft_parallel}, there exist $u_{1} \in \alltraces{P_{1}(\overline{\varTheta})} $, $u_{2} \in \alltraces{P_{2}(\overline{\varTheta})} $, such that $u \in u_{1} \parallel[A] u_{2}$ and, $u_{1} \in \Sigma^{\omega} $ or $u_{2} \in \Sigma^{\omega} $. Let without loss of generality $u_{1} \in \Sigma^{\omega} $. By construction, we have three alternatives for $(U,V)$. 

	\begin{iteMize}{$-$}
		\item Suppose $(U,V) \in \F(P_{1}) \cap \F(P_{2})$. By induction hypothesis, $P_{1}(\overline{\varTheta})$ is $V$-fair. Therefore, $u_{1}$ contains infinitely many events from $V$ and so does $u$.
		
		\item Suppose $(U, V)$ is $(U_{1} \cap U_{2}, V)$ with $(U_{1}, V) \in \F(P_{1})$, $(U_{2}, A) \in \F(P_{2})$. Then $U = U_{1} \cap U_{2} \subseteq U_{1}$ and $\theta_{1}, \ldots, \theta_{n}$ are $U_{1}$-fair. Hence, by induction hypothesis, $P_{1}(\overline{\varTheta})$ is $V$-fair. Then, $u_{1}$ contains infinitely many events from $V$ and so does therefore $u$.

		\item Suppose $(U, V)$ is $(U_{1} \cap U_{2}, V)$ with $(U_{2}, V) \in \F(P_{2})$, $(U_{1}, A) \in \F(P_{1})$. Since $U = U_{1} \cap U_{2}$, we have $U \subseteq U_{1}$, $U \subseteq U_{2}$ and therefore, $\theta_{1}, \ldots, \theta_{n}$ are $U_{1}$-fair and $U_{2}$-fair.  By induction hypothesis for $P_{1}$, $u_{1}$ contains infinitely many events from $A$. Since $u_{1}$ and $u_{2}$ synchronise on the events in $A$, $u_{2}$ contains infinitely many events from $A$. Therefore, $u_{2} \in \Sigma^{\omega} $ and by induction hypothesis for $P_{2}$, $u_{2}$ contains infinitely many events from $V$. Hence,  $u$ contains infinitely many events from $V$.
	\end{iteMize}
	Therefore, $P_{1}(\overline{\varTheta}) \parallel[A] P_{2}(\overline{\varTheta})$ is $V$-fair.
	\qed}

\item $\F(P \hide A) = \set{(U,V) \st (U,V') \in \F(P) \land
                               V' \cap A = \emptyset \land V' \subseteq V}$ 
	{\proof
	Let $(U,V) \in \F(P \hide A)$, $\theta_{1}, \ldots, \theta_{n}$ be a collection of livelock-free $U$-fair processes.

	Since $(U,V) \in \F(P \hide A)$, by construction, there exists $V' \subseteq V$, such that $V' \cap A = \emptyset$ and $(U,V') \in  \F(P)$. Therefore, by induction hypothesis, $P(\overline{\varTheta})$ is livelock-free and $V'$-fair. Suppose $P(\overline{\varTheta}) \hide A$ is not livelock-free. Therefore, since $P(\overline{\varTheta})$ is livelock-free, there  exists 
        $u \in \inftraces{P}$ such that $u \upharpoonright (\Sigma \backslash A)$ is finite. 
        %$u \in \Sigma^{\omega}$, such that $u \upharpoonright (\Sigma \backslash A)$ is finite and for each prefix $s$ of $u$, $s \in \traces{P(\overline{\varTheta})}$. 
        Since $P(\overline{\varTheta})$ is $V'$-fair, $u$ contains infinitely many events from $V'$.	
	Since $V' \cap A = \emptyset$, $u \upharpoonright (\Sigma \backslash A)$ contains infinitely events from $V'$, which is a contradiction with $u \upharpoonright (\Sigma \backslash A)$ being finite. Therefore, $P(\overline{\varTheta}) \hide A$ is livelock-free.

	Let $u \in \inftraces{P(\overline{\varTheta}) \hide A}$. Then, by Lemma~\ref{inft_hide}, there exists $u' \in \inftraces{P(\overline{\varTheta})}$, such that $u = u' \upharpoonright (\Sigma \backslash A)$. 
	Since $P(\overline{\varTheta})$ is $V'$-fair, $u'$ contains infinitely many events from $V'$. Since $V' \cap A = \emptyset$, $u = u' \upharpoonright (\Sigma \backslash A)$ contains infinitely many events from $V' \subseteq V$. Therefore, $u$ contains infinitely many events from $V$. Therefore, $P(\overline{\varTheta}) \hide A$ is $V$-fair.
	\qed}

\item $\F(P \renam{R}) = \set{(U,V) \st (U,V') \in \F(P) \land R(V') \subseteq V }$ 
	{\proof
	Let $(U,V) \in \F(P \renam{R})$, $\theta_{1}, \ldots, \theta_{n}$ be a collection of livelock-free $U$-fair processes.

	Since $(U,V) \in \F(P \renam{R})$, by construction, there exists $V'$, such that $R(V') \subseteq V$ and $(U,V') \in  \F(P)$. Therefore, by induction hypothesis, $P(\overline{\varTheta})$ is livelock-free and $V'$-fair. Hence, $P(\overline{\varTheta}) \renam{R}$ is also livelock-free.

	Let $u \in \inftraces{P(\overline{\varTheta}) \renam{R}}$. Then, by Lemma~\ref{inft_rename}, there  exists $u' \in \inftraces{P(\overline{\varTheta})}$, such that $u' \renamomega u$. Since $P(\overline{\varTheta})$ is $V'$-fair, $u'$ contains infinitely many events from $V'$. Since $u' \renamomega u$, $u$ contains infinitely many events from $R(V') \subseteq V$. Therefore,  $u$ contains infinitely many events from $V$. Hence, $P(\overline{\varTheta}) \renam{R}$ is $V$-fair.
	\qed}

\item $\F(X) = \{ (U, V) \st U \subseteq V \}$ 
	{\proof
	Let $(U,V) \in \F(X)$, $\theta$ be a  livelock-free $U$-fair process. 
	 $X(\theta) = \theta$ is then livelock-free and, since $U \subseteq V $, $X(\theta) = \theta$ is $V$-fair.
	\qed}

\item	$ \F(\mu X \centerdot P) =    \left\{ \begin{array}{l l}
                                    	\set{(U,V) \st (W,W) \in \C_X(P) \cap \F(P) \land U \subseteq W \subseteq V} & \mbox{if $\mu X \centerdot P$ is open } \\

					\phantom{\{} \mathcal{P}(\Sigma) \times  \set{V \st (W,W) \in \C_X(P) \cap \F(P) \land W \subseteq V} & \mbox{otherwise } \\
                                   	\end{array}
			    	\right. $ \\
 				 
	{\proof
		
	Let $P(X, Y_{1}, \ldots, Y_{n} )$ be a CSP term whose free variable are contained within the set $\{ X, Y_{1}, \ldots, Y_{n}  \}$. Let $(U,V) \in \F(\mu X \centerdot P)$ and $(\mu X \centerdot P)$ be open. Let $\theta_{1}, \ldots, \theta_{n}$ be a collection of livelock-free $U$-fair processes.

	Since $(U,V) \in \F(\mu X \centerdot P)$, by construction, there exists $W$, such that $U \subseteq W \subseteq V$ and $(W, W) \in \C_X(P) \cap  \F(P)$. Therefore $\theta_{1}, \ldots, \theta_{n}$ are $W$-fair and, by induction hypothesis:
	\begin{equation}\label{f_inf}
	  P(\xi, \overline{\varTheta}) \mbox{ is livelock-free and } W\mbox{-fair for any livelock-free } W\mbox{-fair process } \xi. 
	\end{equation}

	 Since $(W, W) \in \C_X(P)$,  by Proposition \ref{prop_app_GCF} (\ref{hypo_Cx}), $P(X, \overline{\varTheta})$ is contractive in $X$ with respect to the metric $d_W$. Therefore, from Banach's fixed point theorem, $P(X, \overline{\varTheta})$ has a unique fixed point $(\mu X \centerdot P)(\overline{\theta}) = \bigcup_{n = 0}^{\infty} P^{n} = P^{\ast}$, where $P^{0} = \top =  \STOP$, $P^{n+1} = P( P^{n}, \overline{\varTheta})$. %(any process can be chosen for $P^{0}$, in fact)

	We will prove by induction that for each $n \in \mathbb{N}$, $P^{n}$ is livelock-free  and $W$-fair.
	\begin{iteMize}{$-$}
		\item $n=0$. $\STOP$ is livelock-free and does not contain infinite traces.
		\item Suppose that  $P^{n}$ is  livelock-free and $W$-fair. 
		From (\ref{f_inf}), $P^{n+1} = P( P^{n}, \overline{\varTheta})$ is also livelock-free and $W$-fair.
	\end{iteMize}

\noindent Therefore, for each $n \in \mathbb{N}$, $P^{n}$ is livelock-free. Then, since by Proposition~\ref{prop_app_set_lf_closed} the set of livelock-free processes is closed, $(\mu X \centerdot P)(\overline{\theta}) = \bigcup_{n = 0}^{\infty} P^{n} = P^{\ast}$ is livelock-free. 

	Now, let $u \in \inftraces{(\mu X \centerdot P)(\overline{\varTheta})}$. Then, for each finite prefix $t$ of $u$, $t \in \traces{P^{\ast}}$, \ie there exists some sufficiently large $n_t$, such that $t \in \traces{P^{n_t}}$.

	If there exists $m \in \mathbb{N}$ such that for each prefix $t$ of $u$,  $t \in \traces{P^{m}}$, then $u \in \inftraces{P^{m}}$. In this case, since $P^{m}$ is $W$-fair, $u$ contains infinitely many events from $W$. Then, since $W \subseteq V$, $u$ contains infinitely many events from $V$.

	Otherwise, we can conclude the following:
	\begin{equation}\label{eq_f_rec}
	    \mbox{for each $m \in \mathbb{N}$, there exists a prefix $t$ of $u$, such that $t \not \in \traces{P^{m}}$.}
	\end{equation}

	Let $\varepsilon = 2^{-k}$ for some $k \in \mathbb{N}$. Since the sequence $\langle P^{i} \st i \in   \mathbb{N} \rangle$ converges to $P^{\ast}$ with respect to the metric $d_{W}$, there exists $n_{\varepsilon} \in \mathbb{N}$, such that for each $n \geq n_{\varepsilon}$, $d_{W}(P^{\ast}, P^{n}) < \varepsilon$. From our assumption (\ref{eq_f_rec}), for $n_\varepsilon$ there exists $t_\varepsilon$, such that $t_\varepsilon$ is a prefix of $u$ and $t_\varepsilon \not \in \traces{P^{n_\varepsilon}}$. Then, since $d_{W}(P^{\ast}, P^{n_{\varepsilon}}) < \varepsilon$ and $t_\varepsilon \in \traces{P^{\ast}}$, $\length_{W} t_\varepsilon \geq k$. Since $k$ was arbitrary, we can conclude that $u$ contains infinitely many events from $W$. Then again, since $W \subseteq V$, $u$ contains infinitely many events from $V$. Therefore, $(\mu X \centerdot P)(\overline{\theta})$ is $V$-fair.
	\qed}
\end{iteMize}

\section{Proofs for Section~\ref{sec_static_SFS}}\label{app_static_SFS}

% \begin{prop} \label{prop_fair_cofair}
%  Let $P$ be a structurally finite state process. Let $\Phi(P): \SFS \longrightarrow \mathcal{P}(\mathcal{P}(\Sigma) \times \mathcal{P}(\Sigma))$ and $\delta(P): \SFS \longrightarrow \{ \true, \false \}$ be defined recursively on the structure of $P$ as shown in \autoref{fig_fair_cofair_sets} and \autoref{fig_fair_cofair_div}, respectively. Then, if $\delta(P) = \false$, $P$ is livelock-free. Moreover, if $\Phi(P) = \{ (F_1, C_1), \ldots, (F_k, C_k) \}$, then, for each infinite trace $u$ of $P$, there exists $1 \leq i \leq k$, such that $u$ is fair in $F_i$ and $u$ is co-fair in  $C_i$.
% \end{prop}

\vspace{0.2cm}\noindent \textbf{Proposition~\ref{main2}.} \textit{Let
  $P$ be a structurally finite state process. Let $\Phi: \SFS
  \longrightarrow \mathcal{P}(\mathcal{P}(\Sigma) \times
  \mathcal{P}(\Sigma))$ and $\delta: \SFS \longrightarrow \{ \true,
  \false \}$ be defined recursively on the structure of $P$ as shown
  in Figures \ref{fig_fair_cofair_sets} and \ref{fig_fair_cofair_div},
  respectively. Then, if $\delta(P) = \false$, $P$ is
  livelock-free. Moreover, if in addition $\Phi(P) = \{ (F_1, C_1), \ldots, (F_k,
  C_k) \}$, then, for each infinite trace $u$ of $P$, there exists $1
  \leq i \leq k$, such that $u$ is fair in $F_i$ and $u$ is co-fair in
  $C_i$.  }
\vspace{0.2cm}

\proof

Induction on the structure of the $\SFS$ process $P$.

Note that by construction, all fair/co-fair pairs of sets thus
generated remain disjoint, \ie for each $(F,C) \in \Phi(P)$, $F \inter C = \emptyset$. This is key in the rule for parallel composition, where the fair/co-fair data of individual sub-components enables one to rule out certain pairs for the resulting parallel process. We prove this property only for the case of renaming as for all other cases it follows trivially from the induction hypothesis and the specific construction.

Let us also remark that it might be the case that $\delta(P) = \false$ and $\Phi(P) = \emptyset$ and this indicates that $P$ is livelock free but exhibits only finite traces. We note, however, that if $\delta(P) = \false$ and $\Phi(P) \neq \emptyset$, then for every $(F,C) \in \Phi(P)$, $F \neq \emptyset$. This is true for sequential $\SFS$ processes by construction and follows for compound $\SFS$ processes by induction hypothesis and construction. We prove the property only for the cases of hiding and renaming where the argument is more subtle.

\begin{iteMize}{$\bullet$}
  \item For $P$ being a closed sequential process, $\Phi(P)$ and $\delta(P)$ are computed directly from the labelled transition system associated with $P$ as described in \autoref{sec_static_SFS}.

  \item $\delta(a \then P) = \delta(P)$ and $\Phi(a \then P) = \Phi(P)$.
    {\proof
      Let $\delta(a \then P) = \false$. By construction, $\delta(P) = \false$ and, therefore, by induction hypothesis, $P$ is livelock-free. Hence, by definition, $a \then P$ is also livelock-free.

      Let $u \in \inftraces{a \then P}$. Then, by Lemma~\ref{inft_prefix}, there exists $u' \in \inftraces{P}$, such that $u = \langle a \rangle \concat u'$. By induction hypothesis for $P$, there exists $(F,C) \in \Phi(P)$, such that $u'$ is fair in $F$ and co-fair in $C$. But then $u$ is also fair in $F$ and co-fair in $C$ and, by construction, $(F,C) \in \Phi(a \then P)$.
    \qed}

 \item $\delta(P_1 \oplus P_2) = \delta(P_1) \vee \delta(P_2)$ and $\Phi(P_1 \oplus P_2) = \Phi(P_1) \cup \Phi(P_2)$ if $\oplus \in \set{\intchoice,\extchoice}$.

    {\proof
	 Let $\delta(P_1 \oplus P_2) = \false$. By construction, $\delta(P_1) = \false$ and $\delta(P_2) = \false$. 
	 Therefore, by induction hypothesis, $P_1$ and $P_2$ are livelock-free. Hence, by definition, $P_1 \oplus P_2$ is livelock-free.

	 Let $u \in \inftraces{P_1 \oplus P_2}$. By Lemma~\ref{inft_choice}, $u \in \inftraces{P_1}$ or $u \in \inftraces{P_2}$. Let without loss of generality the former holds. Then, by induction hypothesis for $P_1$, there exists $(F,C) \in \Phi(P_1)$, such that $u$ is fair in $F$ and co-fair in $C$. 
	 By construction, $\Phi(P_1) \subseteq \Phi(P_1 \oplus P_2)$ and, therefore, $(F,C) \in \Phi(P_1 \oplus P_2)$.
    \qed}

\item $\delta(P_1 \semi P_2) = \delta(P_1) \vee \delta(P_2)$ and $\Phi(P_1 \semi P_2) = \Phi(P_1) \cup \Phi(P_2)$.

    {\proof
	 Let $\delta(P_1 \semi P_2) = \false$. By construction, $\delta(P_1) = \false$ and $\delta(P_2) = \false$. 
	 Therefore, by induction hypothesis, $P_1$ and $P_2$ are livelock-free. Hence, by definition, $P_1 \semi P_2$ is livelock-free.

	 Let $u \in \inftraces{P_1 \semi P_2}$. By Lemma~\ref{inft_seq_comp}, $u \in \inftraces{P_1}$ or $u = t \concat u'$ with  $t \concat \ticks \in \traces{P_1} \cap \Sigma^{\ast \checkmark}$,  $u' \in \inftraces{P_2}$. We consider both alternatives.

	      \begin{iteMize}{$-$}
		    \item
			If $u \in \inftraces{P_1}$, by induction hypothesis for $P_1$, there exists $(F,C) \in \Phi(P_1)$, such that $u$ is fair in $F$ and co-fair in $C$.  By construction, $\Phi(P_1) \subseteq \Phi(P_1 \oplus P_2)$ and, therefore, $(F,C) \in \Phi(P_1 \oplus P_2)$.
		    \item
			Let $u = t \concat u'$ where  $t \concat \ticks \in \traces{P_1} \cap \Sigma^{\ast \checkmark}$ and  $u' \in \inftraces{P_2}$. By induction hypothesis for $P_2$, there exists $(F,C) \in \Phi(P_2)$, such that $u'$ is fair in $F$ and co-fair in $C$. The finite prefix $t$ of $u$ does not affect fairness and co-fairness. Therefore, $u = t \concat u'$ is fair in $F$ and co-fair in $C$ and $(F, C) \in \Phi(P_1 \oplus P_2)$ by construction.
	      \end{iteMize}
    \qed}

\item $\delta(P \hide A) =
 \left\{ \begin{array}{ll}
             \false &
                   \mbox{ if $\delta(P) = \false$ and, for each $(F,C) \in \Phi(P)$, $F - A \neq \emptyset$}\\
             \true & \mbox{ otherwise} 
     \end{array} \right.\\ $
  and $\Phi(P \hide A) = \set{(F - A, C \cup A) \st (F,C) \in \Phi(P)} $.

    {\proof
	 Let $\delta(P \hide A) = \false$. By construction, $\delta(P) = \false$ and for each $(F,C) \in \Phi(P)$, $F - A \neq \emptyset$.
	 Since $\delta(P) = \false$, by induction hypothesis, $P$ is livelock-free. Suppose for the sake of the argument that $P \hide A$ can diverge. Since $P$ is livelock-free, by definition, the only alternative is that there exists $u \in \inftraces{P}$, such that $u \upharpoonright (\Sigma \backslash A)$ is finite. By induction hypothesis for $P$, there exists $(F,C) \in \Phi(P)$, such that $u$ is fair in $F$ and co-fair in $C$. By construction, since $\delta(P \hide A) = \false$, $F - A \neq \emptyset$. Therefore, there exists  $b \in F$ such that $b \notin A$ and $b$ occurs infinitely many times in $u$. But then $b$ should also occur infinitely many times in $u \upharpoonright (\Sigma \backslash A)$, which is a contradiction with $u \upharpoonright (\Sigma \backslash A)$ being finite. Therefore,  $P \hide A$ is livelock-free.

	Now, let $u \in \inftraces{P \hide A}$. Since $P \hide A$ is livelock-free, by Lemma~\ref{inft_hide}, there exists $v \in \inftraces{P}$ such that $u = v \upharpoonright (\Sigma \backslash A)$. By induction hypothesis for $P$, there exists $(F,C) \in \Phi(P)$ such that $v$ is fair in $F$ and co-fair in $C$. Then, since $u$ is obtained by deleting all $A$-events from $v$, $u$ is fair in $F - A$ and co-fair in $C \cup A$. Both $F - A \neq \emptyset$ and $(F - A, C \cup A) \in \Phi(P \hide A)$ are guaranteed by construction.    

	Let $\delta(P \hide A) = \false$ and let $(F,C) \in \Phi(P \hide A)$. We now prove that $F \neq \emptyset$.
        Since $\delta(P \hide A) = \false$, by construction we have the following:
	\begin{equation}\label{eq_prop_fair_cofair_hide_non_empty}
	 \mbox{for each } (F',C') \in \Phi(P), F' - A \neq \emptyset.	
	\end{equation}
        As $(F,C) \in \Phi(P \hide A)$, by construction $F = F' - A$ for some $F'$ with $(F',C') \in \Phi(P)$. By (\ref{eq_prop_fair_cofair_hide_non_empty}), $F' - A \neq \emptyset$ and hence $F \neq \emptyset$.	
        \qed}

\item $\delta(P \renam{R})  = \delta(P)$ and \\
\noindent $\Phi(P \renam{R})  = \set{ (F,C) \st (F',C') \in \Phi(P) \land F' \subseteq R^{-1}(F) \land F \subseteq R(F') \land \\
\phantom{\Phi(P \renam{R}) = \{(F,C) \st} C = \set{b \in \Sigma \st R^{-1}(b) \subseteq C'}}$

    {\proof

     In the proof we use the following notation. For any $A \subseteq \Sigma$, $a, b \in \Sigma$, $R(A) = \set{ b \st \exists a \in A \exst a \mathrel{R} b }$ and $R^{-1}(b) = \set{ a \st a \mathrel{R} b }$. Let us also clarify that in the setting of CSP~\cite{Ros97} renaming relations are assumed to be total. If an event $a \in \Sigma$ is not renamed to any other event $b \in \Sigma$, it is assumed that $a$ is renamed to itself and, hence, $R(\set{a}) \neq \emptyset$.

     Let $\delta(P \renam{R}) = \false$. By construction, $\delta(P) = \false$. Then, by induction hypothesis, $P$ is livelock-free and, hence, by definition, so is $P \renam{R}$.

     Let $u \in \inftraces{P \renam{R}}$. By Lemma~\ref{inft_rename},
     there exists $v \in \inftraces{P}$, such that $v \renamomega u$,
     i.e., for every $i \in \mathbb{N}$, $v(i) \mathrel{R} u(i)$. By induction hypothesis for $P$, there exists $(F', C') \in \Phi(P)$, such that $v$ is fair in $F'$ and co-fair in $C'$. 

     Let $C = \set{b \in \Sigma \st R^{-1}(b) \subseteq C'}$ and let $b \in C$. By construction, for each $a \in R^{-1}(b)$, $a \in C'$ and, therefore, $v$ is co-fair in $a$. Now suppose for the sake of contradiction that $u$ contains infinitely many occurrences of $b$. By definition, there exists $a \in \Sigma$, such that $a \mathrel{R} b$ and $a$ occurs infinitely many times in $v$. Therefore, $a \notin C'$ and $R^{-1}(b) \not \subseteq C'$, which is a contradiction with $R^{-1}(b) \subseteq C'$. Therefore, $u$ contains only finitely many $b$'s and, more generally, $u$ is co-fair in $C$.

     We will construct $F$ from $F'$ such that $F \subseteq R(F')$ (which will bound $F$ from above), $F' \subseteq R^{-1}(F)$ (which will bound $F $ from below and will guarantee $F \neq \emptyset$) and $u$ is fair in $F$. Then $(F,C) \in \Phi(P \renam{R})$ by construction.

     By induction hypothesis, $F' \neq \emptyset$. Let $F' = \set{a_1, \ldots, a_m}$ and for each $1 \leq i \leq m$, $R(\set{a_i}) = \set{b_{i_1}, \ldots, b_{i_{n_i}}}$. As each $a_i$ occurs infinitely many times in $v$ and $v \mathrel{R} u$, for each $1 \leq i \leq m$, there exists $b_{j_i}$, such that $a_i \mathrel{R} b_{j_i}$ and $b_{j_i}$ occurs infinitely many times in $u$. We define $F = \set{b_{j_1}, \ldots, b_{j_m}}$. Since $F' \neq \emptyset$, $F \neq \emptyset$. By the construction of $F$, $u$ is fair in $F$ and $F \subseteq R(F')$. As by construction for each $a_i \in F'$ there exists $b_{j_i} \in F$ with $a_i \mathrel{R} b_{j_i}$, then for every $1 \leq i \leq m$, $a_i \in R^{-1}(b_{j_i})$. Therefore, $F' \subseteq R^{-1}(F)$.

     We will also prove that for any $F$ that satisfies $F \subseteq
     R(F')$ and $F' \subseteq R^{-1}(F)$, the sets $F$ and $C$ are
     disjoint. Suppose there exists $b \in \Sigma$ such that $b \in F
     \inter C$. As $b \in C$, by construction, for each $a$ with $ a
     \mathrel{R} b$, we have $a \in C'$. Since $b \in F$ and by
     construction $F \subseteq R(F')$, there exists $a \in F'$, such
     that $a \mathrel{R} b$. Therefore, $a \in C' \inter F'$. This is
     a contradiction with the induction hypothesis according to which
     $F'$ and $C'$ are disjoint. Therefore, $F \inter C = \emptyset$.
     \qed} 

\item $\delta(P_1 \parallel[A] P_2) = \delta(P_1) \vee \delta(P_2)$ and \\
 %    {\allowdisplaybreaks
 %     \begin{align*}
 %     \Phi(P_1 \parallel[A] P_2) \deq
%\set{(F_1 \cup F_2, (C_1 \cap A) \cup (C_2 \cap A) \cup ((C_1 \backslash A) \cap (C_2 \backslash A)) \st (F_i, C_i) \in \Phi(P_i) \mbox{ for } i = 1,2 } \cup \mbox{}  \\
$\Phi(P_1 \parallel[A] P_2) =
\{(F,C) \st F \inter C = \emptyset \land (F_i, C_i) \in
  \Phi(P_i) \mbox{ for } i = 1,2 \land F = F_1 \union F_2 \land \\ 
   \phantom{\Phi(P_1 \parallel[A] P_2) = \{(F,C) \st} C = (C_1 \inter A) \union (C_2 \inter A) \union
  ((C_1 - A) \inter (C_2 - A)) \} \union \mbox{} \\
  \phantom{\Phi(P_1 \parallel[A] P_2) =} \set{(F, C) \st (F, C) \in \Phi(P_1) \land F \cap A = \emptyset} \cup \mbox{}  \\
  \phantom{\Phi(P_1 \parallel[A] P_2) =} \set{(F, C) \st (F, C) \in \Phi(P_2) \land F \cap A = \emptyset}$
  %    \end{align*}}

      {\proof

	 Let $\delta(P_1 \parallel[A] P_2) = \false$. By construction, $\delta(P_1) = \false$ and $\delta(P_2) = \false$. 
	 Therefore, by induction hypothesis, $P_1$ and $P_2$ are livelock-free. Hence, by definition, \smash{$P_1 \parallel[A] P_2$} is livelock-free.

	 Let $u \in \inftraces{P_1 \parallel[A] P_2}$. From Lemma~\ref{inft_parallel}, there exist $u_1 \in \alltraces{P_1}$ and $u_2 \in \alltraces{P_2}$, such that $u \in u_1 \parallel[A] u_2$ and, $u_1 \in \Sigma^{\omega}$ or $u_2 \in \Sigma^{\omega}$. We will consider three different cases.
	 \begin{iteMize}{$-$}
	      \item Let $u_1 \in \Sigma^{\omega}$ and $u_2 \in \Sigma^{\ast \checkmark}$. By induction hypothesis for $P_1$, there exists $(F,C) \in \Phi(P_1)$ such that $u_1$ is fair in $F$ and co-fair in $C$. Suppose $F \cap A \neq \emptyset$. Then, $u_1$ contains infinitely many occurrences of events from A. Since $P_1$ and $P_2$ synchronise on the events in $A$, $u_2$ must also contain infinitely many events from $A$, which is a contradiction with $u_2 \in \Sigma^{\ast \checkmark}$. Therefore, $F \cap A = \emptyset$ and, by construction, $(F,C) \in \Phi(P_1 \parallel[A] P_2)$. Now, since $u_2$ is finite and does not affect fairness and co-fairness, $u$ is fair in $F$ and co-fair in $C$.
	      \item The case where $u_2 \in \Sigma^{\omega}$ and $u_1 \in \Sigma^{\ast \checkmark}$ is handled in the same way.

	      \item Let $u_1 \in \Sigma^{\omega}$ and $u_2 \in \Sigma^{\omega}$.
	      By induction hypothesis for $P_1$ and $P_2$, there exist $(F_1, C_1) \in \Phi(P_1)$ and $(F_2, C_2) \in \Phi(P_2)$, such that $u_1$ is fair in $F_1$ and co-fair in $C_1$ and $u_2$ is fair in $F_2$ and co-fair in $C_2$.  We note, that for each $a \in A$, the number of occurrences of $a$ in $u_1$, $u_2$ and $u$ is the same due to $P_1$ and $P_2$ synchronising on $a$. Therefore, for each $a \in A$, $u_1$ contains infinitely many occurrences of $a$ if and only if $u_2$ contains infinitely many occurrences of $a$. Hence, $F_1 \inter C_2 \inter A = \emptyset$ and $F_2 \inter C_1 \inter A = \emptyset$. 

	      Let $F = F_1 \union F_2$ and $C = (C_1 \inter A) \union (C_2 \inter A) \union ((C_1 - A) \inter (C_2 - A))$.

	      We will first prove that $F \inter C = \emptyset$. Suppose for the sake of the argument that there exists $b \in \Sigma$ such that $b \in F \inter C$.
	      Since $b \in F$, by construction, $b \in F_1$ or $b \in F_2$. Let without loss of generality $b \in F_1$.
	      We will consider the cases $b \in A$ and $b \notin A$.

	      \begin{iteMize}{$*$}
		  \item Suppose $b \in A$. Since $b \in F_1$, $u_1$ is fair in $b$ and, therefore, $b \notin C_1$. Since $b \in C$ and $b \in A$, $b \in C_1 \inter A$ or $b \in C_2 \inter A$. As $b \notin C_1$, $b \in C_2 \inter A$. Therefore, $b \in F_1 \inter C_2$ which is a contradiction with $F_1 \inter C_2 \inter A = \emptyset$. Therefore, this case is not possible.
		  \item Suppose $b \notin A$. Since $b \in C$, $b \in C_1$ and $b \in C_2$. Therefore, $b \in F_1 \inter C_1$ which is a contradiction with the induction hypothesis by which $F_1$ and $C_1$ are disjoint. Therefore, this case is not possible either.
	      \end{iteMize}

\noindent Therefore, $F \inter C = \emptyset$.	     

  Now, for any event $b \in \Sigma$, if $b \in F_1$ or $b \in F_2$,
  i.e., $b$ has infinitely many occurrences in $u_1$ or $u_2$, then
  $b$ has infinitely many occurrences in \smash{$u \in u_1 \parallel[A] u_2$}
  as well. Therefore, $u$ is fair in $F_1 \union F_2$.

  Let for some $a \in A$, $a \in C_1$ or $a \in C_2$ and let without
  loss of generality the former holds. Then, $a$ occurs only finitely
  many times in $u_1$ and, since $P_1$ and $P_2$ synchronise on $a$,
  $a$ occurs only finitely many times in $u_2$ and $u$ as
  well. Therefore, $u$ is co-fair in $a$ and, more generally, in $(C_1
  \cap A) \cup (C_2 \cap A)$. Now let $b \in (C_1 \backslash A) \cap
  (C_2 \backslash A)$. Therefore, $b \notin A$, $b \in C_1$ and $b \in
  C_2$. Therefore, since $b$ occurs only finitely often in both $u_1$
  and $u_2$, $b$ occurs only finitely often in $u$ as well. Therefore,
  $u$ is also co-fair in $(C_1 \backslash A) \cap (C_2 \backslash
  A)$. Hence, $u$ is co-fair in $(C_1 \cap A) \cup (C_2 \cap A) \cup
  ((C_1 \backslash A) \cap (C_2 \backslash A))$ and $(F_1 \cup F_2,
  (C_1 \cap A) \cup (C_2 \cap A) \cup ((C_1 \backslash A) \cap (C_2
  \backslash A)) \in \Phi(P_1 \parallel[A] P_2)$ by
  construction.\qed \end{iteMize}}
\end{iteMize}

\vspace{0.2cm}\noindent \textbf{Proposition~\ref{stronger}.} \textit{
For any structurally finite-state process $P$, if $\F(P) \neq
\emptyset$ then $\delta(P) = \false$.}
\vspace{0.2cm}

\proof (Sketch.) One shows by structural induction on the $\SFS$
process $P$ the stronger statement that if $\F(P) \neq \emptyset$ then
(i)~$\delta(P) = \false$, and (ii)~for any $(U,V) \in \F(P)$ and any 
$(F,C) \in \Phi(P)$, it is the case that $F \inter V \neq \emptyset$.

All cases are relatively straightforward. Note that, since $P \in
\SFS$, recursion does not need to be handled, as it falls within the
`sequential $\SFS$' case. It is worth pointing out that, in carrying
out the inductive proof, it turns out that it is never necessary to
take account of any information regarding either $U$ or $C$; they can
be ignored entirely.  \qed

\section{Case Study: an Abstracted Version of the Alternating Bit Protocol}\label{app_abp}

In this section, we briefly describe an abstracted version of a network communication protocol called the Alternating Bit Protocol. We use the abstracted version only to illustrate our concepts. For the experimental evaluation in \autoref{experiments} we use a modelling of the authentic protocol, the script for which can be found on the website associated with~\cite{Ros11}.

The process $\abssend$ (see Figures~\ref{fig_abp_abstract} and \ref{fig_abp_abstract_lts}) attempts to send messages to itself infinitely often. Those messages, however, have to go through an unreliable network $\absmedium$, which may do an arbitrary (possibly infinite) number of $\cherr$ events before delivering the message back to $\abssend$ in the form of an $\chout$ event. We impose a fairness constraint $\absfair$ on $\absmedium$, forcing it to do at most a single error before delivering the message correctly, i.e., we require that every $\cherr$ event be immediately followed by an $\chout$ event. We construct the system by putting the mutually-recursive processes $\abssend$ and $\absmedium$ in parallel with the process $\absfair$, synchronising  on the set of their shared events $\set{\cherr,\chout}$ and hiding the $\cherr$ event at the top. The resulting process $\system$ is livelock-free and is, in fact, equivalent to the process $B_1 = \chin \then \chout \then B_1$, which implements a single-slot buffer. 
%It is interesting to note, that if we hide on top level the set of events $\set{\cherr,\chout}$, instead of just $\set{\cherr}$, the resulting system is still livelock-free: every infinite execution of $\network \hide \set{\cherr,\chout}$ would contain infinitely many occurrences of the event $\chin$.

\begin{figure}[hb!]
\begin{center}
\boxed{
 \begin{tabular}{l}
   $\abssend = \chin \then \absmedium$ \\
   $\absmedium = \chout \then \abssend \extchoice \cherr \then \absmedium $ \\ \\
   $\absfair = \chout \then \absfair \extchoice \cherr \then \chout \then \absfair $ \\ \vspace{-6 pt}\\
   $\network = \abssend \parallel[\set{\cherr,\chout}] \absfair$ \\ \vspace{-6 pt} \\
   $\system = \network \hide \set{\cherr} $
 \end{tabular}
}
\end{center} 
\caption{ABP: an abstracted version.}\label{fig_abp_abstract}
\end{figure}

\begin{figure}[h]
\begin{center}
 %\beginpgfgraphicnamed{abstracted-abp}
% \input{abp.tex}
  \raisebox{0.75cm}{ $\Bigg($ }
   \beginpgfgraphicnamed{abp1}
    \input{abp1.tex}
   \endpgfgraphicnamed     
   \raisebox{0.75cm}{$\Large\parallel[\set{\small{\mathit{error},\mathit{out}}}]$}
   \beginpgfgraphicnamed{abp2}
   \hbox{\hspace{-0.2cm}\vspace{-1.4cm} \input{abp2.tex}}
    \endpgfgraphicnamed  
   \raisebox{0.75cm}{$\Bigg) \hide \set{ \mathit{ {\small error}}}$}
%  \endpgfgraphicnamed  
 \end{center}
 \caption{Abstracted ABP: transition systems.}\label{fig_abp_abstract_lts}
 \end{figure}

Using the systems of rules presented in \autoref{StaticLivelockAnalysis}, we calculate the sets of fair sets of $\abssend$, $\absfair$, $\network$ and $\system$ as follows (where the operator $\closure$ denotes upper closure on $\powerset{\powerset{\Sigma}}$ and $\Sigma = \set{\chin, \chout, \cherr}$):
\begin{align*}
\F(\abssend) \deq \closure \set{ \set{\chin,\cherr}, \set{\chout, \cherr} } \\
\F(\absfair) \deq \closure \set{ \set{\chout} } \\
\F(\network) \deq \closure \set{ \set{\chout},  \set{\chout,\cherr}, \set{\chin,\cherr}} \\
\F(\system)  \deq \closure \set{ \set{\chout} } 
\end{align*}
Therefore, $\system$ is livelock-free and any infinite trace of  $\system$ contains infinitely many occurrences of the event $\chout$.

An interesting weakness of the general framework is that it fails to establish the fact that $\system$ is also $\set{\chin}$-fair. Indeed, since $\system$ is equivalent to the process $B_1 = \chin \then \chout \then B_1$, any infinite trace of $\system$ should also contain infinitely many occurrences of $\chin$. Therefore, the process $\network \hide \set{\cherr, \chout}$, which is equivalent to the process $\mathit{IN} = \chin \then \mathit{IN}$, is livelock-free and $\set{\chin}$-fair. However, $\F(\network \hide \set{\cherr, \chout}) = \emptyset$ (thanks to the $\F$ rule for hiding) and therefore the general framework would mark $\network \hide \set{\cherr, \chout}$ as potentially divergent.

Let us now illustrate the precision of the system of rules for $\SFS$ processes by trying to establish that the process $\system = \network \hide \set{\cherr, \chout} = (\abssend \parallel[\set{\mathit{error},\mathit{out}}] \absfair)  \hide \set{\cherr, \chout}$ is livelock-free.

The processes $\abssend$ and $\absfair$ depicted in \autoref{fig_abp_abstract_lts} are both sequential $\SFS$ processes---for those we apply the algorithms described in \autoref{section_static_atomic_sfs} to conclude that $\delta(\abssend) = \delta(\absfair) = \false$ and, regarding the set of fair/co-fair pairs,
\begin{align*}
 \Phi(\abssend) \eq \set{ \hspace{0.1cm} (\set{\chin,\chout}, \set{\cherr}), \quad (\set{\cherr}, \set{\chin,\chout}), \quad (\set{\cherr,\chin,\chout}, \emptyset) \hspace{0.1cm} }, \\
 \Phi(\absfair) \eq \set{\hspace{0.1cm} (\set{\chout}, \set{\chin, \cherr}), \quad (\set{\cherr, \chout}, \set{\chin}) \hspace{0.1cm}}.
\end{align*}
Now let us consider the process $\network = \abssend \parallel[\set{\mathit{error},\mathit{out}}] \absfair$.

Since both $\abssend$ and $\absfair$ are livelock-free, there is no way of having a divergence in $\network$, which is confirmed by the rule $\delta(\network) = \delta(\abssend) \vee  \delta(\absfair) =  \false$.

Let us now have a look at the $\Phi$ rule for parallel
composition. Since each of the fair/co-fair pairs $(F,C)$ of
$\abssend$ and $\absfair$ have non-empty intersection with
the synchronisation set $A = \set{\cherr, \chout}$ of the parallel
composition, we conclude that:
\begin{enumerate}[(1)]
 \item  We can only use the first set-comprehension clause for assembling the fair/co-fair pairs of $\absmedium$. 
 \item Both $\abssend$ and $\absfair$ contribute infinite traces in
   any infinite trace $u$ of $\network$, \ie $u = u_1 \parallel[A]
   u_2$, where $u_1$ in $\inftraces{\abssend}$ and $u_2 \in
   \inftraces{\absfair}$.
\end{enumerate}

\noindent Intuitively, every infinite trace of $\absfair$, and in particular
$u_2$, contains infinitely many occurrences of $\chout$. Since
$\abssend$ and $\absfair$ synchronise on $\chout$, $u_1$ also
contains infinitely many occurrences of $\chout$. But in $u_1$,
$\chout$ occurs infinitely often precisely whenever $\chin$ occurs
infinitely often. Therefore, $u_1$, and hence also $u$, both contain
infinitely many occurrences of $\chin$. Therefore $u$ is fair in
$\chin$.

Formally, since both $u_1$ and $u_2$ are infinite, we need to consider every pair $((F_1, C_1),$ $(F_2, C_2))$ in the Cartesian product of $\Phi(\abssend)$ and $\Phi(\absfair)$, decide whether to discard it and, if not, figure out how to merge appropriately the pair of pairs into a single pair $(F,C)$.

One of the crucial observations is the following. For $a \in A = \set{\cherr, \chout}$, the number of occurrences of $a$ in $u_1$, $u_2$ and $u$ is the same. Therefore we can discard all those pairs $((F_1, C_1), (F_2, C_2))$ such that there is $a \in A$ with $a \in F_1 \inter C_2$ or $a \in C_1 \inter F_2$. This leaves us with only two pairs:
\begin{enumerate}
 \item $((\set{\chin,\chout}, \set{\cherr}), (\set{\chout}, \set{\chin, \cherr}))$, and 
 \item $((\set{\cherr,\chin,\chout}, \emptyset), (\set{\cherr, \chout}, \set{\chin}))$.
\end{enumerate}
The important question now is what do we do with the event $\chin$ which does not belong to the synchronisation set $A$. The reasoning we apply is that $u$ is fair in $\chin$ if at least one of $u_1$ and $u_2$ is fair in $\chin$, and $u$ is co-fair in $\chin$ if both $u_1$ and $u_2$ are co-fair in $\chin$. Then from the first pair we obtain $(F,C) = (\set{\chin,\chout}, \set{\cherr})$ and from the second pair we obtain $(F,C) = (\set{\cherr,\chin,\chout,},\emptyset)$. Hence we obtain the following final result for $\Phi(\network)$, which confirms that every infinite trace of $\network$ contains infinitely many occurrences of $\chin$:
\begin{align*}
 \Phi(\network) \eq \set{ \hspace{0.1cm} (\set{\chin,\chout}, \set{\cherr}), \quad (\set{\cherr,\chin,\chout}, \emptyset) \hspace{0.1cm} }
\end{align*}
Now the only thing that remains is to handle the hiding operator, \ie analyse $\system = \network \hide \set{\cherr, \chout} $. Since for all $(F,C) \in \Phi(\network)$, $F - \set{\cherr, \chout} \neq \emptyset$, $\delta(\system)$ $=$ $\false$, \ie we establish, as required, that $\system$ is livelock-free.  As a nice consequence $\Phi(\system) = \set{(\set{\chin}, \set{\cherr,\chout})}$ asserts that every infinite trace of $u$ contains infinitely many occurrences of $\chin$ and only finitely many occurrences of $\chout$ and $\cherr$.

\section{Symbolic Encoding}\label{app_encoding}

\let\oldoverline\overline
\renewcommand{\overline}[1]{#1}

In this section we focus on the details regarding the symbolic part of
our frameworks and algorithms.

In general, because we need to encode sets of sets of events, we use
one-hot Boolean encoding~\cite{logicDesign}, \ie for each $a \in
\Sigma^{\tau \tick}$ we employ a Boolean variable which is also
written $a$. The Boolean formula $a$ then encodes all sets of events
$\set{A \subseteq \Sigma^{\tau \tick} \st a \in A}$. For the $\SFS$
framework we use a single vector $\overline{y}$ of
$\modulus{\Sigma^{\tau \tick}}$ Boolean variables, whereas for the
general framework---the one described in
\autoref{StaticLivelockAnalysis}---we employ two copies: one vector
$\overline{x}$ for modelling the $U$ component and another
$\overline{y}$ for modelling the $V$ component (see Propositions
\ref{prop_non_exp}, \ref{prop_contr_sets}, and
\ref{prop_fair_sets}). In addition, we use auxiliary copies of
variables for constructing more complex expressions using quantifiers
and substitution. For those we use primed versions of $\overline{x}$
and $\overline{y}$.

It is important to note that SAT techniques enable us to find a
\emph{single} fair set or fair/co-fair pair of sets for a process. An
advantage of this approach is the efficiency of modern SAT
solvers. However, we need to introduce fresh vectors of variables for
each instance of (even the same) subprocess. This is required because
it might be necessary to generate different fair or fair/co-fair sets
for a given term, depending on the context in which it appears.

Using BDDs~\cite{Bryant} enables us to find \emph{all possible} fair
or fair/co-fair sets that the system of rules is capable of
detecting. Hence we do not need to duplicate subprocess encodings, but
we need to take care of variable orderings which can dramatically
influence the size of the resulting BDD. We use a variable ordering
similar to the ones proposed in~\cite{Par02} and adopted by the
probabilistic model checker
PRISM~\cite{HintonKNP06,KwiatkowskaNP11}. BDDs generally generate more
compact representations than SAT encodings due to their canonicity and
capacity to capture regularities.

\subsection{The $\SFS$ Framework}

\subsubsection{Computing Fair/Co-Fair Sets for Sequential $\SFS$ Processes}

Let $P$ be a sequential $\SFS$ process and let us suppose that we have
already established that $\delta(P) = \false$, \ie that $P$ is
livelock-free.  As described in \autoref{section_static_atomic_sfs},
we then generate a collection of fair/co-fair pairs of disjoint sets
$\Phi(P)$ = $\set{(F_1, C_1), \ldots,(F_k,C_k)} \subseteq
\mathcal{P}(\Sigma) \times \mathcal{P}(\Sigma)$ such that for every $1
\leq i \leq k$,
\begin{equation}\label{eq_prop_fair_cofair}
 (F_i,C_i) \in \Phi(P) \longleftrightarrow  \exists u \in \inftraces{P} \exst \mbox{$u$ is fair in $F_i$ and co-fair in $C_i$}.
\end{equation}
The computation of $\Phi(P)$ is carried out directly on the labelled
transition system $M_P$ associated with $P$ (in which unreachable
states have been excised). Let us fix $M_P = \langle S, \init,
\Sigma_P, \then \rangle$ and let us suppose that $P$ is a subcomponent
of a system with alphabet $\Sigma$.

\begin{algorithm}
\caption[Static analysis: computing fair/co-fair sets of events.]{Computing $\Phi(P)$}
\label{alg_atomic_fair_cofair}
\begin{algorithmic}[1]
 \STATE $\Phi(P) = \emptyset$
 \FOR{\textbf{every} non-empty set $L \subseteq \Sigma_P$}
	\STATE construct a labelled graph $G_{L}$ from $P$'s LTS
        (having pruned unreachable states)\\
by deleting all ($\Sigma-L$)-labelled transitions
	\IF{$G_{L}$ contains an SCC which comprises \emph{every} event in $L$}
	      \STATE include $(L, \Sigma-L)$ in $\Phi(P)$
	\ENDIF
 \ENDFOR
 \STATE return $\Phi(P)$
\end{algorithmic}
\end{algorithm}

For a particular non-empty $L \subseteq \Sigma_P$, deciding whether or
not to include $(L, \Sigma-L)$ in $\Phi(P)$ (lines 3--4,
Algorithm~\ref{alg_atomic_fair_cofair}) can be carried out in
PTIME. More specifically, after obtaining $G_{L}$, we can check
whether there exists $s \in S$ such that for every $a \in L$, there
exists a transition $\src \thena{a} \dest$, such that there are paths
from $s$ to $\src$ and from $\dest$ back to $s$, as illustrated
in \autoref{fig_scc_fair_cofair} for $L =
\set{\chin,\chout,\cherr}$. Note that such paths necessarily consist
entirely of events in $L \union \{ \tau \}$.

\begin{figure}[h]
 \begin{center}
  \beginpgfgraphicnamed{curly}
 \input{curly_fair.tex}
  \endpgfgraphicnamed 
\end{center}
\caption{Calculating fair/co-fair sets for sequential $\SFS$ processes.}\label{fig_scc_fair_cofair}
\end{figure}

\begin{figure}
\centering
\begin{minipage}{0.3 \linewidth}
\centering
\beginpgfgraphicnamed{matrix-ts}
\input{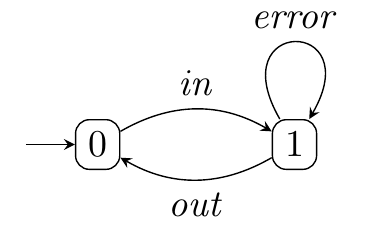}
\endpgfgraphicnamed

\end{minipage}
\begin{minipage}{0.3 \linewidth}
\centering
	 \[\mathsf{A} = 
	 \left[ \begin{array}{c c} 
		      \false & \chin \\[2ex] 
		      \chout & \cherr 
		\end{array}
	 \right]
	 \]

\end{minipage}
\vspace{0.5cm}
\begin{minipage}{0.7 \linewidth}
\centering

	\[ \gpath = \left[
	 \begin{array}{c c}
               (\chin \wedge \chout) & (\chin) \vee (\chin \wedge \cherr) \phantom{bb} \\[2ex] 
	       (\chout) \vee (\cherr \wedge \chout) & (\cherr) \vee (\chout \wedge \chin) \vee (\cherr \wedge \cherr) \;
	 \end{array}  
	 \right]
	\]

\end{minipage}
\caption[A symbolic representation of the adjacency matrix and the path matrix of a sequential $\SFS$ process]{A symbolic representation of the adjacency matrix and the path matrix of the process $\abssend$. The path matrix is computed using successive matrix multiplications.} \label{fig_symbolic_path_matrix}
\end{figure}

In fact, we can encode this symbolically for all possible subsets of
$\Sigma_P$ via the following Boolean formula:

\begin{equation}\label{formula_max_scc} 
  \maxscc = \bigvee_{s \in S} \Big\{   \bigwedge_{a \in \Sigma_P} \big[ \neg a \svee
	 \bigvee_{\src \thena{a} \dest} \big( \gpath(s, \src) \swedge \gpath(\dest, s) \big)   \big]  \Big\},
\end{equation}
\noindent
where:
\begin{enumerate}[(1)]
 \item For all $s,t \in S$, $\gpath(s,t)$ encodes all symbolic traces
   over $\Sigma_P$ from $s$ to $t$ of length at most $\modulus{S}$,
   \ie all symbolic traces of length at most the longest simple path
   in $M_P$. In order to compute $\gpath(s,t)$ for all $s,t \in S$
   simultaneously, we extend standard algorithms for computing the
   transitive closure of the adjacency matrix of the transition
   relation of $M_P$, such as Floyd-Warshall, iterative squaring, or
   successive adjacency-matrix multiplications.  Since the order of
   events on those traces is irrelevant to fairness and co-fairness,
   we do not employ symbolic state variables and use just a single
   copy of event variables to carry out the computation, as
   illustrated in \autoref{fig_symbolic_path_matrix}.  We note that in
   those algorithms we do not check whether we reach a fixed point in
   the computation. As a consequence, if using a SAT encoding, the
   resulting formulas may contain redundancies.
 \item The Boolean formula \eqref{formula_max_scc} contains an implicit iterator over all possible subsets $L$ of $\Sigma_P \union \set{\tau}$. In order to exclude the options of $L = \emptyset$ and $L = \set{\tau}$, we conjoin the formula with the restriction $\bigvee_{a \in \Sigma_P} a$.
 \item We need to also declare all infinite traces of $P$ as co-fair in $\Sigma - \Sigma_P$. To do so, we add another Boolean conjunct $\bigwedge_{a \in (\Sigma - \Sigma_P)} \neg a$.
\end{enumerate}
The Boolean encoding of $\Phi(P)$ is then as follows:
\begin{equation}\label{formula_phi_p}
 \Phi(P) = (\bigvee_{a \in \Sigma_P} a) \swedge (\bigwedge_{a \in (\Sigma - \Sigma_P)} \neg a) \swedge \maxscc.
\end{equation}

\paragraph{\textbf{The Key: PTIME Algorithms and Circuits.}}

As stated earlier, given a process $P$ and a non-empty set of events
$L \subseteq \Sigma_P$, deciding whether or not to include $(L,
\Sigma-L)$ in $\Phi(P)$ can be carried out in PTIME\@. Therefore, for
the particular $P$ and $L$, there exists a polynomial-size
variable-free Boolean circuit that outputs true if and only if the
pair $(L, \Sigma-L)$ is a fair/co-fair pair for $P$\footnote{This
  follows from the PTIME-hardness of \textsc{circuit value}.}.

Let us fix $P$ and let us observe, moreover, that the construction of
the variable-free circuit does not depend on the particular choice of
$L$ (see \eqref{formula_max_scc} and
\eqref{formula_phi_p}). Therefore, we can leave the $\Sigma_P$ input
gates of the circuit as Boolean
variables~\cite{papadimitriou2003computational}. What we obtain is a
compact circuit of size polynomial in the syntax of $P$ that encodes
the computation of $\Phi(P)$ once and for all possible inputs, i.e.,
for all exponentially many subsets of $\Sigma_P$. We remark that the
size of the circuit is polynomial in the size of $P$'s LTS, which in
turn is polynomial in $P$'s syntactic description, since we are
dealing with sequential $\SFS$ processes.

Since the circuit is of polynomial size, it can be turned into a
polynomial-size (equisatisfiable) Boolean formula using, \eg Tseitin's
encoding~\cite{Tse68}. The circuit can be also turned into a BDD, in
which case the size of the BDD could potentially blow up; however
this is usually not the case in practice.  Consequently, the
Boolean formula or the BDD encoding of $\Phi(P)$ can be plugged into
our compositional rules and be queried on demand when necessary, which
fits very nicely into our symbolic framework.

\subsubsection{Encoding Compositional Rules}

The encoding of the rules for computing the livelock flag $\delta(P)$ and the collections of fair/co-fair pairs $\Phi(P)$ of a compound $\SFS$ process $P$ (see Theorem~\ref{main2}) are given in Figures~\ref{enc_sat_phi} and \ref{enc_sat_delta} for Boolean formulas (\ie for SAT) and in Figures~\ref{enc_bdd_phi} and \ref{enc_bdd_delta} for BDDs.

\begin{figure}[h]
\centering 
\begin{boxedminipage}{0.95 \linewidth}
{\allowdisplaybreaks
 \begin{align*}
\Phi(a \then P)(\overline{y}) \deq \Phi(P)(\overline{y}) \\
\Phi(P_1 \oplus P_2)(y) \deq \Phi(P_1)(\overline{y'}) \swedge \Phi(P_2)(\overline{y''}) \swedge 
\big[ \bigwedge_{a \in \Sigma} a(y) \leftrightarrow a(y') \svee \bigwedge_{a \in \Sigma} a(y) \leftrightarrow a(y'')  \big] \ \
\mbox{if $\oplus \in \set{\intchoice,\extchoice,\semi}$} \\
\Phi(P_1 \smash{\parallel[A]} P_2)(\overline{y}) \deq \Phi(P_1)(\overline{y'}) \swedge \Phi(P_2)(\overline{y''}) \swedge \\
\pheq \big[ \{ \bigwedge_{a \in A} \neg a(y') \swedge \bigwedge_{a \in \Sigma} a(y) \leftrightarrow a(y')  \} \svee \\
\pheq \phantom{\big[ } \{ \bigwedge_{a \in A} \neg a(y'') \swedge \bigwedge_{a \in \Sigma} a(y) \leftrightarrow a(y'')  \} \svee \\ 
\pheq \phantom{\big[ } \{ \bigwedge_{a \in \Sigma}  a(y) \leftrightarrow ( a(y') \vee a(y'')) \swedge \bigwedge_{a \in A} \neg a(y) \leftrightarrow ( \neg a(y') \vee \neg a(y'')) \\
\pheq \phantom{\big[ \{ \bigwedge_{a \in \Sigma}  a(y) \leftrightarrow ( a(y') \vee a(y''))} \swedge \bigwedge_{a \in \Sigma \backslash A} \neg a(y) \leftrightarrow ( \neg a(y') \wedge \neg a(y'')) \}
\big] \\
\Phi(P \hide A)(\overline{y}) \deq \Phi(P)(\overline{y'}) \swedge  \bigwedge_{a \in \Sigma \backslash A} a(y) \leftrightarrow a(y') \swedge \bigwedge_{a \in A} \neg a(y)   \\
\Phi(P \renam{R})(\overline{y})  \deq\Phi(P)(\overline{y'}) \swedge \bigwedge_{a \in \Sigma} [ a(y') \rightarrow (\bigvee_{a \sprel b} b(y))] \swedge \bigwedge_{b \in \Sigma} [ (\bigwedge_{c \sprel b} \neg c(y')) \rightarrow \neg b(y) ] 
 \end{align*}
} 
\end{boxedminipage}
\caption{SAT encoding of $\Phi(P)$.}\label{enc_sat_phi}
\end{figure}

\begin{figure}[h]
\centering 
\begin{boxedminipage}{0.7 \linewidth}
{\allowdisplaybreaks
 \begin{align*}
\delta(P \hide A) \deq \delta(P) \svee \bigg(  \neg \big[ \Phi(P)(\overline{y}) \rightarrow (\bigvee_{b \in \Sigma \backslash A} b(y)) \big]  \mbox{ is SAT } \bigg)
 \end{align*}
} 
\end{boxedminipage}
\caption{SAT encoding of $\delta(P)$.}\label{enc_sat_delta}
\end{figure}

\begin{figure}[h]
\centering 
\begin{boxedminipage}{0.95 \linewidth}
{\allowdisplaybreaks
 \begin{align*}
\Phi(a \then P)(\overline{y}) \deq \Phi(P)(\overline{y}) \\
\Phi(P_1 \oplus P_2)(y) \deq   \Phi(P_1)(\overline{y}) \swedge \Phi(P_2)(\overline{y})  \ \
\mbox{if $\oplus \in \set{\intchoice,\extchoice,\semi}$} \\
\Phi(P_1 \smash{\parallel[A]} P_2)(\overline{y}) \deq \exists \overline{y'} \exists \overline{y''} \exst \Phi(P_1)(\overline{y'}) \swedge \Phi(P_2)(\overline{y''}) \swedge \\
\pheq \big[ \{ \bigwedge_{a \in A} \neg a(y') \swedge \bigwedge_{a \in \Sigma} a(y) \leftrightarrow a(y')  \} \svee \\
\pheq \phantom{\big[ } \{ \bigwedge_{a \in A} \neg a(y'') \swedge \bigwedge_{a \in \Sigma} a(y) \leftrightarrow a(y'')  \} \svee \\ 
 \pheq \phantom{\big[ } \{ \bigwedge_{a \in \Sigma}  a(y) \leftrightarrow ( a(y') \vee a(y'')) \swedge \bigwedge_{a \in A} \neg a(y) \leftrightarrow ( \neg a(y') \vee \neg a(y'')) \\ 
\pheq \phantom{\big[ \{ \bigwedge_{a \in \Sigma}  a(y) \leftrightarrow ( a(y') \vee a(y''))}
\swedge \bigwedge_{a \in \Sigma \backslash A} \neg a(y) \leftrightarrow ( \neg a(y') \wedge \neg a(y'')) \}
\big] \vspace{-6 pt}\\
\Phi(P \hide A)(\overline{y}) \deq \big[ \exists \overline{y_A} \exst  \Phi(P)(\overline{y}) \big] \swedge \bigwedge_{a \in A} \neg a(y)   \\
\Phi(P \renam{R})(\overline{y})  \deq  \exists \overline{y'} \exst
\Phi(P)(\overline{y'}) \swedge \bigwedge_{a \in \Sigma} [ a(y') \rightarrow (\bigvee_{a \sprel b} b(y))] \swedge \bigwedge_{b \in \Sigma} [ (\bigwedge_{c \sprel b} \neg c(y')) \rightarrow \neg b(y) ] 
 \end{align*}
}
\end{boxedminipage} 
\caption{BDD encoding of $\Phi(P)$.} \label{enc_bdd_phi}
\end{figure}

\begin{figure}[h]
\centering
\begin{boxedminipage}{0.7 \linewidth} 
{\allowdisplaybreaks
 \begin{align*}
\delta(P \hide A) \deq \delta(P) \svee \bigg(   \big[ \Phi(P)(\overline{y}) \rightarrow (\bigvee_{b \in \Sigma \backslash A} b(y)) \big]  \mbox{ is not valid } \bigg)
 \end{align*}
} 
\end{boxedminipage}
\caption{BDD encoding of $\delta(P)$.}\label{enc_bdd_delta}%\vspace{2 cm}
\end{figure}

\newpage
\subsection{The General Framework}

The BDD and SAT encodings of the rules for computing the nonexpansive,
guard, contractive, and fair sets of CSP terms (see
Theorems~\ref{prop_non_exp}, \ref{prop_guard_sets},
\ref{prop_contr_sets}, and \ref{prop_fair_sets}) are formalised
similarly to the ones for the structurally finite-state processes. We
illustrate the scheme and the use of two vectors of event variables by
providing the BDD encoding of the rules for computing nonexpansive
sets in \autoref{enc_bdd_nonexp}.

In the encoding, the vectors of Boolean variables $x$ and $y$ model,
respectively, the $U$ and $V$ components of the pairs of sets of
events. To understand the meaning of the encoding operators $\ucl,
\dcl$, and $\udcl$, suppose the formula $\varphi(\ol{x}, \ol{y})$
encodes the set of pairs of sets of events $A = \{ (U,V) \st \ldots
\ \}$ and the formula $\psi (\ol{y})$ encodes the set of sets of
events $B = \{ V \st \ldots \ \}$. Then the formulas $\ucl(A)(x,y)$,
$\udcl(A)(x,y)$, $\ucl(B)(y)$, and $\dcl(B)(y)$ encode, respectively,
the sets $\{ (U, V) \st (U, V') \in A \land V' \subseteq V \}$, $\{
(U, V) \st (U', V') \in A \land U \subseteq U' \land V' \subseteq V
\}$, $ \{ V \st V' \in B \land V' \subseteq V \}$ and $ \{ V \st V'
\in B \land V \subseteq V' \}$:
\begin{align*}
 \ucl(A)(\ol{x}, \ol{y}) \deq \exists \ol{y'} \centerdot \varphi(\ol{x}, \ol{y'}) \wedge \bigwedge_i (y'_i \rightarrow y_i ) \\
 \udcl(A)(\ol{x}, \ol{y}) \deq \exists \ol{x'} \ol{y'} \centerdot \varphi(\ol{x'}, \ol{y'})
	\wedge \bigwedge_i (x_i \rightarrow x_i' ) \wedge \bigwedge_i (y'_i \rightarrow y_i ) \\
 \ucl(B)(\ol{y}) \deq \exists \ol{y'} \centerdot \psi(\ol{y'}) \wedge \bigwedge_i (y'_i \rightarrow y_i ) \\
 \dcl(B)(\ol{y}) \deq \exists \ol{y'} \centerdot \psi(\ol{y'}) \wedge \bigwedge_i (y_i \rightarrow y'_i ) 
\end{align*}

\begin{figure}[h]
\centering 
\begin{boxedminipage}{0.95 \linewidth}
{\allowdisplaybreaks
 \begin{align*}
 \N_X(P)(x,y) \deq \true \ \ 
         \mbox{\textbf{whenever $X$ is not free in $P$; otherwise:}} \\[1ex]
 \N_X(a \then P)(x, y) \deq \N_X(P)(x,y)  \\
 \N_X(P_1 \oplus P_2)(x,y) \deq \N_X(P_1)(x,y) \swedge \N_X(P_2)(x,y) \ \
         \mbox{\textbf{if $\oplus \in \set{\intchoice,\extchoice, \semi, \parallel[A]}$}} \\
 \N_X(P \hide A)(x,y) \deq \ucl(\N_X(P)(x,y) \swedge \chi(\{ V \st V \subseteq \Sigma - A\})(y)) \\
 	    \deq \ucl(\N_X(P)(x,y) \swedge \dcl (\Sigma - A)(y)) \\
 \N_X(P\renam{R})(x,y) \deq \ucl(  \exists y' \exst \N_X(P)( x, y') \swedge \rho(y',y)) \\
 \N_X(X)(x,y) \deq \bigwedge_i (x_i \rightarrow y_i) \\
 \N_X(\mu Y \centerdot P)(x,y) \deq \udcl(\N_X(P)(x,y) \swedge
 	\exists x' \centerdot ( \N_Y(P)(x', y) \swedge \bigwedge_i (x'_i \leftrightarrow y_i)))  \ \
                                        \mbox{\textbf{if $Y \neq X$}} \enspace
 \end{align*}
}
\end{boxedminipage} 
\caption{BDD encoding of $\N_X(P)$.} \label{enc_bdd_nonexp}
\end{figure}
    \insert\copyins{\hsize.57\textwidth
\vbox to 0pt{\vskip12 pt%
      \fontsize{6}{7 pt}\normalfont\upshape
      \everypar{}%
      \noindent\fontencoding{T1}%
  \textsf{\color{black}This work is licensed under the Creative Commons
  Attribution-NoDerivs License. To view a copy of this license, visit
  \texttt{http://creativecommons.org/licenses/by-nd/2.0/} or send a
  letter to Creative Commons, 171 Second St, Suite 300, San Francisco,
    CA 94105, USA, or Eisenacher Strasse 2, 10777 Berlin, Germany}\vss}
      \par
      \kern 0 pt}%

\let\overline\oldoverline
\color{white}

\end{document}